\documentclass[titlepage]{article}
\usepackage{graphicx}
\usepackage{amsmath}
\usepackage{amssymb,amsfonts,amsthm}
\usepackage{bm,textcomp}
\usepackage{enumitem}
\usepackage{color}
\usepackage{booktabs}
\usepackage{todonotes}
\usepackage{siunitx}
\usepackage[labelfont=bf]{caption}
\usepackage{epstopdf}
\usepackage[utf8]{inputenc}
\usepackage{tikz}
\usetikzlibrary{shapes,arrows,automata,arrows.meta,decorations.pathreplacing}
\usepackage{circuitikz}
\usepackage{rotating}
\usepackage{mathtools}
\usepackage{float}
\usepackage{xr}
\usepackage[a4paper,top=3cm]{geometry}
\usepackage[numbers]{natbib}
\usepackage{titlesec}

\usepackage[pdftex,bookmarks]{hyperref}

\hypersetup{colorlinks=true, linkcolor=blue, citecolor=blue, urlcolor=blue}

\setlist{itemsep=0pt,leftmargin=*}

\titlespacing{\section}{0pt}{0pt}{3pt}
\titlespacing{\subsection}{0pt}{0pt}{0pt}
\titlespacing{\subsubsection}{0pt}{0pt}{-5pt}

\DeclareMathOperator*{\rlh}{\rightleftharpoons}

\def\imagetop#1{\vtop{\null\hbox{#1}}}

\DeclareMathOperator*{\argmin}{arg\,min}

\makeatletter
\def\XR@[#1]#2{{%
		\makeatletter
		\def\XR@prefix{#1}%
		\XR@next"#2.aux"\relax\\}} %added quotes
\makeatother 

\begin{document}
\title{\textbf{Bond graph modelling of the cardiac action potential: Implications for drift and non-unique steady states}}

\author{Michael Pan$^1$, Peter J. Gawthrop$^1$, Kenneth Tran$^2$, Joseph Cursons$^{3,4}$, \\ Edmund J. Crampin$^{1,5,6,*}$}

\date{$^1$Systems Biology Laboratory, School of Mathematics and Statistics, and Department of Biomedical Engineering, Melbourne School of Engineering, University of Melbourne, Parkville, Victoria 3010, Australia \\[0.3cm]
	$^2$Auckland Bioengineering Institute, University of Auckland \\[0.3cm]
	$^3$Bioinformatics Division, Walter and Eliza Hall Institute of Medical Research, Parkville, Victoria 3052, Australia \\[0.3cm]
	$^4$Department of Medical Biology, School of Medicine, University of Melbourne, Parkville, Victoria 3010, Australia  \\[0.3cm]
	$^5$ARC Centre of Excellence in Convergent Bio-Nano Science and Technology, Melbourne School of Engineering, University of Melbourne, Parkville, Victoria 3010, Australia \\[0.3cm]
	$^6$School of Medicine, University of Melbourne, Parkville, Victoria 3010, Australia \\[0.3cm]
	*Corresponding author. Email: edmund.crampin@unimelb.edu.au}
\maketitle

\textbf{Abstract}\\
Mathematical models of cardiac action potentials have become increasingly important in the study of heart disease and pharmacology, but concerns linger over their robustness during long periods of simulation, in particular due to issues such as model drift and non-unique steady states. Previous studies have linked these to violation of conservation laws, but only explored those issues with respect to charge conservation in specific models. Here, we propose a general and systematic method of identifying conservation laws hidden in models of cardiac electrophysiology by using bond graphs, and develop a bond graph model of the cardiac action potential to study long-term behaviour. Bond graphs provide an explicit energy-based framework for modelling physical systems, which makes them well-suited for examining conservation within electrophysiological models. We find that the charge conservation laws derived in previous studies are examples of the more general concept of a ``conserved moiety''. Conserved moieties explain model drift and non-unique steady states, generalising the results from previous studies. The bond graph approach provides a rigorous method to check for drift and non-unique steady states in a wide range of cardiac action potential models, and can be extended to examine behaviours of other excitable systems.

\section{Introduction}
Models of the cardiac action potential have been developed to study cardiac diseases such as arrhythmia \cite{luo_dynamic_1994,luo_dynamic_1994-1,faber_action_2000}, ischaemia \cite{terkildsen_balance_2007} and acidosis \cite{crampin_dynamic_2006}. Increasing model complexity has led to concerns over the occurrence of drift and non-unique steady states \cite{guan_discussion_1997,fraser_quantitative_2007,kneller_time-dependent_2002}, particularly for extensions of the DiFrancesco and Noble \cite{difrancesco_model_1985} and Luo-Rudy \cite{luo_dynamic_1994,faber_action_2000} models. While solutions to these issues have been proposed using conservation principles \cite{hund_ionic_2001,livshitz_uniqueness_2009}, they have not been universally applied for more recent models, many of which still use nonconservative stimulus currents that predispose them to drift \cite{aslanidi_mechanisms_2009,carro_human_2011,grandi_novel_2010}. More recently, the Food and Drug Administration (FDA) has initiated plans to use cardiac action potential models to assess potential drug side-effects on cardiac instability through the human \textit{ether-\`{a}-go-go}-related gene (hERG) K\textsuperscript{+} channel. Thus, with an increasing emphasis on model robustness and accuracy, there is a renewed incentive to resolve the issues of drift and non-unique steady states \cite{sager_rechanneling_2014,colatsky_comprehensive_2016}.

Drift is the failure of a model to reach a consistent limit cycle when simulated over long periods, and is often caused by a nonconservative stimulus containing current with no charge carrier \cite{guan_discussion_1997,hund_ionic_2001}. Hund \textit{et al.} \cite{hund_ionic_2001} derived a charge conservation law, and found that nonconservative stimulus currents violate this conservation law, hence they proposed K\textsuperscript{+} ions as the current charge carrier to resolve this. A related issue in many models where drift has been resolved is that steady state limit cycles under constant pacing depend upon the initial conditions and are therefore non-unique \cite{fraser_quantitative_2007,hund_ionic_2001,kneller_time-dependent_2002}. Thus, depending on the initial conditions, the same model may lead to different conclusions. Like drift, authors have suggested that charge conservation can constrain initial conditions such that they lead to the same steady state \cite{hund_ionic_2001,kneller_time-dependent_2002,livshitz_uniqueness_2009}.

While the studies by Hund \textit{et al.} \cite{hund_ionic_2001} and Livshitz and Rudy \cite{livshitz_uniqueness_2009} suggest measures to eliminate drift and attain a unique steady state by using conservation laws, their analyses are limited in their scope and not a comprehensive solution for all models. Because existing studies \cite{hund_ionic_2001,livshitz_uniqueness_2009} explore charge conservation only in specific models, and the conservation laws were derived from physical intuition rather than a principled mathematical approach, it is difficult to generalise their findings to other models. Furthermore, because these studies focus only on conservation of charge, they may miss other conservation laws relevant for long-term behaviour, such as those corresponding to ions, ion channels and buffers. A general approach is, therefore, desirable to deal with the issues of drift and steady states in a more systematic manner and for a broader range of models.

To facilitate a general approach, we propose the use of bond graphs which explicitly model energy transfer across physical systems to ensure compliance with conservation principles. Bond graphs were initially invented to model hydroelectric systems \cite{paynter_analysis_1961} and they have subsequently been extended to model chemical \cite{borutzky_advances_1995}, biochemical \cite{oster_network_1973,gawthrop_energy-based_2014} and electrochemical systems \cite{gawthrop_bond_2017-1}. As with all physical systems, biological processes must obey the fundamental principles of physics and thermodynamics \cite{omholt_human_2016}, therefore bond graphs are well-suited for constraining models of biological systems to physically plausible solutions \cite{gawthrop_hierarchical_2015}, and also for inferring the energetic cost of biological processes \cite{gawthrop_bond_2017-1,gawthrop_bond_2017,gawthrop_hierarchical_2015,gawthrop_energy-based_2017}. Because the bond graph representation emphasises analogies between different physical domains, electrophysiological systems can be analysed as an analogous biochemical system with a stoichiometric matrix that describes the stoichiometry of each reaction within its columns \cite{beard_energy_2002,beard_thermodynamic_2004,gawthrop_energy-based_2014,van_der_schaft_mathematical_2013}. In this context, the ``conservation principle'' described in earlier studies is an example of the more general principle of a conserved moiety in metabolic and bond graph analysis \cite{haraldsdottir_identification_2016,gawthrop_hierarchical_2015}.

In this study, we develop a bond graph model of the cardiac action potential and outline a general approach to study the effects of conserved moieties on drift and steady-state behaviour. Our bond graph model simulates physiological action potentials, and because bond graphs are energy-based this easily provides an estimate of the energetic cost (in Joules) of the cardiac action potential. Our analysis reveals conservation of charge as one of the conserved moieties of our model, along with other conserved moieties corresponding to ions, channels, transporters and buffers. We observed that our model solution was subject to drift when the stimulus current violated any conservation laws corresponding to the conserved moieties, and that changes to the initial conditions led to different steady states if the value of any conserved moiety was changed. To demonstrate that our approach is general, we analyse variants of our bond graph model where different ions have been fixed at a constant concentration (corresponding to ``chemostats''). It should be noted that fixing an ion concentration can change the conserved moieties of a system, therefore influencing a model's susceptibility to drift and non-unique steady states. The bond graph approach is a useful and general method to identify and interpret conservation principles, and it can link conserved moieties to individual steady states. We build upon existing reports \cite{hund_ionic_2001,livshitz_uniqueness_2009} to propose solutions for drift and non-unique steady states which work for all cardiac action potential models that can be represented using bond graphs.

\section{Methods}
\label{sec:methods}
\subsection{Model components}
\begin{figure}
	\centering
	\includegraphics[width=0.9\linewidth]{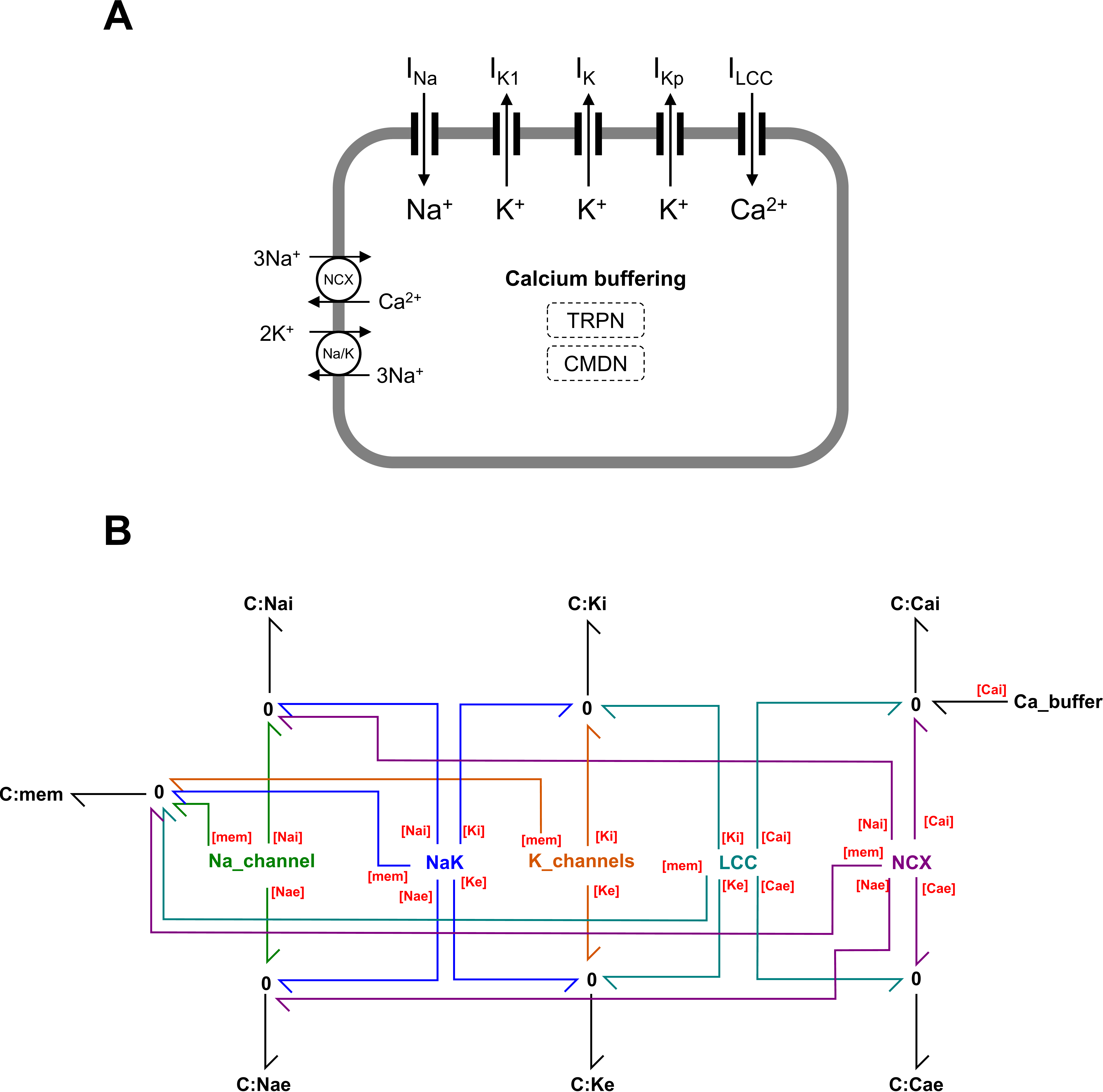}
	\caption{Action potential model. \textbf{(A)} Cell schematic; \textbf{(B)} Overall bond graph structure. The bond graph modules Na{\_}channel, NaK, K{\_}channels, LCC, NCX and Ca{\_}buffer contain more detailed aspects of the bond graph structure which are described further in the Supporting Material. Coloured bonds link bond graph modules to the appropriate chemical species. Definitions: $I_\mathrm{Na}$, sodium current; $I_\mathrm{K1}$, time-independent K\textsuperscript{+} current; $I_\mathrm{K}$, time-dependent K\textsuperscript{+} current; $I_\mathrm{Kp}$, plateau K\textsuperscript{+} current; $I_\mathrm{LCC}$, L-type Ca\textsuperscript{2+} current; NCX, Na\textsuperscript{+}-Ca\textsuperscript{2+} exchanger; Na/K, Na$^+$/K$^+$ ATPase; TRPN, troponin; CMDN, calmodulin.}
	\label{fig:cardiac_AP_model}
\end{figure}

To study the issues of drift and non-unique steady-states, we built a bond graph model of the cardiac action potential, with the minimal number of channels and pumps required to simulate a physiological action potential, and maintain ionic concentrations over long periods of simulation. Accordingly, our model was based on the Luo-Rudy 1994 dynamic model \cite{luo_dynamic_1994}, although it is possible to use other models and/or model more sub-cellular processes. Model components are shown in \autoref{fig:cardiac_AP_model}A, together with the overall bond graph structure (\autoref{fig:cardiac_AP_model}B). Ion channels and Ca\textsuperscript{2+} buffering components were based upon their representations in Luo and Rudy \cite{luo_dynamic_1994}. The L-type Ca\textsuperscript{+} channel in the Luo-Rudy model is permeable to calcium, sodium and potassium, but we neglected its sodium conductance as this has a relatively small contribution to the action potential. The Na$^+$/K$^+$ ATPase model was based on the model by Terkildsen \textit{et al.} \cite{terkildsen_balance_2007}, with modifications suggested by Pan \textit{et al.} \cite{pan_cardiac_2017} to allow conversion into a bond graph model. The equation for the Na\textsuperscript{+}-Ca\textsuperscript{2+} exchanger (NCX) current in Luo and Rudy did not have an obvious correspondence to a bond graph structure, thus we modelled this component using a simple bond graph module that was fitted to experimental data \cite{kimura_identification_1987,beuckelmann_sodium-calcium_1989}. Further detail on the modelling of each component is given in the Supporting Material.

\subsection{Bond graph modelling}
Here we briefly outline bond graph components as used in electrophysiological modelling. For a more comprehensive introduction, the texts by Gawthrop and Smith \cite{gawthrop_metamodelling:_1996} and Borutzky \cite{borutzky_bond_2010} provide detailed descriptions of bond graph theory, and Gawthrop and Bevan \cite{gawthrop_bond-graph_2007} provide a short tutorial for engineers. Theory for bond graph modelling of biochemical systems can be found in \cite{oster_network_1973,gawthrop_energy-based_2014,gawthrop_hierarchical_2015,gawthrop_bond-graph_2017}.

Bond graphs consist of components (representing physical objects and processes), bonds (representing the transfer of energy), and junctions (representing network structure). Each bond carries two variables: an effort $e$ and a flow $f$, such that their product determines the power of the bond (i.e.~$p=ef$). Thus bond graphs explicitly account for energy transfer, and are thermodynamically consistent. Because effort and flow are generalised variables, they can represent quantities from a variety of physical systems, including mechanical ($e=\text{force}$ [N], $f=\text{velocity}$ [m/s]), electrical ($e=\text{voltage}$ [V], $f=\text{current}$ [A]) and hydraulic systems ($e=\text{pressure}$ [Pa], $f=\text{volumetric flow rate}$ [$\si{m^3/s}$]) \cite{borutzky_bond_2010}.

The network structure of a bond graph is specified by 0 and 1 junctions. The 0 (or effort) junctions specify that efforts of all connected bonds are equal, and thus to ensure conservation of energy through this junction, the flows of the bonds must sum to zero. In the electrical and hydraulic domains, 0 junctions represent parallel connections, whereas they represent series connections in the mechanical domain. By a similar principle, 1 (or flow) junctions specify that the flows of all connected bonds are equal, ensuring that their efforts sum to zero. Thus, 1 junctions correspond to series connections in the electrical and hydraulic domains, and parallel connections in the mechanical domain.

\begin{figure}
\centering
\begin{turn}{90}
\begin{minipage}{0.7\paperheight}
\includegraphics[width=\linewidth]{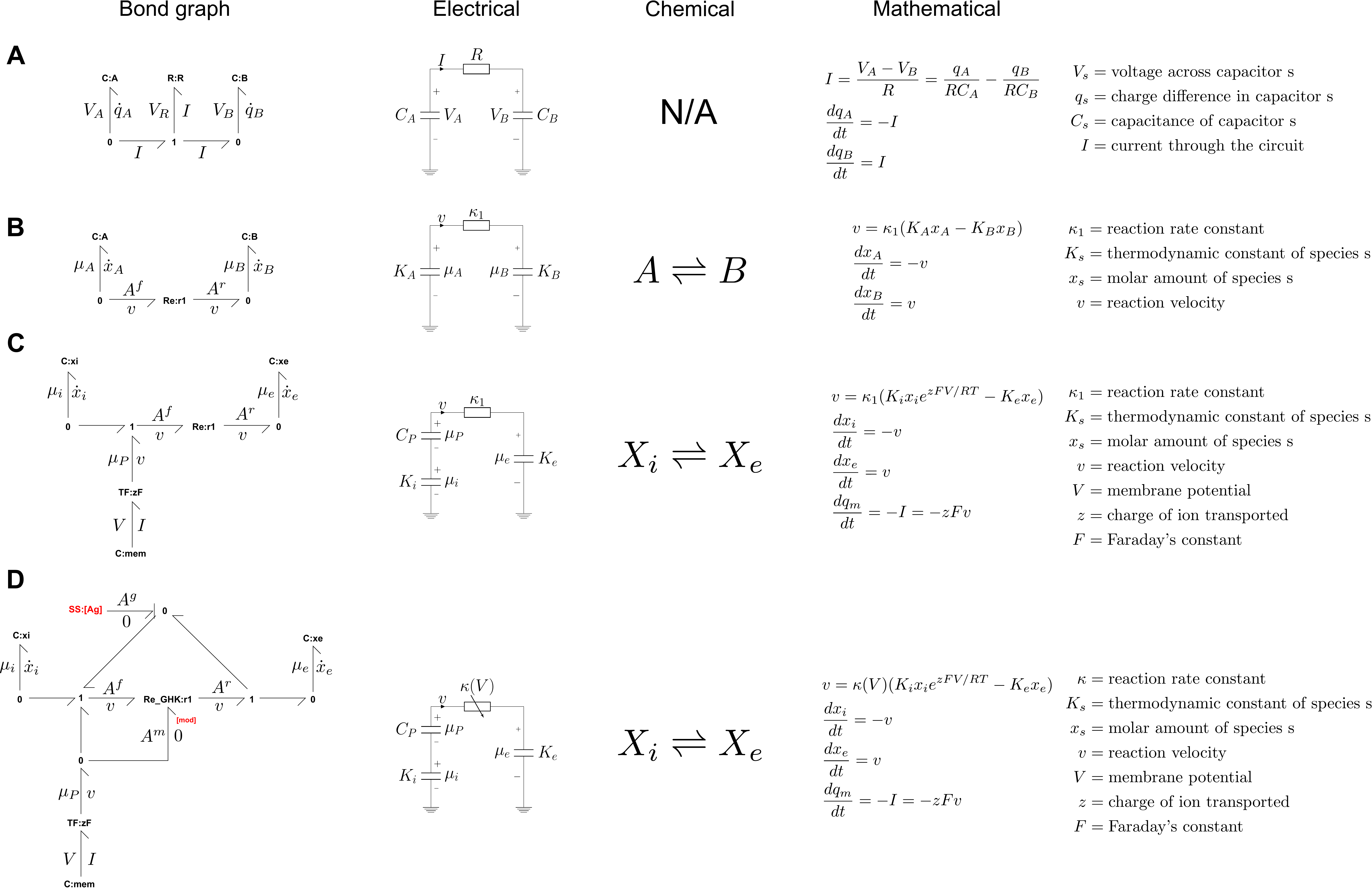}
\caption{Conceptual representations of physical systems. \textbf{(A)} A bond graph for the illustrated simple electric circuit with two capacitors and a resistor in series. \textbf{(B)} A bond graph analogous to the electric circuit in (A) can also represent the chemical reaction $A \rlh B$. \textbf{(C)} Bond graphs can also model interaction of components in both the chemical and physical domains, such as the transport of an ion across a membrane. \textbf{(D)} Transport of an ion across a membrane through an ion channel involves gating which modulates the rate of reaction. Thus the ion channel is analogous to a potentiometer.}
\label{fig:representations}
\end{minipage}
\end{turn}
\end{figure}

To illustrate the use of a bond graph for electric circuit analysis, we consider the electric circuit where two capacitors are connected to a resistor in series (\autoref{fig:representations}A). All components are linear, described by the equations:
\begin{align}
	V_A &= \frac{q_A}{C_A} \qquad \text{(capacitor)} \\
	V_B &= \frac{q_B}{C_B} \qquad \text{(capacitor)} \\
	I &= \frac{V_R}{R} \qquad \text{(resistor)}
\end{align}
The 1 junction enforces Kirchhoff's voltage law, such that:
\begin{align}
	V_R = V_A - V_B
\end{align}
Combining these equations gives rise to a system of first-order differential equations:
\begin{align}
	\frac{dq_A}{dt} &= -I = \frac{V_B - V_A}{R} = \frac{q_B}{RC_B} - \frac{q_A}{RC_A} \\
	\frac{dq_B}{dt} &= I = \frac{V_A - V_B}{R} = \frac{q_A}{RC_A} - \frac{q_B}{RC_B}
\end{align}

More recently, bond graphs have been extended to model biochemical systems \cite{oster_network_1973,gawthrop_energy-based_2014} where the chemical potential $\mu$ [$\si{J/mol}$] is the effort variable, and molar flow rate $v$ [$\si{mol/s}$] is the flow variable. Each chemical species is represented as a capacitor. However, in contrast to the electrical domain, the constitutive equation for the capacitor representing each species is nonlinear:
\begin{align}
	\mu = RT\ln(Kx)
	\label{eq:cp}
\end{align}
where $x$ [mol] is the molar amount of the species, $K$ [$\si{mol^{-1}}$] is a species thermodynamic constant, $R=8.314\ \si{J\cdot  mol^{-1} \cdot K^{-1}}$ is the gas constant and $T$ is the absolute temperature of the system. Reactions are modelled as two-port resistors using the Marcelin-de Donder equation as the constitutive equation:
\begin{align}
	v = \kappa (e^{A^f/RT} - e^{A^r/RT})
	\label{eq:MdD}
\end{align}
where $\kappa$ [$\si{mol/s}$] is a reaction rate constant and $A^f$ [$\si{J/mol}$] and $A^r$ [$\si{J/mol}$] are the forward and reverse affinities, respectively. The two affinities represent the potential energies present in the reactants and products, and the reaction proceeds in the direction of decreasing potential. As illustrated by the example in \autoref{fig:representations}B, the reaction $A \rlh B$ has a physical analogy to \autoref{fig:representations}A, with the same equivalent electric circuit. By using the constitutive equations in Eqs. \ref{eq:cp} and \ref{eq:MdD}, the reaction velocity for the bond graph model follows mass-action kinetics:
\begin{align}
	v = \kappa_1 (e^{A^f/RT} - e^{A^r/RT})
	= \kappa_1 (e^{\mu_a/RT} - e^{\mu_b/RT})
	= \kappa_1 (K_a x_a - K_b x_b)
	= k^+ x_a - k^- x_b
\end{align}
where the forward and reverse rate constants are $k^+ = \kappa_1 K_a$ and $k^- = \kappa_1 K_b$. For more general chemical reaction networks, 1 junctions describe the presence of multiple reactants or products in a single reaction, whereas 0 junctions describe the involvement of a single species in multiple reactions \cite{gawthrop_energy-based_2014}. For some models, we may wish to keep the amount $x$ of a species constant and this is achieved by defining the species as a ``chemostat'' \cite{polettini_irreversible_2014}. Because chemostats can be interpreted as an external flow that balances internal flows, they require energy to be pumped into (or out of) the system \cite{gawthrop_hierarchical_2015}.

The bond graph framework for biochemistry can be extended to electrochemical systems \cite{gawthrop_bond_2017-1} as demonstrated in \autoref{fig:representations}C, which models the transport of a positively charged species $X$ across a membrane. It should be noted that chemical species are described with C components that have a logarithmic association, whereas the C component corresponding to the (electric) membrane potential has a linear constitutive relationship. A transformer (TF) is used to convert the membrane voltage into an equivalent chemical potential through Faraday's constant ${F = 96485\ \si{C/mol}}$, such that:
\begin{align}
	\mu_P &= FV \\
	I &= Fv
\end{align}
Thus, the reaction velocity is:
\begin{align}
	v = \kappa_1 (e^{A^f/RT} - e^{A^r/RT})
	= \kappa_1 (e^{(\mu_i + \mu_P)/RT} - e^{\mu_e/RT})
	= \kappa_1 (K_i x_i e^{zFV/RT} - K_e x_e)
\end{align}
By setting $v=0$ the familiar Nernst equation can be derived \cite{gawthrop_bond_2017-1}. However, where electrical circuit representations of the membrane Nernst potential use voltage sources, the bond graph approach necessarily accounts for possible changes in ionic concentrations, and thus this ``voltage source'' is split into two capacitors that provide an equivalent voltage difference.

We chose to represent ion channels such that conductance was modulated by membrane voltage, both directly and indirectly through gating processes. A bond graph representation for this relationship is given in \autoref{fig:representations}D. As shown, this model has the same electrical representation as \autoref{fig:representations}C however it uses a variable resistor. The bond graph representation contains the same states, with C:xi, C:xe, and C:mem (with a transformer) connected through 0 junctions. In this case however, the Re components that describe the constitutive relation have been changed, such that Re{\_}GHK:r1 is connected to an additional effort that modulates its velocity, and the gating affinity $A^g$ is added to both the forward and reverse affinities to describe changes in permeability due to gating. Further detail on modelling ion channels using bond graphs is given in the Supporting Material.

\subsection{Modelling approach}
\label{sec:modelling_approach}

Because bond graphs constrain the equations of a model to ensure thermodynamic consistency, many existing models do not have a direct bond graph representation \cite{gawthrop_hierarchical_2015}. For the example here, equations representing ion channels in the Luo-Rudy model could not be directly translated into a bond graph model due to difficulties with simultaneously modelling open-channel currents and channel gating, and due to thermodynamic inconsistencies in the time-dependent K\textsuperscript{+} and L-type Ca\textsuperscript{2+} channels (see Supporting Material). Therefore rather than attempting to reproduce the Luo-Rudy equations exactly, we built a bond graph structure as implied by the equations in Luo and Rudy model, and chose parameters of our bond graph model to fit aspects of the Luo-Rudy model as closely as possible, specifically the current-voltage (I-V) curves and gating parameters. For all other components conversion into a bond graph model was more straightforward, and we used the methods of Gawthrop \textit{et al.} \cite{gawthrop_hierarchical_2015}. Further information on the bond graph model, and parameter identification is given in the Supporting Material.

\subsection{Finding conserved moieties}
Within a biochemical model, conserved moieties are chemical structures that are neither created, removed nor broken down. A common example in energy-dependent metabolic networks is the adenosine moiety found in AMP, ADP and ATP \cite{haraldsdottir_identification_2016,gawthrop_hierarchical_2015}. Mass balance specifies that the total amount of each conserved moiety remains constant, and if information on the molecular structure of each species of a reaction network is available, these conservation laws can be derived by counting the number of moieties across all species \cite{haraldsdottir_identification_2016}. In practice, many models do not contain this structural information and this approach cannot be used, however the conservation laws still hold. Here we outline a method to find conserved moieties using stoichiometric information rather than chemical structures.

Models of cardiac electrophysiology can be represented by the differential equation
\begin{align}
	\dot{X} = NV
\end{align}
where $X(t)$ is a vector of each state (such as species, or charge difference across a membrane), $N$ is the stoichiometric matrix \cite{beard_energy_2002,beard_thermodynamic_2004,gawthrop_energy-based_2014,van_der_schaft_mathematical_2013}, and $V$ is a vector of fluxes (such as reaction velocities, or ion channel currents) \cite{gawthrop_energy-based_2014,liebermeister_modular_2010,beard_energy_2002}. If the model contains chemostats, the entries of $X$, and rows of $N$ corresponding to the chemostats are deleted prior to performing the above analysis \cite{polettini_irreversible_2014}. Using results from biochemical systems \cite{gawthrop_energy-based_2014}, if $g$ is a row vector in the left nullspace of $N$, i.e.~$gN=0$, then
\begin{align}
	g\dot{X} = gNV = 0
\end{align}
Therefore the linear combination $gX$ is constant for the duration of the simulation. We call the linear combination of species represented by $g$ a conserved moiety. The space of all conserved moieties can be described by a left nullspace matrix $G$, whereby linear combinations of the rows of $G$ give all possible conserved moieties $g$ \cite{palsson_systems_2006,klipp_systems_2009}. Advantages of using the left nullspace are that it does not require information on chemical structures and it accounts for all conservation laws. The left nullspace matrix for any given system is generally not unique, however there are many well-established techniques for calculating nullspace matrices \cite{anton_elementary_2014}, specialised approaches for finding meaningful conserved moieties in biochemical networks \cite{schuster_what_1995,schuster_determining_1991,haraldsdottir_identification_2016} as well as methods for finding conserved moieties from the junction structure of a bond graph \cite{gawthrop_bond-graph_2017}. In this study, we chose conserved moieties with clear physical interpretations \cite{schuster_determining_1991}, but our conclusions hold regardless of our choice of the left nullspace matrix.

\subsection{Stimulus currents}
The cardiac action potential model was stimulated using a constant current stimulus that contained enough charge to raise the membrane potential by 30 mV over 0.1 ms. As recommended by Kneller \textit{et al.} \cite{kneller_time-dependent_2002}, conservative stimulus currents contained K\textsuperscript{+} ions as the charge carrier.

\begin{figure}
	\centering
	\includegraphics[width=0.9\linewidth]{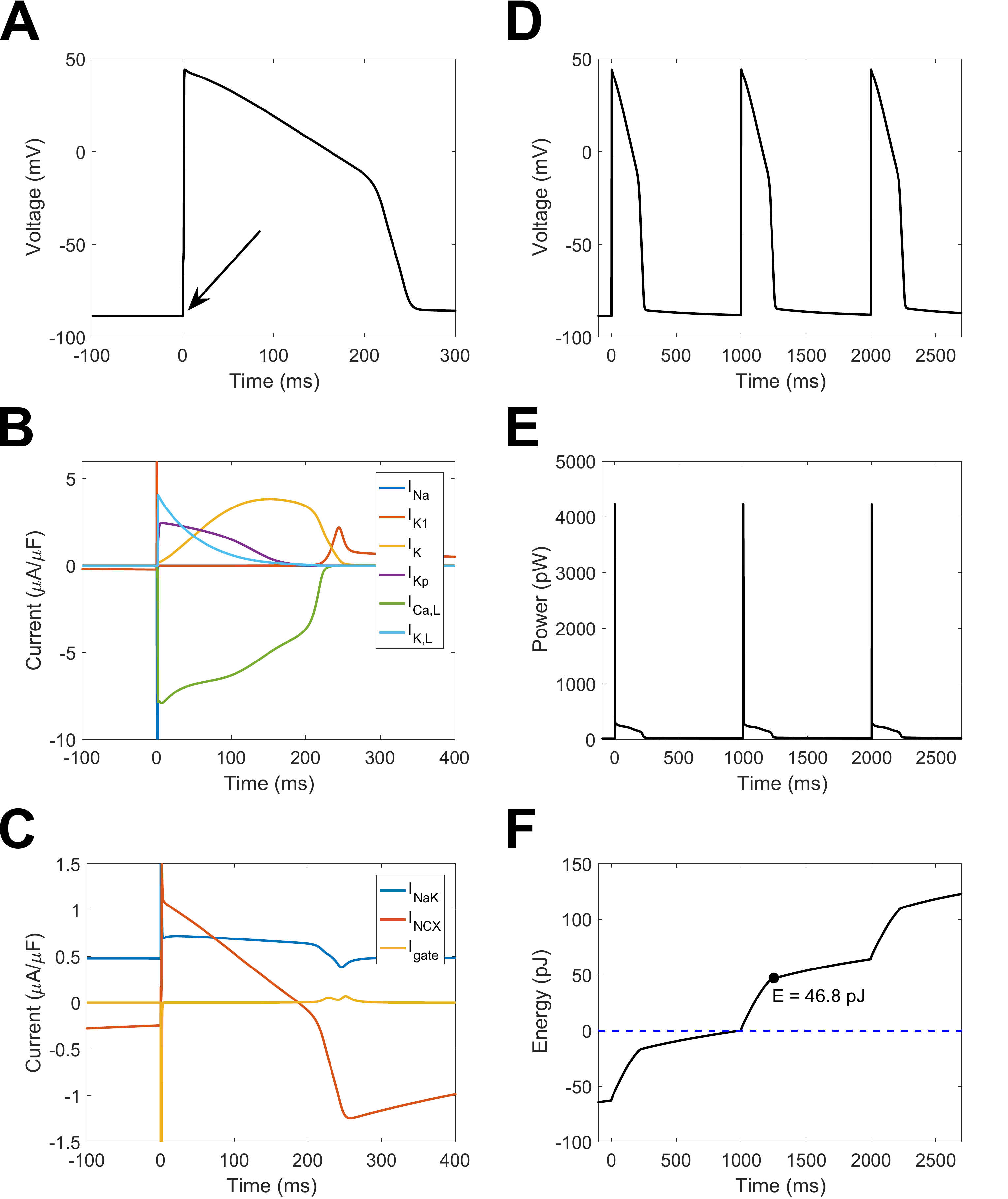}
	\caption{A simulation of the cardiac action potential using a bond graph model. \textbf{(A)} Membrane voltage, following stimulation with a conservative stimulus current (arrow); \textbf{(B)} Ion channel currents; \textbf{(C)} Transporter and gating currents; \textbf{(D)} Membrane voltage over three cycles, for comparison with (E) and (F); \textbf{(E)} Power consumption; \textbf{(F)} Energy dissipated, with the variable $E$ representing the energy consumption over the duration of the action potential. The model was run initially for 300 ms to allow the membrane potential and channel gates to stabilise. The intracellular ion concentrations were dynamic variables with initial concentrations $\mathrm{[Na_i^+]} = 10\ \si{mM}$, $\mathrm{[K_i^+]} = 145\ \si{mM}$ and $\mathrm{[Ca_i^+]} = 0.12\ \si{\micro M}$. Constant concentrations were $\mathrm{[Na_e^+]} = 140\ \si{mM}$, $\mathrm{[K_e^+]} = 5.4\ \si{mM}$, $\mathrm{[Ca_e^+]} = 1.8\ \si{mM}$,  $\mathrm{[MgATP] = 6.95\ \si{mM} }$, $\mathrm{[MgADP] = 0.035\ \si{mM} }$, $\mathrm{[P_i] = 0.3971\ \si{mM} }$ and $\mathrm{pH = 7.095 }$. $T = 310\ \si{K}$.}
	\label{fig:single_AP}
\end{figure}

\section{Results}
\label{sec:results}
\subsection{Simulation of a single action potential}
To verify that our bond graph model reproduced a typical action potential we simulated the model over a single beat (\autoref{fig:single_AP}A-C). The membrane potential (\autoref{fig:single_AP}A, with stimulation indicated by the arrow) resembled a typical cardiac action potential, with a distinct peak and plateau phase. The contributions of ion channel currents reproduce some common features of cardiac action potentials (\autoref{fig:single_AP}B). Once the action potential is initiated by a stimulus current, the sodium current $I_\mathrm{Na}$ briefly activates to give rise to a voltage spike. Following this, the plateau phase occurs where depolarising L-type Ca\textsuperscript{2+} currents oppose the repolarising K\textsuperscript{+} currents $I_\mathrm{K}$ and $I_\mathrm{Kp}$. Towards the end of the action potential, $I_\mathrm{K1}$ activates to restore the resting potential \cite{noble_models_2001}. Our model also simulates the reversal of NCX current across the action potential, and the consistent outward current of the Na$^+$/K$^+$ ATPase to maintain ionic gradients (\autoref{fig:single_AP}C). A consequence of modelling ion channels using bond graphs is that transitions between channel states are associated with a gating current resulting from charged residues moving in an electric field \cite{hodgkin_quantitative_1952}. Our model reveals that the total gating current across all channels $I_\mathrm{gate}$ has minimal contribution to total current (\autoref{fig:single_AP}C).

\autoref{fig:single_AP}E shows the power consumption of the membrane model over three cardiac cycles which was integrated to estimate the energetic cost of the cardiac action potential (\autoref{fig:single_AP}F). Note that energy continues to be consumed even during the resting state due the presence of currents associated with ion transporters. Thus while energy is predominantly consumed during the action potential, there is a rising gradient between action potentials (\autoref{fig:single_AP}F). By setting the energy consumption at the start of the second action potential to zero (\autoref{fig:single_AP}F, dotted blue line), we calculated the energetic cost over the duration of the action potential to be $46.8\ \si{pJ}$. Since the capacitive area of membrane for this model is $1.534 \times 10^{-4}\ \si{cm^2}$, the energy consumed per unit membrane area is $305\ \si{nJ/cm^{2}}$. When compared to Gawthrop et al.'s \cite{gawthrop_bond_2017-1} estimate of $173\ \si{nJ/cm^{2}}$ for the energetic cost of an action potential in the giant axon of a squid, the cardiac action potential uses 76\% more energy. The main reason for this difference is that in contrast to a neuron, the cardiac action potential contains a plateau phase with opposing currents. Despite the relatively slow rate of change in voltage, the Ca\textsuperscript{2+} and K\textsuperscript{+} currents remain relatively high, therefore a large amount of energy is dissipated during the plateau phase.

\begin{table}
	\caption{Conserved moieties associated with chemostat selection. Across some biochemical subgroups (``Moiety''), there are constraints (``Conserved quantity'') on a corresponding sum of species representing the total of the moiety. The conserved quantities remain constant over the course of a simulation. Q represents contributions of other species to charge imbalance across the membrane. The symbol $\Sigma$ represents charge contributions from Markov states of channels and transporters. The definition of $\Sigma$, and all species can be found in the Supporting Material and code.}
	\centering
	\bgroup
	\def\arraystretch{1.3}
	\begin{tabular}{r l p{0.7\linewidth}}
		\toprule
		& Moiety & Conserved quantity \\ \midrule
		\multicolumn{3}{l}{\textit{Conserved moieties common to all variants (A,B,C)}} \\ \midrule
		1 & K1 channel & $\mathrm{C_{K1} + O_{K1}}$ \\
		2 & K channel & $\mathrm{S_{00,K} + S_{10,K} + S_{20,K} + S_{01,K} + S_{11,K} + S_{21,K}}$ \\
		3 & Kp channel & $\mathrm{C_{Kp} + O_{Kp}}$ \\
		4 & Na channel & $\mathrm{S_{000,Na} + S_{100,Na} + S_{200,Na} + S_{300,Na} + S_{010,Na} + S_{110,Na} + S_{210,Na}}$ $\mathrm{ + S_{310,Na} + S_{001,Na} + S_{101,Na} + S_{201,Na} + S_{301,Na} + S_{011,Na}}$ $\mathrm{ + S_{111,Na} + S_{211,Na} + S_{311,Na}}$\\
		5 & LCC& $\mathrm{S_{000,LCC} + S_{010,LCC} + S_{020,LCC} + S_{100,LCC} + S_{110,LCC} + S_{120,LCC}}$ $\mathrm{ + S_{001,LCC} + S_{011,LCC} + S_{021,LCC} + S_{101,LCC} + S_{111,LCC} + S_{121,LCC}}$  \\
		6 & $\mathrm{Na^+}/\mathrm{K^+}$ ATPase& $\mathrm{P1_{NaK} + P2_{NaK} + P3_{NaK} + P4_{NaK} + P5_{NaK} + P6_{NaK} + P7_{NaK}}$ $\mathrm{ + P8_{NaK} + P9_{NaK} + P10_{NaK} + P11_{NaK} + P12_{NaK} + P13_{NaK}}$ $\mathrm{ + P14_{NaK} + P15_{NaK}}$ \\
		7 & NCX& $\mathrm{P1_{NCX} + P2_{NCX} + P3_{NCX} + P4_{NCX} + P5_{NCX} + P6_{NCX}}$\\
		8 & Troponin & $\mathrm{TRPN + TRPNCa}$ \\
		9 & Calmodulin & $\mathrm{CMDN + CMDNCa}$ \\ \midrule
		\multicolumn{3}{l}{\textit{Dynamic ion concentrations (A)}} \\ \midrule
		& Chemostats & MgADP, MgATP, Pi, $\mathrm{H^+}$\\
		10 & $\mathrm{K^+}$ ion & $\mathrm{K^+_i + K^+_e + 2P1_{NaK} + P2_{NaK} + P12_{NaK}+ 2P13_{NaK}+ 2P14_{NaK}}$ $+ \mathrm{2P15_{NaK}}$ \\
		11 & $\mathrm{Na^+}$ ion & $\mathrm{Na^+_i + Na^+_e + P4_{NaK} + 2P5_{NaK} + 3P6_{NaK}+ 3P7_{NaK}+ 3P8_{NaK}}$ $\mathrm{ + 2P9_{NaK}+ P10_{NaK} + 3P1_{NCX} + 3P6_{NCX}}$ \\
		12 & $\mathrm{Ca^{2+}}$ ion & $\mathrm{Ca^{2+}_i + Ca^{2+}_e + 2S_{001,LCC} + 2S_{011,LCC} + 2S_{021,LCC} + 2S_{101,LCC}}$ $\mathrm{ + 2S_{111,LCC} + 2S_{121,LCC} + P3_{NCX} + P4_{NCX} + TRPNCa}$ $\mathrm{ + CMDNCa}$\\
		13 & Charge & $\mathrm{Q - K^+_i - Na^+_i - 2Ca^{2+}_i + 2\mathrm{TRPN} + 2\mathrm{CMDN} + \Sigma}$ \\ \midrule
		\multicolumn{3}{l}{\textit{Dynamic intracellular ion concentrations (B)}} \\ \midrule
		& Chemostats & MgADP, MgATP, Pi, $\mathrm{H^+}$, $\mathrm{K^+_e}$, $\mathrm{Na^+_e}$, $\mathrm{Ca^{2+}_e}$\\
		10 & Charge & $\mathrm{Q - K^+_i - Na^+_i - 2Ca^{2+}_i + 2\mathrm{TRPN} + 2\mathrm{CMDN} + \Sigma}$ \\ \midrule
		\multicolumn{3}{l}{\textit{Constant ion concentrations (C)}} \\ \midrule
		& Chemostats & MgADP, MgATP, Pi, $\mathrm{H^+}$, $\mathrm{K^+_i}$, $\mathrm{K^+_e}$,  $\mathrm{Na^+_i}$, $\mathrm{Na^+_e}$, $\mathrm{Ca^{2+}_i}$, $\mathrm{Ca^{2+}_e}$\\	\bottomrule
	\end{tabular}
	\egroup
	\label{tab:cm}
\end{table}

\subsection{Chemostats influence the conserved moieties of cardiac action potential models}

Because the earliest models of the cardiac action potential did not include active transporters, they used constant intracellular concentrations to maintain ionic gradients across multiple cardiac cycles \cite{difrancesco_model_1985,luo_model_1991}. Later models incorporated ion transporters, allowing them to represent physiological conditions with dynamic intracellular ion concentrations, and constant extracellular ion concentrations to model washout from the circulatory system \cite{luo_dynamic_1994,faber_action_2000}. Under ischaemic conditions, washout is greatly inhibited, thus models of ischaemia use dynamic extracellular ion concentrations \cite{terkildsen_balance_2007}. We investigated the issue of drift in three classes of model: those with (A) dynamic ion concentrations on both sides of the membrane, representing models of myocytes under ischaemic conditions; (B) dynamic intracellular ion concentrations but constant extracellular ion concentrations, representing models of myocytes under physiological conditions; and (C) constant ion concentrations, representing models without transporters.

We used our bond graph model to represent these classes of models, selecting ions to fix at constant concentrations that resulted in three variants representative of the classes listed above. Conserved moieties of each variant were found using the left nullspace matrix of the stoichiometric matrix (\autoref{tab:cm}), and these include for example, the total amount of K1 channel (moiety 1). Because the channel is neither synthesised nor degraded in our model, the total amount of channel, i.e.~the sum of its closed ($\mathrm{C_{K1}}$) and open ($\mathrm{O_{K1}}$) states, remains constant over the course of a simulation.

Similarly, moiety 10 for variant (A) represents the total amount of K\textsuperscript{+} ions, which includes intracellular K\textsuperscript{+}, extracellular K\textsuperscript{+} and the K\textsuperscript{+} ions bound to Na\textsuperscript{+}/K\textsuperscript{+} ATPase. The total amount of K\textsuperscript{+} is constant when ion concentrations are dynamic. However, because fixing the concentration of K\textsuperscript{+} requires an additional external flux, the conservation law is broken in variants (B) and (C). Because the membrane capacitance is included in the stoichiometry of the system, our method automatically identifies a charge conservation law (moiety 13 for variant (A), and moiety 10 for variant (B)). 

Finally, the overall amount of intracellular charge can be described as a sum of contributions from intracellular K\textsuperscript{+}, Na\textsuperscript{+}, Ca\textsuperscript{2+} (and its buffers) and Markov states from ion channels and transporters ($\Sigma$), similar to forms found in previous studies \cite{hund_ionic_2001,varghese_conservation_1997}. It should be noted, however, that when all ion concentrations were held constant charge conservation was broken, as indicated by the absence of a conserved charge moiety in the bottom partition of \autoref{tab:cm}. In general, holding the concentration of a species constant breaks conservation laws \cite{polettini_irreversible_2014} and the number of conserved moieties progressively decreases as more ions concentrations are modelled as chemostats. We discuss the consequences of this in later sections.

\begin{figure}
	\centering
	\includegraphics[width=0.9\linewidth]{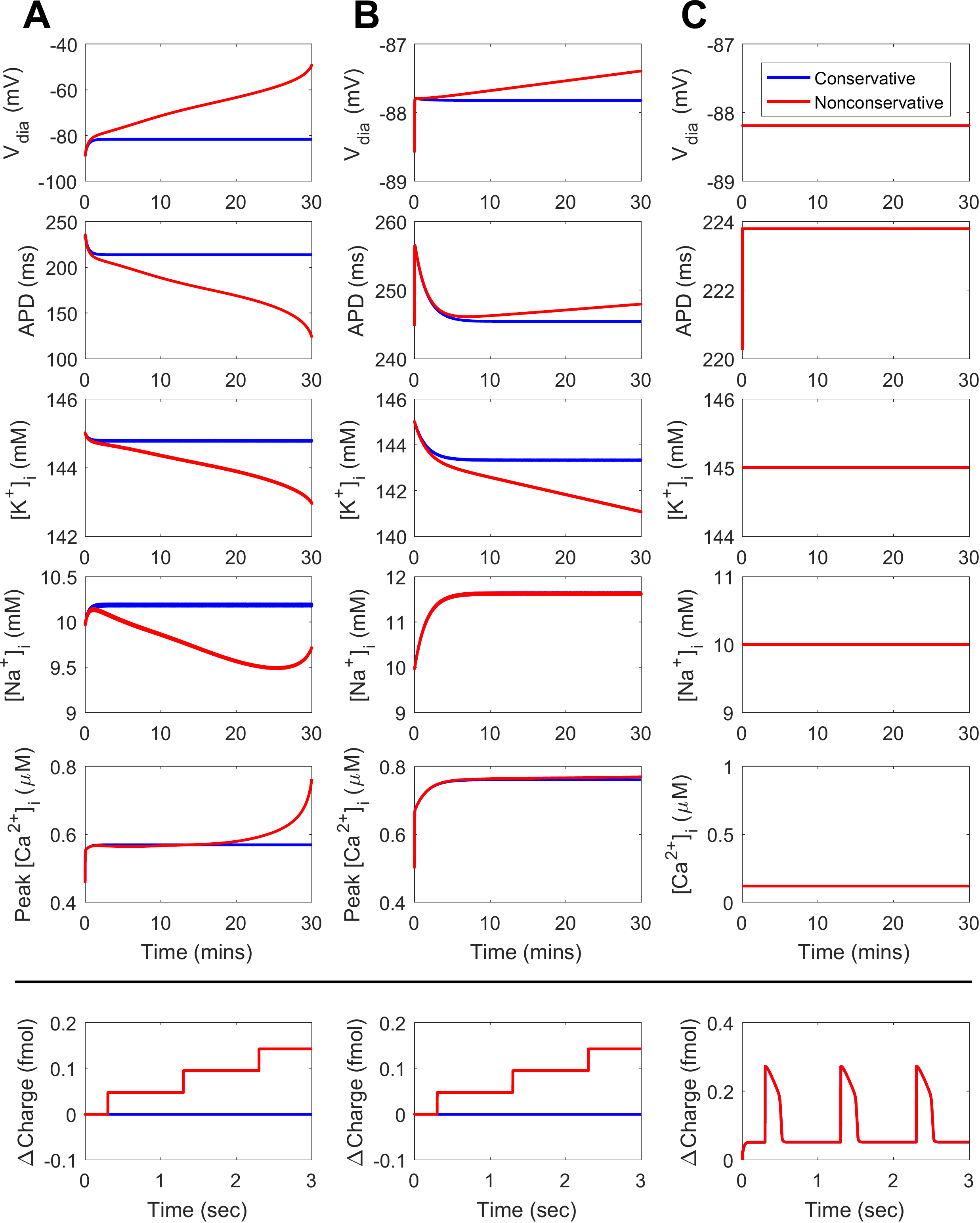}
	\caption{Effect of stimulus type and variable ion concentrations on model drift. \textbf{(A)} Dynamic ion concentrations; \textbf{(B)} Dynamic intracellular ion concentrations; \textbf{(C)} Constant ion concentrations. Results are shown for stimuli that conserve overall charge (blue) and those that do not conserve charge (red). Charge values are given as differences from the initial value of $-$5882.2 fmol. $T = \text{310 K}$. Definitions: $V_\text{dia}$, diastolic membrane potential; APD, action potential duration at 90\% repolarisation.}
	\label{fig:stim_currents}
\end{figure}

\subsection{Nonconservative stimulus currents cause drift in models with a charge conservation law}

An important feature of cardiac electrophysiology models is that they must be simulated for extended periods to examine physiologically relevant changes in behaviour, thus we tested how the type of stimulus current affected each variant of the cardiac action potential model by pacing at 1 Hz for 30 minutes. As illustrated (\autoref{fig:stim_currents}A,B), a nonconservative stimulus resulted in drift when the model had dynamic ion concentrations either for all compartments, or only within the intracellular compartment. The drift was particularly pronounced when all ion concentrations were dynamic (\autoref{fig:stim_currents}A), as extracellular concentrations changed faster than intracellular concentrations. In contrast, the model was resistant to drift from a nonconservative stimulus when all ion concentrations were held constant (\autoref{fig:stim_currents}C).

These results suggested that drift arose due to violations of the conserved charge moiety. Charge is a conserved moiety (\autoref{tab:cm}) in model variants where drift occured with a nonconservative stimulus. In this situation nonconservative stimulus currents cause drift because every stimulus causes a stepwise increase in the value of the conserved charge moiety (\autoref{fig:stim_currents}A,B bottom panels). However, because conservation laws are broken as more species are represented as chemostats \cite{polettini_irreversible_2014}, charge is no longer a conserved moiety when all ion concentrations are constant (\autoref{tab:cm}). Thus an observation which may not be obvious to intuition is that under these conditions charge is no longer constant between stimuli, and therefore free to return to its original value after each stimulus (\autoref{fig:stim_currents}C, bottom panel), allowing such models to achieve a steady-state limit cycle.

\begin{figure}
	\centering
	\includegraphics[width=\linewidth]{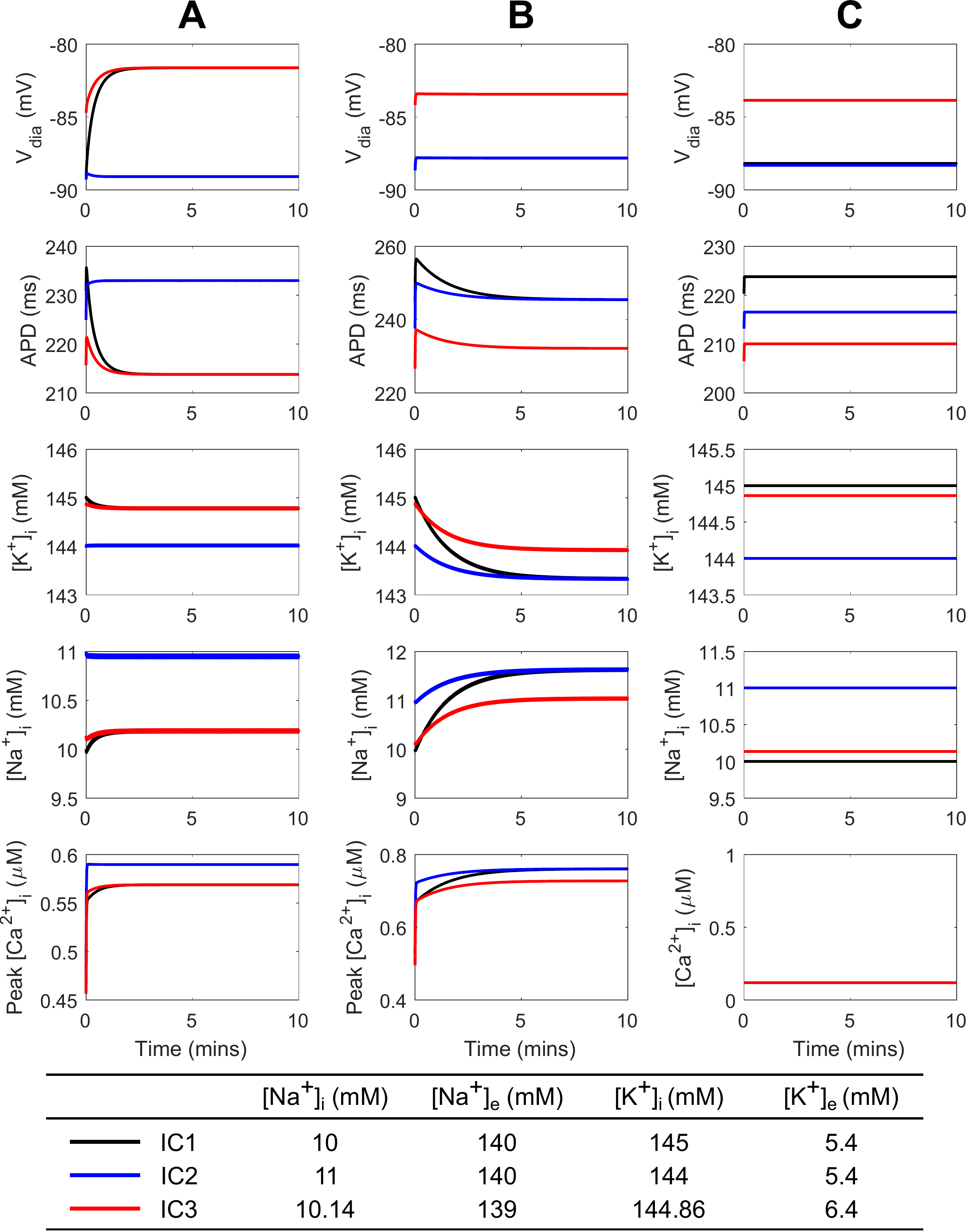}
	\caption{Effect of initial conditions on steady-state behaviour. \textbf{(A)} Dynamic ion concentrations; \textbf{(B)} Dynamic intracellular ion concentrations; \textbf{(C)} Constant ion concentrations. The models were paced at 1Hz for 30 minutes using a conservative stimulus current. $\mathrm{[MgATP] = 6.95\ \si{mM} }$, $\mathrm{[MgADP] = 0.035\ \si{mM} }$, $\mathrm{[P_i] = 0.3971\ \si{mM} }$, $\mathrm{pH = 7.095 }$, $T = 310\ \si{K}$. Definitions: $V_\text{dia}$, diastolic membrane potential; APD, action potential duration at 90\% repolarisation.}
	\label{fig:ic_ss}
\end{figure}

\subsection{Initial conditions influence steady states through conserved moieties and chemostats}
Next, for different sets of conserved moieties (as determined by constrained/dynamic ionic concentrations) we tested how the steady-state behaviour of the cardiac action potential was altered under three different initial conditions (\autoref{fig:ic_ss}). The first set of initial conditions (IC1) are common values for comparison (Fig. 5; \textit{at bottom}). IC2 is the same as IC1 but with 1mM intracellular K\textsuperscript{+} exchanged for 1mM of intracellular Na\textsuperscript{+}, such that charge is conserved but K\textsuperscript{+} and Na\textsuperscript{+} are not conserved. Similarly, IC3 is IC1, but with some K\textsuperscript{+} extruded and an equal amount of Na\textsuperscript{+} moved into the cell such that charge, Na\textsuperscript{+}, and K\textsuperscript{+} are all conserved. When all ion concentrations are dynamic IC1 and IC3 lead to the same steady state, but IC2 results in a different steady state (\autoref{fig:ic_ss}A). If only intracellular ion concentrations are dynamic, however, IC1 and IC2 result in identical steady states, but IC3 leads to a different steady state (\autoref{fig:ic_ss}B). Finally, keeping all ion concentrations constant leads to different steady states for all initial conditions (\autoref{fig:ic_ss}C).

These results demonstrate that the summed amount for each conserved moiety and/or chemostat value determines the steady-state behaviour of cardiac action potential models. To investigate this further, we calculated the values for conserved moieties and chemostats that resulted from each initial condition (\autoref{tab:ic_cm}; \textit{differences from IC1 indicated in bold}). For two sets of initial conditions to achieve identical steady states, all conserved moieties and chemostats must have the same value. Thus under dynamic ion concentrations (\autoref{fig:ic_ss}A), IC3 results in the same steady state as IC1 because all conserved moieties have been preserved (\autoref{tab:ic_cm}), whereas, IC2 causes a different steady state because the $\mathrm{K^+}$ and $\mathrm{Na^+}$ conserved moieties take on different values. Similarly, when only intracellular ion concentrations are dynamic, IC2 preserves the value of all conserved moieties and chemostats, but IC3 changes the values of the chemostats corresponding to extracellular Na\textsuperscript{+} and K\textsuperscript{+} concentrations (\autoref{tab:ic_cm}), hence the different steady state. When all ion concentrations were held constant, changes in the chemostat values (\autoref{tab:ic_cm}) were associated with different steady states for all three initial conditions (\autoref{fig:ic_ss}C).

\begin{table}
	\caption{The values of conserved moieties and chemostats under different initial conditions. All values are in fmol. Chemostats are indicated with (cs). Values different from IC1 are shown in bold.}
	\centering
	\begin{tabular}{c c c c}
		\toprule
		& \multicolumn{3}{c}{Value} \\
		Moiety/Chemostat & IC1 & IC2 & IC3 \\ \midrule
		\multicolumn{4}{l}{\textit{Dynamic ion concentrations (A)}} \\ \midrule
		$\mathrm{K^+}$ & 5538.1 & \textbf{5500.1} & 5538.1 \\
		$\mathrm{Na^+}$ & 1105.6 & \textbf{1143.6} & 1105.6 \\
		Charge & $-$5882.2 & $-$5882.2 & $-$5882.2 \\  \midrule
		\multicolumn{4}{l}{\textit{Dynamic intracellular ion concentrations (B)}} \\ \midrule
		$\mathrm{K_e^+}$ (cs) & 27.98 & 27.98 & \textbf{38.35} \\
		$\mathrm{Na_e^+}$ (cs)  & 5510 & 5510 & \textbf{715.12} \\
		Charge & $-$5882.2 & $-$5882.2 & $-$5882.2 \\ \midrule
		\multicolumn{4}{l}{\textit{Constant ion concentrations (C)}} \\ \midrule
		$\mathrm{K_e^+}$ (cs)  & 27.98 & 27.98 & \textbf{38.35} \\
		$\mathrm{Na_e^+}$ (cs)  & 725.48 & 725.48 & \textbf{715.12} \\
		$\mathrm{K_i^+}$ (cs)  & 5510  & \textbf{5472} & \textbf{5500} \\
		$\mathrm{Na_i^+}$ (cs)  & 380 & \textbf{418} & \textbf{390.36} \\ \bottomrule
	\end{tabular}
	\label{tab:ic_cm}
\end{table}

\section{Discussion}
In this study, we developed a bond graph model of the cardiac action potential with the aim of resolving the issues of drift and non-unique steady states. Analysis using conserved moieties enabled the discovery of all conservation laws within the model. In addition to the conservation of charge law from previous studies \cite{varghese_conservation_1997,hund_ionic_2001,endresen_theory_2000}, we found conservation laws corresponding to ions, states of Markov models of channels and transporters, and buffers. Two key advantages of our approach over existing analyses are that it reveals all conservation laws in a comprehensive and systematic manner, and that it is general for all models of the cardiac action potential that can be represented as bond graphs. When simulated over long periods with a nonconservative stimulus our bond graph model displayed solution drift, but it became resistant to drift when ion concentrations were held constant, demonstrating that changes in the value of a conserved charge moiety drive model drift. We also found that two sets of initial conditions can lead to different steady states if the values of their corresponding conserved moieties and chemostats are different, suggesting a strong link between conserved moieties and the steady-state limit cycles of cardiac action potential models. To demonstrate that our approach is general, we tested how the selection of chemostats (i.e. fixed concentrations) influenced drift and steady-states by using variants of our model that were representative of existing models in the literature. Our approach shows that holding ion concentrations constant changes the conserved moieties of the model, which in turn has an effect on the susceptibility of a model to drift and non-unique steady states.

\subsection{Drift}
When paced with a nonconservative stimulus, variants of the model with a charge conservation law underwent drift (\autoref{fig:stim_currents}A,B) consistent with previous studies \cite{hund_ionic_2001,livshitz_uniqueness_2009}. By observing changes in the charge conserved moiety, the bond graph approach attributes drift to regular perturbations in charge that cannot be restored due to the presence of a conservation law. Whereas previous analyses relied solely on intuition to derive a conservation law corresponding to charge \cite{hund_ionic_2001,livshitz_uniqueness_2009}, we note that our approach automatically derives conservation laws and can detect other conservation laws that may be relevant for drift.

As demonstrated, the bond graph method requires construction of a stoichiometric matrix, providing a simple approach to check whether a stimulus current will cause drift. Let $v_s$ be a row matrix representing the stoichiometry of the stimulus current (with chemostats removed), $N^{cd}$ be the stoichiometric matrix after removing rows corresponding to chemostats, and $G$ be the left nullspace matrix of $N^{cd}$. To avoid altering any of the conserved moieties, the stimulus current must have zero contribution to them, i.e.~ $Gv_s = 0$ (or equivalently, $v_s$ needs to lie in the image of $N^{cd}$). Thus the model drifts if $Gv_s \neq 0$. While it is common practice to use K\textsuperscript{+} as the charge carrier for stimulus currents, it is likely that multiple species contribute to the current \cite{hund_ionic_2001,kneller_time-dependent_2002}. Thus the automated approach suggested here is well-suited for checking whether more complex stimulus currents satisfy conservation of charge, as well as other conservation laws within the model. It should be noted however, that while a model satisfying $Gv_s = 0$ will not drift due to violating conservation laws, drift may still occur due to an imbalance of currents throughout the action potential, for instance, in the absence of Na$^+$/K$^+$ ATPase, the ionic gradients would gradually disappear in a model with dynamic ion concentrations. 

Finally, we believe that this analysis provides a link between the issues of drift and steady states. Our models show that drift due to a nonconservative stimulus current can be attributed to changes in the value of the charge conserved moiety with every stimulus, and accordingly the steady state of the model changes. Model drift then occurs as the solution continually chases a moving steady state.

\subsection{Effects of initial conditions on steady states}
We also found that initial conditions of cardiac action potential models change their steady states through the values of chemostats and conserved moieties (\autoref{fig:ic_ss}, \autoref{tab:ic_cm}). Accordingly, the same perturbation to initial conditions can have different effects on the steady state depending on which species are held constant. Therefore in addition to ensuring that the concentration of ions are physiological, care should be taken to correctly initialise each state of buffers and Markov models of ion channels and ion transporters, as they may contain a significant fraction of total ion abundance. For example, Ca\textsuperscript{2+} buffers and SERCA can sequester a significant amount of Ca\textsuperscript{2+} and they should be initialised with the correct amount of bound Ca\textsuperscript{2+} when multi-state models are used \cite{higgins_buffering_2006}. We note that the difficulty of manually deriving conservation laws increases exponentially as models of cardiac electrophysiology become more complex, and we believe that our approach extends on existing analyses \cite{hund_ionic_2001,livshitz_uniqueness_2009} to provide a general method for assessing steady-state behaviour by comparing the values of conserved moieties and chemostats that result from each initial condition.

In the field of biochemical network analysis, there is a well-established dependence of quiescent steady states on conserved moieties, and numerous mathematical techniques for assessing the uniqueness and stability of these steady states have been developed \cite{gross_algebraic_2016,feliu_variable_2012}. However, the influence of conserved moieties on limit cycles in an oscillating system that is regularly stimulated has yet to be investigated. Our results hint at similarities between these two fields, and while we only tested the uniqueness of steady states using relatively small perturbations to the initial conditions, it is possible that a set of conserved moieties may have multiple steady states, and greater perturbations may lead to other limit cycles.

\subsection{The ``differential'' and ``algebraic'' methods}
The discovery of conservation principles in cardiac electrophysiology has lead to a debate over whether to use the differential or algebraic methods of simulation \cite{hund_ionic_2001,fraser_quantitative_2007,livshitz_uniqueness_2009,varghese_conservation_1997,endresen_theory_2000}. The differential method is the calculation of membrane voltage by integrating total current, and the algebraic method is the calculation of membrane voltage using an algebraic relationship derived from charge conservation. We chose the differential method over the algebraic method since it better supports model reuse and modularity - in particular it is easier to modify the equations to select different species as chemostats, and to combine equations when two models are coupled. We note, however, that the algebraic method may reduce computational complexity \cite{gawthrop_energy-based_2014,hund_ionic_2001}. In bond graph modelling, the algebraic method can be implemented by using conserved moieties to turn the system of ordinary differential equations (ODEs) into an index-0 differential algebraic equation (DAE) (Eq. 3.48 of \cite{gawthrop_energy-based_2014}). This method generalises existing algebraic methods to reduce the system of differential equations by using all conserved moieties and not just the conserved charge moiety. While we did not use the algebraic approach, we emphasise that the choice of method relates to numerical approaches for model simulation rather than the underlying physics of the system \cite{hund_ionic_2001}. Therefore, the differential and algebraic methods are equivalent in conservative systems provided that the initial conditions and values of conserved moieties are consistent.

\subsection{Integration into whole-cell models}
Our bond graph model of the cardiac action potential is the first step towards a fully-integrated whole-cell bond graph model of a cardiomyocyte that couples electrophysiology, signalling, metabolism and mechanics. Modelling studies for the energetic regulation of a cardiac cell exist across the literature \cite{tran_regulation_2015}, but while some components used in these models are thermodynamically consistent \cite{tran_thermodynamic_2009,smith_development_2004}, existing whole-cell models are neither energy-based nor thermodynamically consistent throughout the entire model. Furthermore, because existing experimental and modelling studies use ATP consumption as a proxy for energy consumption, they can only estimate the energy consumption of major energy sinks: the Na$^+$/K$^+$ ATPase, SERCA, and crossbridge cycling \cite{schramm_energy_1994,tran_regulation_2015}. A bond graph approach may thus provide more detailed insights into how a cardiac cell uses energy downstream of ATP hydrolysis processes, and help to identify energy-consuming processes. Because the bond graph approach is energy-based it allows us to directly assess energy consumption of the model (in Joules). We found that when normalised against membrane area the cardiac action potential consumes approximately 76\% more energy than an action potential in the axon of a giant squid. To the authors' knowledge, this is the first account of energy consumed by electrochemical processes during the cardiac action potential. 

\subsection{Limitations}
Because of physical restrictions imposed by the framework not all model components can be directly converted into a bond graph form. Accordingly, we were forced to model ion channels and transporters using Markov states to faithfully represent their underlying physics, however, this produced a model that had numerous states compared to the number of biological processes. While it is reassuring to find that our method of identifying conserved moieties remained robust despite this complexity, simulation of the model was computationally expensive. For the purpose of integrating this action potential model into a larger whole-cell model, it would be useful to have simple model components that reduce computational cost. While current methods for reducing biochemical models in the bond graph framework are not advanced enough to apply to the biological components in this study, we note that bond graphs provide a useful foundation for applying model simplification while ensuring that thermodynamic consistency is maintained \cite{gawthrop_energy-based_2014}.

We also decided to limit the transport processes included in our model to those considered essential for producing a cardiac action potential, while maintaining a limit cycle using dynamic ion concentrations. Our bond graph model omitted many ionic currents due to their small amplitudes, however these channels may have greater contributions under conditions which vary from those tested here. Thus an obvious extension of this work would be the integration of other electrogenic processes within the cardiac membrane. It would be interesting to investigate whether coupling other models requires further tuning of parameters \cite{babtie_how_2017}, and whether the presence of physical bond graph parameters changes this process.

When formulating the structure and parameters for a bond graph model of the cardiac action potential (or most other processes), it is possible to either fit against existing mathematical models or the underlying experimental measurements. For all processes in this study excluding the NCX, we developed our bond graph model to reproduce the behaviour of an existing model, in an attempt to re-use existing knowledge about these processes. This approach poses constraints on the bond graph structure used, especially for gating structure. Therefore it would be interesting to develop an approach that assesses bond graph structures as well as bond graph parameters, based on their fits to data \cite{babtie_how_2017}. Such an approach may provide a better fit to the data, and uncover insights into the physical mechanisms of ion channels.

\section{Conclusion}
In this study we have developed a bond graph model of the cardiac action potential and used this to explore the issues of drift and non-unique steady states. We demonstrate that the analysis of conserved moieties generalises the concept of charge conservation used in earlier studies, and found that changes in conserved moieties can explain drift as well as changes in steady state behaviour. Importantly, holding ion concentrations constant can have significant consequences on both drift and steady states as they change the conserved moieties in the model. Our approach is sufficiently general that it can be applied to any cardiac action potential model which can be represented as a bond graph. We hope that the bond graph approach outlined here will prove useful for the development of future cardiac electrophysiology models, and eventually whole-cell models of the cardiomyocyte. 

\textbf{Data access:}
The code associated with this study is available from GitHub (\url{https://github.com/uomsystemsbiology/bond_graph_cardiac_AP}), and archived on Zenodo (\url{https://doi.org/10.5281/zenodo.1172205}) \cite{pan_supporting_2018}. The code contains MATLAB (The MathWorks, Natick, MA) that generate the figures, CellML code containing parameters, initial conditions and equations of the model, and full details of the bond graph structure.

\textbf{Author contributions:}
M.P., P.J.G., J.C. and E.J.C. developed the theory. M.P. performed the research. K.T. provided conceptual advice and helped interpret the results. All authors contributed to the text of the manuscript and gave final approval for publication.

\textbf{Competing interests:}
We have no competing interests.

\textbf{Funding:}
M.P. would like to acknowledge financial support provided by an Australian Government Research Training Program Scholarship. P.J.G. would like to thank the Melbourne School of Engineering for its support via a Professorial Fellowship. K.T. is supported by the Heart Foundation of New Zealand (Research Fellowship 1692) and the Marsden Fund Council from Government funding, managed by Royal Society Te Apārangi (Marsden Fast-Start 17-UOA-300).

\bibliographystyle{RS}
\small
\bibliography{bibliography,bibliography2}{}

\begin{thebibliography}{99}

\bibitem{luo_dynamic_1994}
Luo CH, Rudy Y. 1994a  A dynamic model of the cardiac ventricular action
  potential. {I}. {Simulations} of ionic currents and concentration changes..
  {\em Circ. Res.} \textbf{74}, 1071--1096.
\href{http://dx.doi.org/10.1161/01.RES.74.6.1071}{doi:10.1161/01.RES.74.6.1071}.

\bibitem{luo_dynamic_1994-1}
Luo CH, Rudy Y. 1994b  A dynamic model of the cardiac ventricular action
  potential. {II}. {Afterdepolarizations}, triggered activity, and
  potentiation.. {\em Circ. Res.} \textbf{74}, 1097--1113.
\href{http://dx.doi.org/10.1161/01.RES.74.6.1097}{doi:10.1161/01.RES.74.6.1097}.

\bibitem{faber_action_2000}
Faber GM, Rudy Y. 2000  Action {Potential} and {Contractility} {Changes} in
  [{Na}$^+$]$_\text{i}$ {Overloaded} {Cardiac} {Myocytes}: {A} {Simulation}
  {Study}. {\em Biophys. J.} \textbf{78}, 2392--2404.
\href{http://dx.doi.org/10.1016/S0006-3495(00)76783-X}{doi:10.1016/S0006-3495(00)76783-X}.

\bibitem{terkildsen_balance_2007}
Terkildsen JR, Crampin EJ, Smith NP. 2007  The balance between inactivation and
  activation of the {Na}\textsuperscript{+}-{K}\textsuperscript{+} pump
  underlies the triphasic accumulation of extracellular {K}\textsuperscript{+}
  during myocardial ischemia. {\em Am. J. Physiol-heart. C.} \textbf{293},
  H3036--H3045.
\href{http://dx.doi.org/10.1152/ajpheart.00771.2007}{doi:10.1152/ajpheart.00771.2007}.

\bibitem{crampin_dynamic_2006}
Crampin EJ, Smith NP. 2006  A {Dynamic} {Model} of {Excitation}-{Contraction}
  {Coupling} during {Acidosis} in {Cardiac} {Ventricular} {Myocytes}. {\em
  Biophys. J.} \textbf{90}, 3074--3090.
\href{http://dx.doi.org/10.1529/biophysj.105.070557}{doi:10.1529/biophysj.105.070557}.

\bibitem{guan_discussion_1997}
Guan S, Lu Q, Huang K. 1997  A Discussion About the {DiFrancesco}–{Noble}
  Model. {\em J. Theor. Biol.} \textbf{189}, 27--32.
\href{http://dx.doi.org/10.1006/jtbi.1997.0486}{doi:10.1006/jtbi.1997.0486}.

\bibitem{fraser_quantitative_2007}
Fraser JA, Huang CLH. 2007  Quantitative techniques for steady-state
  calculation and dynamic integrated modelling of membrane potential and
  intracellular ion concentrations. {\em Prog. Biophys. Mol. Biol.}
  \textbf{94}, 336--372.
\href{http://dx.doi.org/10.1016/j.pbiomolbio.2006.10.001}{doi:10.1016/j.pbiomolbio.2006.10.001}.

\bibitem{kneller_time-dependent_2002}
Kneller J, Ramirez RJ, Chartier D, Courtemanche M, Nattel S. 2002
  Time-dependent transients in an ionically based mathematical model of the
  canine atrial action potential. {\em American Journal of Physiology - Heart
  and Circulatory Physiology} \textbf{282}, H1437--H1451.
\href{http://dx.doi.org/10.1152/ajpheart.00489.2001}{doi:10.1152/ajpheart.00489.2001}.

\bibitem{difrancesco_model_1985}
DiFrancesco D, Noble D. 1985  A model of cardiac electrical activity
  incorporating ionic pumps and concentration changes. {\em Philosophical
  Transactions of the Royal Society of London B: Biological Sciences}
  \textbf{307}, 353--398.

\bibitem{hund_ionic_2001}
Hund TJ, Kucera JP, Otani NF, Rudy Y. 2001  Ionic {Charge} {Conservation} and
  {Long}-{Term} {Steady} {State} in the {Luo}–{Rudy} {Dynamic} {Cell}
  {Model}. {\em Biophys. J.} \textbf{81}, 3324--3331.
\href{http://dx.doi.org/10.1016/S0006-3495(01)75965-6}{doi:10.1016/S0006-3495(01)75965-6}.

\bibitem{livshitz_uniqueness_2009}
Livshitz L, Rudy Y. 2009  Uniqueness and Stability of Action Potential Models
  during Rest, Pacing, and Conduction Using Problem-Solving Environment. {\em
  Biophys. J.} \textbf{97}, 1265--1276.
\href{http://dx.doi.org/10.1016/j.bpj.2009.05.062}{doi:10.1016/j.bpj.2009.05.062}.

\bibitem{aslanidi_mechanisms_2009}
Aslanidi OV, Boyett MR, Dobrzynski H, Li J, Zhang H. 2009  Mechanisms of
  {Transition} from {Normal} to {Reentrant} {Electrical} {Activity} in a
  {Model} of {Rabbit} {Atrial} {Tissue}: {Interaction} of {Tissue}
  {Heterogeneity} and {Anisotropy}. {\em Biophys. J.} \textbf{96}, 798--817.
\href{http://dx.doi.org/10.1016/j.bpj.2008.09.057}{doi:10.1016/j.bpj.2008.09.057}.

\bibitem{carro_human_2011}
Carro J, Rodríguez JF, Laguna P, Pueyo E. 2011  A human ventricular cell model
  for investigation of cardiac arrhythmias under hyperkalaemic conditions. {\em
  Phil. Trans. R. Soc. A} \textbf{369}, 4205--4232.
\href{http://dx.doi.org/10.1098/rsta.2011.0127}{doi:10.1098/rsta.2011.0127}.

\bibitem{grandi_novel_2010}
Grandi E, Pasqualini FS, Bers DM. 2010  A novel computational model of the
  human ventricular action potential and {Ca} transient. {\em J. Mol. Cell.
  Cardiol.} \textbf{48}, 112--121.
\href{http://dx.doi.org/10.1016/j.yjmcc.2009.09.019}{doi:10.1016/j.yjmcc.2009.09.019}.

\bibitem{sager_rechanneling_2014}
Sager PT, Gintant G, Turner JR, Pettit S, Stockbridge N. 2014  Rechanneling the
  cardiac proarrhythmia safety paradigm: {A} meeting report from the {Cardiac}
  {Safety} {Research} {Consortium}. {\em Am. Heart J.} \textbf{167}, 292--300.
\href{http://dx.doi.org/10.1016/j.ahj.2013.11.004}{doi:10.1016/j.ahj.2013.11.004}.

\bibitem{colatsky_comprehensive_2016}
Colatsky T, Fermini B, Gintant G, Pierson JB, Sager P, Sekino Y, Strauss DG,
  Stockbridge N. 2016  The {Comprehensive} in {Vitro} {Proarrhythmia} {Assay}
  ({CiPA}) initiative — {Update} on progress. {\em J. Pharmacol. Toxicol.}
  \textbf{81}, 15--20.
\href{http://dx.doi.org/10.1016/j.vascn.2016.06.002}{doi:10.1016/j.vascn.2016.06.002}.

\bibitem{paynter_analysis_1961}
Paynter HM. 1961 {\em Analysis and design of engineering systems}.
MIT press.

\bibitem{borutzky_advances_1995}
Borutzky W, Dauphin-Tanguy G, Thoma JU. 1995  Advances in bond graph modelling:
  theory, software, applications. {\em Math. Comput. Simulat.} \textbf{39},
  465--475.
\href{http://dx.doi.org/10.1016/0378-4754(95)00106-6}{doi:10.1016/0378-4754(95)00106-6}.

\bibitem{oster_network_1973}
Oster GF, Perelson AS, Katchalsky A. 1973  Network thermodynamics: dynamic
  modelling of biophysical systems. {\em Q. Rev. Biophys.} \textbf{6}, 1--134.
\href{http://dx.doi.org/10.1017/S0033583500000081}{doi:10.1017/S0033583500000081}.

\bibitem{gawthrop_energy-based_2014}
Gawthrop PJ, Crampin EJ. 2014  Energy-based analysis of biochemical cycles
  using bond graphs. {\em Proceedings of the Royal Society of London A:
  Mathematical, Physical and Engineering Sciences} \textbf{470}, 20140459.
\href{http://dx.doi.org/10.1098/rspa.2014.0459}{doi:10.1098/rspa.2014.0459}.

\bibitem{gawthrop_bond_2017-1}
Gawthrop PJ, Siekmann I, Kameneva T, Saha S, Ibbotson MR, Crampin EJ. 2017
  Bond graph modelling of chemoelectrical energy transduction. {\em IET Syst.
  Biol.} \textbf{11}, 127--138.
\href{http://dx.doi.org/10.1049/iet-syb.2017.0006}{doi:10.1049/iet-syb.2017.0006}.

\bibitem{omholt_human_2016}
Omholt SW, Hunter PJ. 2016  The {Human} {Physiome}: a necessary key for the
  creative destruction of medicine. {\em Interface Focus} \textbf{6}, 20160003.
\href{http://dx.doi.org/10.1098/rsfs.2016.0003}{doi:10.1098/rsfs.2016.0003}.

\bibitem{gawthrop_hierarchical_2015}
Gawthrop PJ, Cursons J, Crampin EJ. 2015  Hierarchical bond graph modelling of
  biochemical networks. {\em Proc. R. Soc. A} \textbf{471}, 20150642.
\href{http://dx.doi.org/10.1098/rspa.2015.0642}{doi:10.1098/rspa.2015.0642}.

\bibitem{gawthrop_bond_2017}
Gawthrop PJ. 2017  Bond {Graph} {Modeling} of {Chemiosmotic} {Biomolecular}
  {Energy} {Transduction}. {\em Ieee T. Nanobiosci.} \textbf{16}, 177--188.
\href{http://dx.doi.org/10.1109/TNB.2017.2674683}{doi:10.1109/TNB.2017.2674683}.

\bibitem{gawthrop_energy-based_2017}
Gawthrop PJ, Crampin EJ. 2017  Energy-based analysis of biomolecular pathways.
  {\em Proc. R. Soc. A} \textbf{473}, 20160825.
\href{http://dx.doi.org/10.1098/rspa.2016.0825}{doi:10.1098/rspa.2016.0825}.

\bibitem{beard_energy_2002}
Beard DA, Liang Sd, Qian H. 2002  Energy {Balance} for {Analysis} of {Complex}
  {Metabolic} {Networks}. {\em Biophys. J.} \textbf{83}, 79--86.
\href{http://dx.doi.org/10.1016/S0006-3495(02)75150-3}{doi:10.1016/S0006-3495(02)75150-3}.

\bibitem{beard_thermodynamic_2004}
Beard DA, Babson E, Curtis E, Qian H. 2004  Thermodynamic constraints for
  biochemical networks. {\em J. Theor. Biol.} \textbf{228}, 327--333.
\href{http://dx.doi.org/10.1016/j.jtbi.2004.01.008}{doi:10.1016/j.jtbi.2004.01.008}.

\bibitem{van_der_schaft_mathematical_2013}
van~der Schaft A, Rao S, Jayawardhana B. 2013  On the {Mathematical}
  {Structure} of {Balanced} {Chemical} {Reaction} {Networks} {Governed} by
  {Mass} {Action} {Kinetics}. {\em Siam J. Appl. Math.} \textbf{73}, 953--973.
\href{http://dx.doi.org/10.1137/11085431X}{doi:10.1137/11085431X}.

\bibitem{haraldsdottir_identification_2016}
Haraldsdóttir HS, Fleming RMT. 2016  Identification of {Conserved} {Moieties}
  in {Metabolic} {Networks} by {Graph} {Theoretical} {Analysis} of {Atom}
  {Transition} {Networks}. {\em Plos Comput. Biol.} \textbf{12}.
\href{http://dx.doi.org/10.1371/journal.pcbi.1004999}{doi:10.1371/journal.pcbi.1004999}.

\bibitem{pan_cardiac_2017}
Pan M, Gawthrop PJ, Cursons J, Tran K, Crampin EJ. 2017  The cardiac
  {Na}$^+$/{K}$^+$ {ATPase}: {An} updated, thermodynamically consistent model.
  {\em arXiv:1711.00989 [q-bio]}.

\bibitem{kimura_identification_1987}
Kimura J, Miyamae S, Noma A. 1987  Identification of sodium-calcium exchange
  current in single ventricular cells of guinea-pig.. {\em The Journal of
  Physiology} \textbf{384}, 199--222.
\href{http://dx.doi.org/10.1113/jphysiol.1987.sp016450}{doi:10.1113/jphysiol.1987.sp016450}.

\bibitem{beuckelmann_sodium-calcium_1989}
Beuckelmann DJ, Wier WG. 1989  Sodium-calcium exchange in guinea-pig cardiac
  cells: exchange current and changes in intracellular
  {Ca}\textsuperscript{2+}.. {\em The Journal of Physiology} \textbf{414},
  499--520.
\href{http://dx.doi.org/10.1113/jphysiol.1989.sp017700}{doi:10.1113/jphysiol.1989.sp017700}.

\bibitem{gawthrop_metamodelling:_1996}
Gawthrop P, Smith L. 1996 {\em Metamodelling: for bond graphs and dynamic
  systems}.
Prentice {Hall} international series in systems and control engineering.
  London, New York: Prentice Hall.

\bibitem{borutzky_bond_2010}
Borutzky W. 2010 {\em Bond {Graph} {Methodology}}.
Springer.

\bibitem{gawthrop_bond-graph_2007}
Gawthrop P, Bevan G. 2007  Bond-graph modeling. {\em IEEE Control Syst.}
  \textbf{27}, 24--45.
\href{http://dx.doi.org/10.1109/MCS.2007.338279}{doi:10.1109/MCS.2007.338279}.

\bibitem{gawthrop_bond-graph_2017}
Gawthrop PJ. 2017  Bond-{Graph} {Modelling} and {Causal} {Analysis} of
  {Biomolecular} {Systems}. In {\em Bond {Graphs} for {Modelling}, {Control}
  and {Fault} {Diagnosis} of {Engineering} {Systems}} pp. 587--623. Springer,
  Cham.
DOI: 10.1007/978-3-319-47434-2\_16.

\bibitem{polettini_irreversible_2014}
Polettini M, Esposito M. 2014  Irreversible thermodynamics of open chemical
  networks. {I}. {Emergent} cycles and broken conservation laws. {\em The
  Journal of Chemical Physics} \textbf{141}, 024117.
\href{http://dx.doi.org/10.1063/1.4886396}{doi:10.1063/1.4886396}.

\bibitem{liebermeister_modular_2010}
Liebermeister W, Uhlendorf J, Klipp E. 2010  Modular rate laws for enzymatic
  reactions: thermodynamics, elasticities and implementation. {\em Method.
  Biochem. Anal.} \textbf{26}, 1528--1534.
\href{http://dx.doi.org/10.1093/bioinformatics/btq141}{doi:10.1093/bioinformatics/btq141}.

\bibitem{palsson_systems_2006}
Palsson B. 2006 {\em Systems biology: properties of reconstructed networks}.
Cambridge University Press.

\bibitem{klipp_systems_2009}
Klipp E. 2009 {\em Systems biology: a textbook}.
Wiley-VCH.

\bibitem{anton_elementary_2014}
Anton H, Rorres C. 2014 {\em Elementary linear algebra : applications version.}
Hoboken, NJ : John Wiley \& Sons Inc.

\bibitem{schuster_what_1995}
Schuster S, Hilgetag C. 1995  What {Information} about the {Conserved}-{Moiety}
  {Structure} of {Chemical} {Reaction} {Systems} {Can} be {Derived} from
  {Their} {Stoichiometry}?. {\em The Journal of Physical Chemistry}
  \textbf{99}, 8017--8023.
\href{http://dx.doi.org/10.1021/j100020a026}{doi:10.1021/j100020a026}.

\bibitem{schuster_determining_1991}
Schuster S, Höfer T. 1991  Determining all extreme semi-positive conservation
  relations in chemical reaction systems: a test criterion for conservativity.
  {\em J. Chem. Soc., Faraday Trans.} \textbf{87}, 2561--2566.
\href{http://dx.doi.org/10.1039/FT9918702561}{doi:10.1039/FT9918702561}.

\bibitem{noble_models_2001}
Noble D, Rudy Y. 2001  Models of cardiac ventricular action potentials:
  iterative interaction between experiment and simulation. {\em Philos. T. Roy.
  Soc. A.} \textbf{359}, 1127--1142.
\href{http://dx.doi.org/10.1098/rsta.2001.0820}{doi:10.1098/rsta.2001.0820}.

\bibitem{hodgkin_quantitative_1952}
Hodgkin AL, Huxley AF. 1952  A quantitative description of membrane current and
  its application to conduction and excitation in nerve. {\em The Journal of
  Physiology} \textbf{117}, 500--544.
\href{http://dx.doi.org/10.1113/jphysiol.1952.sp004764}{doi:10.1113/jphysiol.1952.sp004764}.

\bibitem{luo_model_1991}
Luo CH, Rudy Y. 1991  A model of the ventricular cardiac action potential.
  {Depolarization}, repolarization, and their interaction.. {\em Circ. Res.}
  \textbf{68}, 1501--1526.
\href{http://dx.doi.org/10.1161/01.RES.68.6.1501}{doi:10.1161/01.RES.68.6.1501}.

\bibitem{varghese_conservation_1997}
Varghese A, Sell GR. 1997  A {Conservation} {Principle} and its {Effect} on the
  {Formulation} of {Na}–{Ca} {Exchanger} {Current} in {Cardiac} {Cells}. {\em
  J. Theor. Biol.} \textbf{189}, 33--40.
\href{http://dx.doi.org/10.1006/jtbi.1997.0487}{doi:10.1006/jtbi.1997.0487}.

\bibitem{endresen_theory_2000}
Endresen LP, Hall K, Høye JS, Myrheim J. 2000  A theory for the membrane
  potential of living cells. {\em Eur. Biophys. J.} \textbf{29}, 90--103.
\href{http://dx.doi.org/10.1007/s002490050254}{doi:10.1007/s002490050254}.

\bibitem{higgins_buffering_2006}
Higgins ER, Cannell MB, Sneyd J. 2006  A {Buffering} {SERCA} {Pump} in {Models}
  of {Calcium} {Dynamics}. {\em Biophys. J.} \textbf{91}, 151--163.
\href{http://dx.doi.org/10.1529/biophysj.105.075747}{doi:10.1529/biophysj.105.075747}.

\bibitem{gross_algebraic_2016}
Gross E, Harrington HA, Rosen Z, Sturmfels B. 2016  Algebraic {Systems}
  {Biology}: {A} {Case} {Study} for the {Wnt} {Pathway}. {\em B. Math. Biol.}
  \textbf{78}, 21--51.
\href{http://dx.doi.org/10.1007/s11538-015-0125-1}{doi:10.1007/s11538-015-0125-1}.

\bibitem{feliu_variable_2012}
Feliu E, Wiuf C. 2012  Variable {Elimination} in {Chemical} {Reaction}
  {Networks} with {Mass}-{Action} {Kinetics}. {\em Siam J. Appl. Math.}
  \textbf{72}, 959--981.
\href{http://dx.doi.org/10.1137/110847305}{doi:10.1137/110847305}.

\bibitem{tran_regulation_2015}
Tran K, Loiselle DS, Crampin EJ. 2015  Regulation of cardiac cellular
  bioenergetics: mechanisms and consequences. {\em Physiological Reports}
  \textbf{3}, e12464.
\href{http://dx.doi.org/10.14814/phy2.12464}{doi:10.14814/phy2.12464}.

\bibitem{tran_thermodynamic_2009}
Tran K, Smith NP, Loiselle DS, Crampin EJ. 2009  A {Thermodynamic} {Model} of
  the {Cardiac} {Sarcoplasmic}/{Endoplasmic} {Ca}2+ ({SERCA}) {Pump}. {\em
  Biophys. J.} \textbf{96}, 2029--2042.
\href{http://dx.doi.org/10.1016/j.bpj.2008.11.045}{doi:10.1016/j.bpj.2008.11.045}.

\bibitem{smith_development_2004}
Smith NP, Crampin EJ. 2004  Development of models of active ion transport for
  whole-cell modelling: cardiac sodium–potassium pump as a case study. {\em
  Prog. Biophys. Mol. Biol.} \textbf{85}, 387--405.
\href{http://dx.doi.org/10.1016/j.pbiomolbio.2004.01.010}{doi:10.1016/j.pbiomolbio.2004.01.010}.

\bibitem{schramm_energy_1994}
Schramm M, Klieber HG, Daut J. 1994  The energy expenditure of
  actomyosin-{ATPase}, {Ca}\textsuperscript{2+}-{ATPase} and
  {Na}\textsuperscript{+},{K}\textsuperscript{+}-{ATPase} in guinea-pig cardiac
  ventricular muscle.. {\em J Physiol} \textbf{481}, 647--662.

\bibitem{babtie_how_2017}
Babtie AC, Stumpf MPH. 2017  How to deal with parameters for whole-cell
  modelling. {\em J. R. Soc. Interface} \textbf{14}, 20170237.
\href{http://dx.doi.org/10.1098/rsif.2017.0237}{doi:10.1098/rsif.2017.0237}.

\bibitem{pan_supporting_2018}
Pan M, Gawthrop PJ, Tran K, Cursons J, Crampin EJ. 2018  Supporting code for
  ``{Bond} graph modelling of the cardiac action potential: {Implications} for
  drift and non-unique steady states''.  Zenodo.
DOI: 10.5281/zenodo.1172205.

\bibitem{keener_mathematical_2009}
Keener J, Sneyd J. 2009 {\em Mathematical {Physiology}} vol. 8/1{\em
  Interdisciplinary {Applied} {Mathematics}}.
New York, NY: Springer New York.

\bibitem{rudy_computational_2006}
Rudy Y, Silva JR. 2006  Computational biology in the study of cardiac ion
  channels and cell electrophysiology. {\em Q. Rev. Biophys.} \textbf{39},
  57--116.
\href{http://dx.doi.org/10.1017/S0033583506004227}{doi:10.1017/S0033583506004227}.

\bibitem{fink_markov_2009}
Fink M, Noble D. 2009  Markov models for ion channels: versatility versus
  identifiability and speed. {\em Philos. T. Roy. Soc. A.} \textbf{367},
  2161--2179.
\href{http://dx.doi.org/10.1098/rsta.2008.0301}{doi:10.1098/rsta.2008.0301}.

\bibitem{rasmusson_mathematical_1990}
Rasmusson RL, Clark JW, Giles WR, Robinson K, Clark RB, Shibata EF, Campbell
  DL. 1990  A mathematical model of electrophysiological activity in a bullfrog
  atrial cell. {\em Am. J. Physiol-heart. C.} \textbf{259}, H370--H389.

\bibitem{giladi_structure-functional_2016}
Giladi M, Shor R, Lisnyansky M, Khananshvili D. 2016  Structure-{Functional}
  {Basis} of {Ion} {Transport} in {Sodium}–{Calcium} {Exchanger} ({NCX})
  {Proteins}. {\em Int. J. Mol. Sci.} \textbf{17}.
\href{http://dx.doi.org/10.3390/ijms17111949}{doi:10.3390/ijms17111949}.

\bibitem{hilgemann_steady-state_1992-1}
Hilgemann DW, Matsuoka S, Nagel GA, Collins A. 1992  Steady-state and dynamic
  properties of cardiac sodium-calcium exchange. {Sodium}-dependent
  inactivation.. {\em The Journal of General Physiology} \textbf{100},
  905--932.
\href{http://dx.doi.org/10.1085/jgp.100.6.905}{doi:10.1085/jgp.100.6.905}.

\bibitem{reuter_ion_1984}
Reuter H. 1984  Ion channels in cardiac cell membranes. {\em Annu. Rev.
  Physiol.} \textbf{46}, 473--484.

\bibitem{sakmann_conductance_1984}
Sakmann B, Trube G. 1984  Conductance properties of single inwardly rectifying
  potassium channels in ventricular cells from guinea-pig heart.. {\em J
  Physiol} \textbf{347}, 641--657.

\bibitem{shibasaki_conductance_1987}
Shibasaki T. 1987  Conductance and kinetics of delayed rectifier potassium
  channels in nodal cells of the rabbit heart.. {\em The Journal of Physiology}
  \textbf{387}, 227.

\bibitem{yue_characterization_1996}
Yue L, Feng J, Li GR, Nattel S. 1996  Characterization of an ultrarapid delayed
  rectifier potassium channel involved in canine atrial repolarization.. {\em J
  Physiol} \textbf{496}, 647--662.

\bibitem{hinch_simplified_2004}
Hinch R, Greenstein JL, Tanskanen AJ, Xu L, Winslow RL. 2004  A {Simplified}
  {Local} {Control} {Model} of {Calcium}-{Induced} {Calcium} {Release} in
  {Cardiac} {Ventricular} {Myocytes}. {\em Biophys. J.} \textbf{87},
  3723--3736.
\href{http://dx.doi.org/10.1529/biophysj.104.049973}{doi:10.1529/biophysj.104.049973}.

\end{thebibliography}

\newpage
\appendix
\renewcommand{\thetable}{S\arabic{table}}   
\renewcommand{\thefigure}{S\arabic{figure}}
\renewcommand{\theequation}{S\arabic{equation}}   
\setcounter{figure}{0}
\setcounter{table}{0}
\setcounter{equation}{0}

\normalsize
\section{Ion channel modelling}
\subsection{Bond graph structure}
\begin{figure}[H]
	\centering
	\begin{tabular}{l l} 
		\textbf{\textsf{(A) channel{\_}Kp}}&  \textbf{\textsf{(B) gate{\_}en{\_}Kp}} \\
		\imagetop{\includegraphics[width=0.4\linewidth]{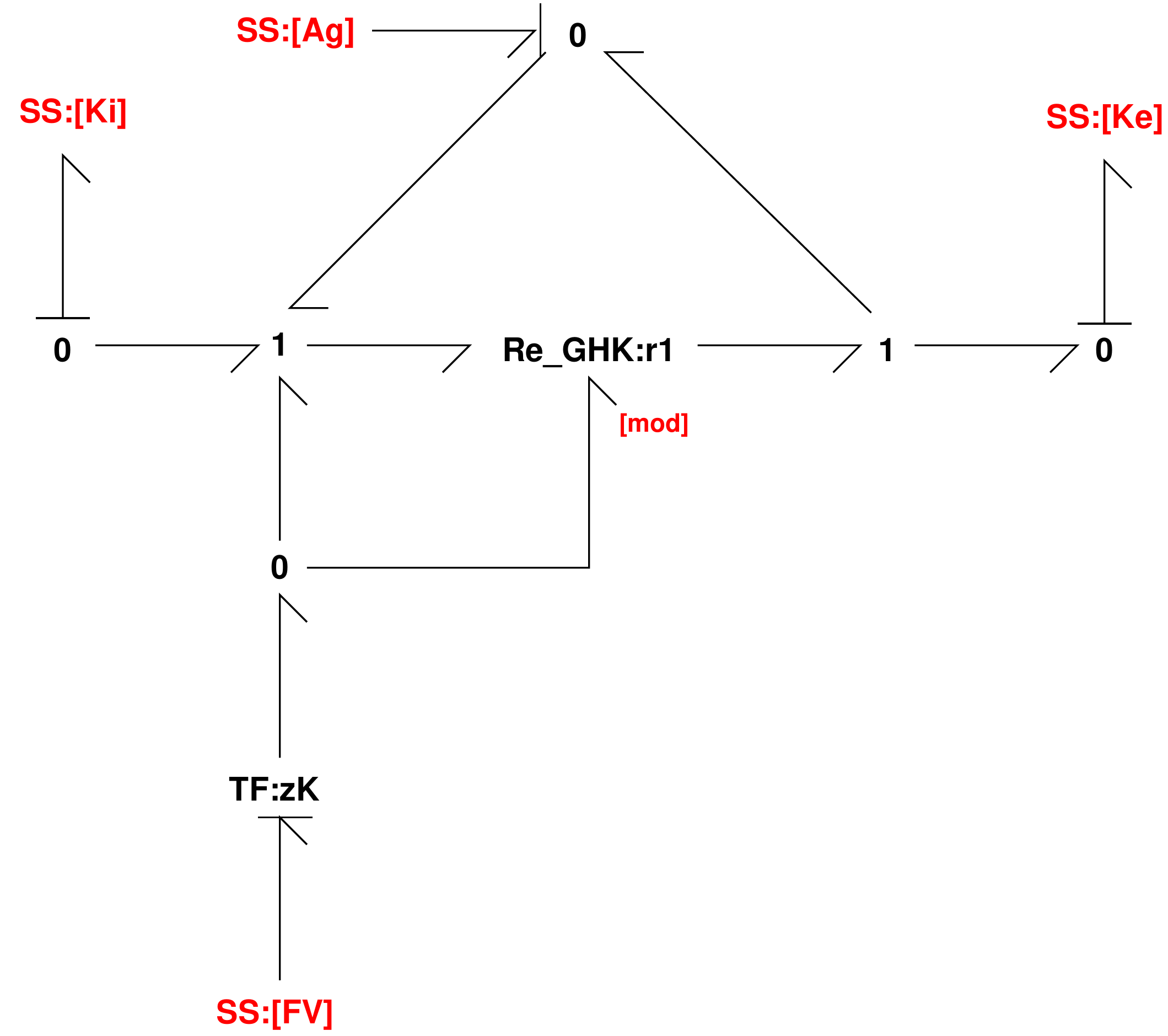}} &
		\imagetop{\includegraphics[width=0.4\linewidth]{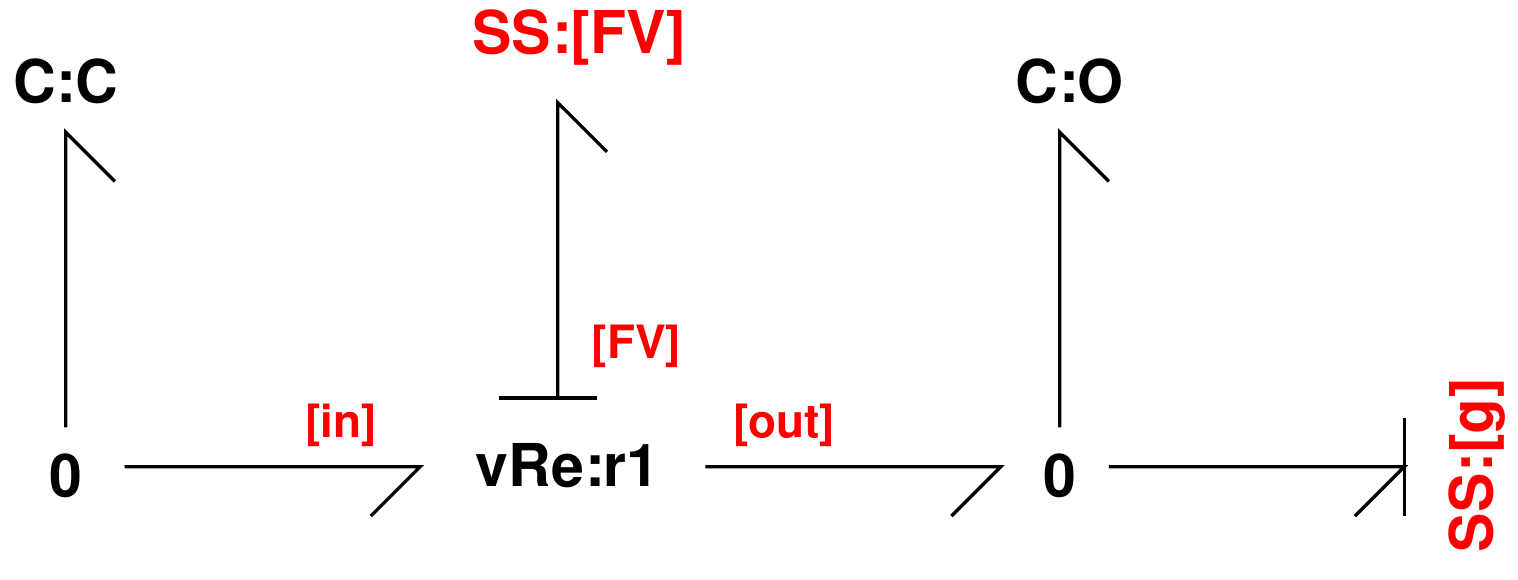}} 
	\end{tabular}\\[1cm]
	\begin{tabular}{l l}
		\textbf{\textsf{(C) vRe}} & 
		\textbf{\textsf{(D) Kp{\_}channel}} \\
		\imagetop{\includegraphics[width=0.5\linewidth]{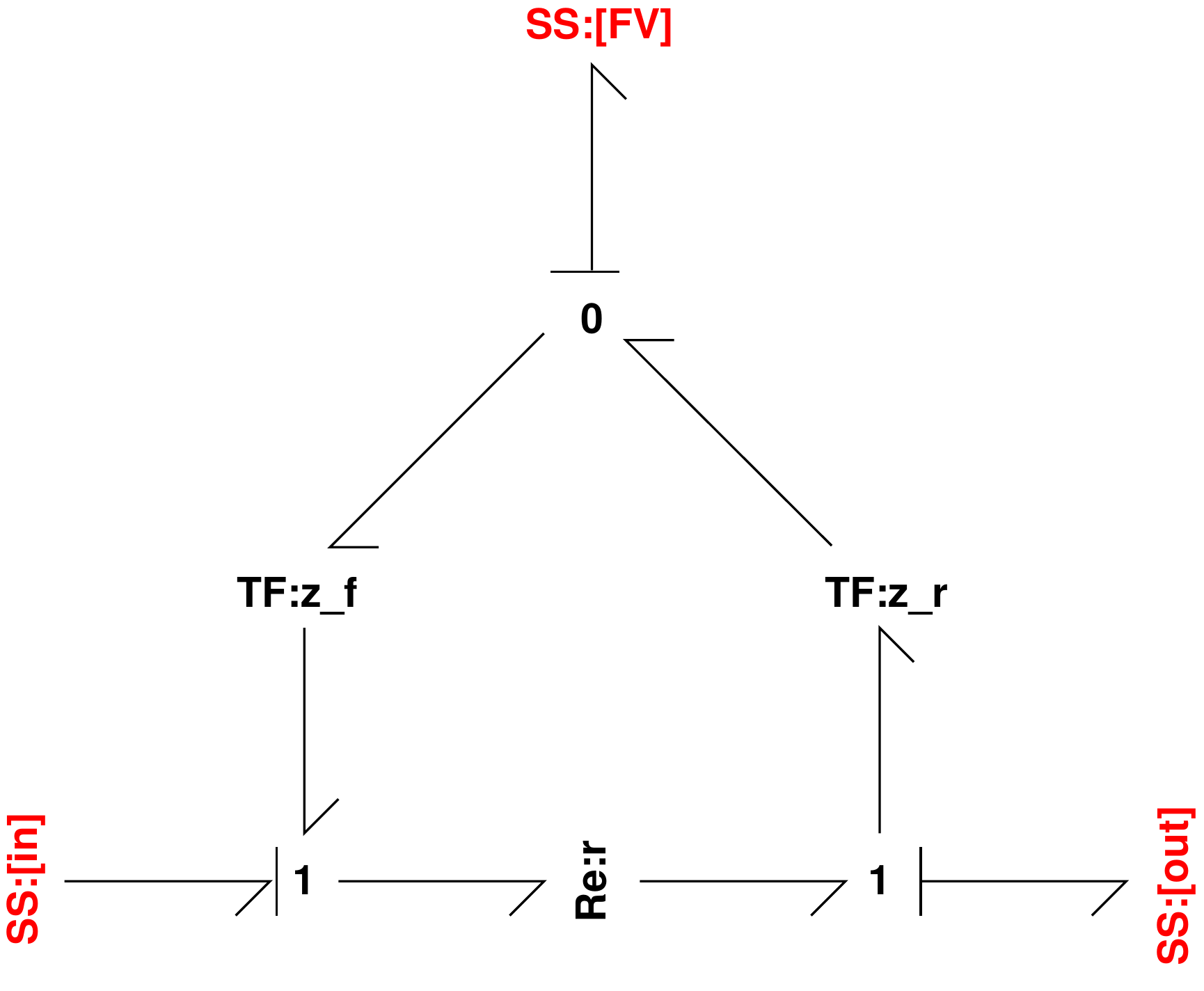}} &
		\imagetop{\includegraphics[width=0.35\linewidth]{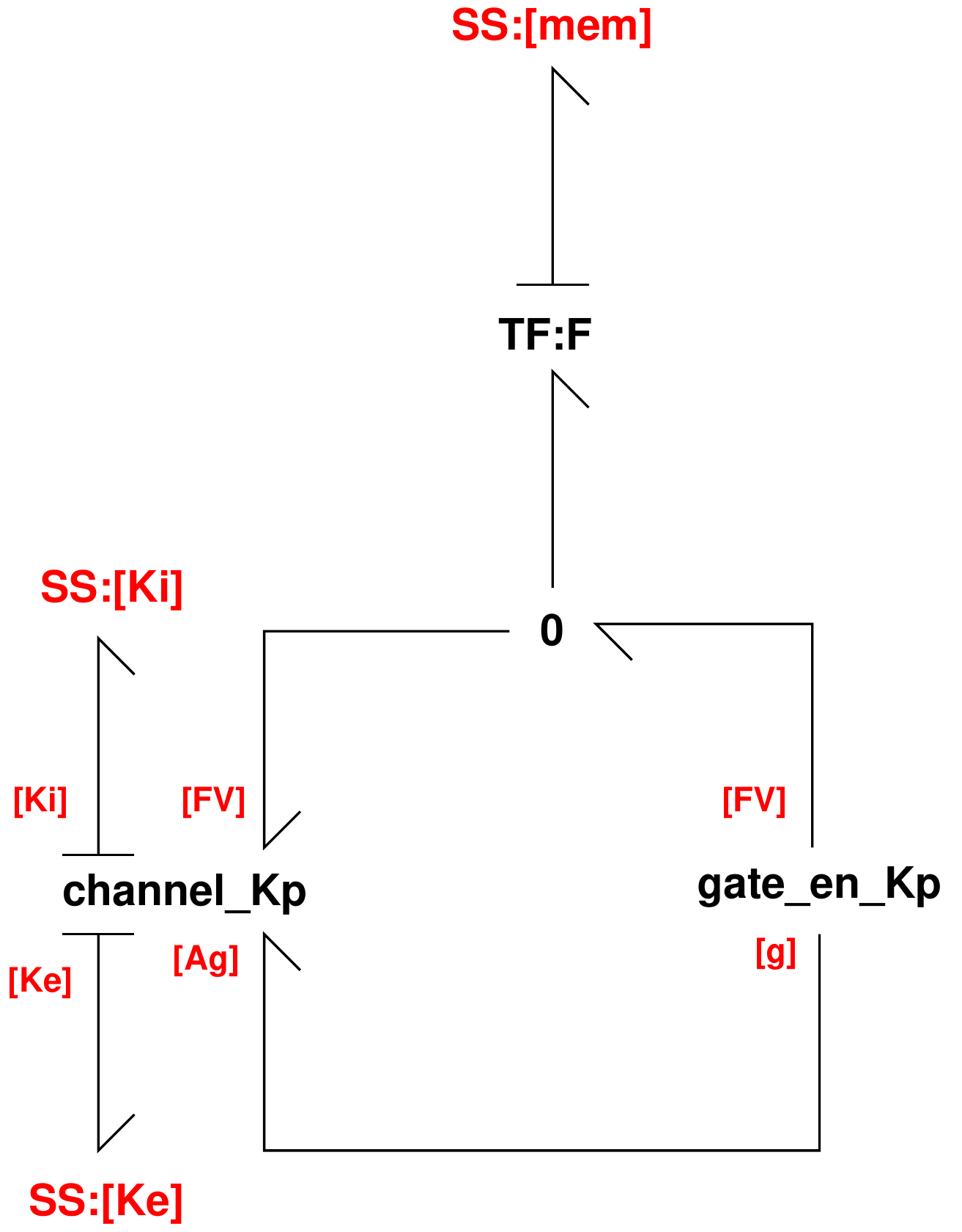}}
	\end{tabular}
	\caption{\textbf{The bond graph model of a plateau K\textsuperscript{+} channel.} \textbf{(A)} The \textbf{\textsf{channel{\_}Kp}} module describes the current through the ion channel. \textbf{(B)} The \textbf{\textsf{gate{\_}en{\_}Kp}} contains the states required for gating. \textbf{(C)} The \textbf{\textsf{vRe}} module contains a voltage-dependent reaction used to describe channel state transitions. \textbf{(D)} The channel current and gating modules are combined into an ion channel model (\textbf{\textsf{Kp{\_}channel}}).}
	\label{fig:Kp_channel}
\end{figure}

In this section, we discuss decisions made in developing models of ion channels. The bond graph structure for the Kp channel is shown in \autoref{fig:Kp_channel}. The other channels have similar structures that follow from the discussion in this section.

\subsection{Current-voltage relations}
\label{sec:GHK}
While thermodynamic properties can be used to determine how membrane voltage and ionic concentrations relate at equilibrium, they do not specify behaviour away from equilibrium. For this purpose, the current-voltage (I-V) relationship defines how the membrane voltage relates to the current through a specific channel. Using bond graphs, it is difficult to incorporate the effects of gating using a linear I-V equation. Therefore we use the Goldman-Hodgkin-Katz (GHK) equation to model ion channels, as it enables relatively simple incorporation of ion channel gating as a physics-based biochemical module \cite{gawthrop_bond_2017-1}. The GHK equation defines a non-linear relationship between current $I$ and membrane voltage $V$:
\begin{align}
I = P\frac{z^2F^2}{RT} V \left(
\frac{c_i - c_e e^{-zFV/RT}}{1 - e^{-zFV/RT}}
\right)
\label{eq:GHK}
\end{align}
where $c_i$ and $c_e$ are the ion's intracellular and extracellular concentrations respectively \cite{keener_mathematical_2009}. In a bond graph, the GHK equation for current can be described by a modulated Re component with a single modulator (see \autoref{fig:Kp_channel}A, left panel), using the constitutive equation from Gawthrop \textit{et al.} \cite{gawthrop_bond_2017-1}:
\begin{align}
v = \begin{cases}
\kappa \frac{ \frac{A^{m}}{RT} }
{\exp\left(\frac{A^{m}}{RT} \right) - 1} \left[
\exp\left(
\frac{A^f}{RT}
\right)
-
\exp\left(
\frac{A^r}{RT}
\right)
\right], & A^m \neq 0 \\
\kappa  \left[
\exp\left(
\frac{A^f}{RT}
\right)
-
\exp\left(
\frac{A^r}{RT}
\right)
\right], & A^m = 0
\end{cases}
\end{align}
As discussed in Gawthrop \textit{et al.} \cite{gawthrop_bond_2017-1}, setting
\begin{align}
A^f &= \mu_i + zFV \\
A^r &= \mu_e \\
A^m &= zFV
\end{align}
gives rise to the GHK equation. Since many ion channels in the Luo-Rudy model are described using a linear I-V relationship, the use of GHK equations requires some approximations.

\subsection{Modulation}
\label{sec:modulation}
While the I-V curves describe currents through open ion channels, a formulation for gating is required to describe the number of open ion channels at any given time. In the Hodgkin-Huxley framework, gating is modelled as differential equations that give the proportion of open gates at any given time. We incorporated the effects of gating through a gating affinity $A^g$, which is added to both the forward and reverse affinities of a reaction (\autoref{fig:Kp_channel}A) to modulate its rate without changing the equilibrium \cite{gawthrop_bond_2017-1}.

\subsection{State models}
\label{sec:channel_states}
\begin{figure}
	\centering
	\begin{tikzpicture}[auto, outer sep=3pt, node distance=2cm,>=latex']
	\node (S00) {$S_{00}$};
	\node [right of = S00] (S10) {$S_{10}$};
	\node [right of = S10] (S20) {$S_{20}$};
	\node [right of = S20] (S30) {$S_{30}$};
	\node [below of = S00] (S01) {$S_{01}$};
	\node [below of = S10] (S11) {$S_{11}$};
	\node [below of = S20] (S21) {$S_{21}$};
	\node [below of = S30] (S31) {$S_{31}$};
	
	\begin{scope}[every node/.style={font= \scriptsize}]
	\draw[transform canvas={yshift=0.3ex},-left to] (S00) -- node[above]{$3\alpha_m$}  (S10);
	\draw[transform canvas={yshift=-0.3ex},left to-] (S00) -- node[below] {$\beta_m$} (S10); 
	
	\draw[transform canvas={yshift=0.3ex},-left to] (S10) -- node[above]{$2\alpha_m$}  (S20);
	\draw[transform canvas={yshift=-0.3ex},left to-] (S10) -- node[below] {$2\beta_m$} (S20);
	
	\draw[transform canvas={yshift=0.3ex},-left to] (S20) -- node[above]{$\alpha_m$}  (S30);
	\draw[transform canvas={yshift=-0.3ex},left to-] (S20) -- node[below] {$3\beta_m$} (S30);
	
	\draw[transform canvas={yshift=0.3ex},-left to] (S01) -- node[above]{$3\alpha_m$}  (S11);
	\draw[transform canvas={yshift=-0.3ex},left to-] (S01) -- node[below] {$\beta_m$} (S11);
	
	\draw[transform canvas={yshift=0.3ex},-left to] (S11) -- node[above]{$2\alpha_m$}  (S21);
	\draw[transform canvas={yshift=-0.3ex},left to-] (S11) -- node[below] {$2\beta_m$} (S21);
	
	\draw[transform canvas={yshift=0.3ex},-left to] (S21) -- node[above]{$\alpha_m$}  (S31);
	\draw[transform canvas={yshift=-0.3ex},left to-] (S21) -- node[below] {$3\beta_m$} (S31);
	
	\draw[transform canvas={xshift=0.3ex},-left to] (S00) -- node[right]{$\alpha_h$}  (S01);
	\draw[transform canvas={xshift=-0.3ex},left to-] (S00) -- node[left] {$\beta_h$} (S01);
	
	\draw[transform canvas={xshift=0.3ex},-left to] (S10) -- node[right]{$\alpha_h$}  (S11);
	\draw[transform canvas={xshift=-0.3ex},left to-] (S10) -- node[left] {$\beta_h$} (S11);
	
	\draw[transform canvas={xshift=0.3ex},-left to] (S20) -- node[right]{$\alpha_h$}  (S21);
	\draw[transform canvas={xshift=-0.3ex},left to-] (S20) -- node[left] {$\beta_h$} (S21);
	
	\draw[transform canvas={xshift=0.3ex},-left to] (S30) -- node[right] {$\alpha_h$}  (S31);
	\draw[transform canvas={xshift=-0.3ex},left to-] (S30) -- node[left] {$\beta_h$} (S31);
	\end{scope}
	
	\end{tikzpicture}
	
	%\begin{tikzpicture}[auto, outer sep=3pt, node distance=2cm,>=latex']
	%\node (Cm) {$\mathrm{C_m}$};
	%\node [right of = Cm] (Om) {$\mathrm{O_m}$};
	%\node [below of = Cm] (Ch) {$\mathrm{C_h}$};
	%\node [right of = Ch] (Oh) {$\mathrm{O_h}$};
	%
	%
	%\begin{scope}[every node/.style={font= \scriptsize}]
	%\draw[transform canvas={yshift=0.3ex},-left to] (Cm) -- node[above]{$\alpha_m$}  (Om);
	%\draw[transform canvas={yshift=-0.3ex},left to-] (Cm) -- node[below] {$\beta_m$} (Om); 
	%
	%\draw[transform canvas={yshift=0.3ex},-left to] (Ch) -- node[above]{$\alpha_h$}  (Oh);
	%\draw[transform canvas={yshift=-0.3ex},left to-] (Ch) -- node[below] {$\beta_h$} (Oh); 
	%
	%\end{scope}
	%
	%\end{tikzpicture}
	
	\caption{\textbf{Channel states of a Na\textsuperscript{+} channel.}}
	\label{fig:gating_comparison}
\end{figure}
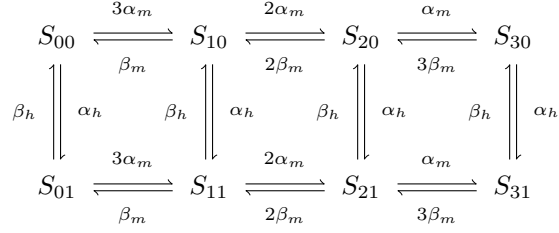

Ion channel models must account for gating and bond graphs require the use of physical components to achieve this. We model gating as transitions between channel states, known in the literature as Markov models \cite{rudy_computational_2006,fink_markov_2009}. To illustrate, we use the example of a typical Na\textsuperscript{+} channel in which the current $I$ is described by the equation
\begin{align}
I = m^3 h \bar{I}
\end{align}
where $\bar{I}$ is the current when all channels are open. This can be described using the reaction scheme in \autoref{fig:gating_comparison}, where $S_{31}$ represents the open channel. Because individual channel states are modelled, the current depends only on the amount of $S_{31}$ and not any of the other closed states. Thus, incorporation into the gating framework described above is intuitive; each state represents a structural conformation of the ion channel and the number of channels in each state are explicitly tracked, facilitating a simple approach to account for the energetics of gating under varying ion channel densities.

\subsection{Voltage dependence of state transitions}
\label{sec:V_dependence}
The transition rates between open and closed states are voltage-dependent for ion channels. Hodgkin-Huxley models describe state transitions using ODEs of the form
\begin{align}
\frac{dg}{dt} = \alpha(V) (1-g) - \beta(V) g
\end{align}
where $g$ is a gating variable such as $m$, $h$ or $n$. In typical Hodgkin-Huxley models, these rates $\alpha(V)$ and $\beta(V)$ are modelled using mathematical expressions on the basis of empirical fits to data. However, due to the physics-based nature of bond graphs, the open and closed states of channels must be explicitly modelled as physical components with a restricted set of constitutive equations. Because common expressions for $\alpha(V)$ and $\beta(V)$ do not obey the laws of thermodynamics, bond graphs are unable to perfectly replicate existing mathematical expressions used for ion channel transition rates. We chose to model state transitions by using the module described in \autoref{fig:Kp_channel}B, which results in an exponential voltage-dependence for both the forward and reverse reactions. In the case of the plateau K\textsuperscript{+} channel, the rate of transition from the closed state (C) to the open state (O) is:
\begin{align}
v &=  \alpha_0 \exp\left(\frac{z_f FV}{RT} \right) x_A - \beta_0 \exp\left(\frac{z_r FV}{RT} \right)x_B
\label{eq:gate_trans_rate}
\end{align}
where
\begin{align}
\alpha_0 = \kappa K_C \\
\beta_0 = \kappa K_O
\end{align}
The parameters $\alpha_0$, $z_f$, $\beta_0$ and $z_r$ are fitted against mathematical equations in the original model, and then used to determine the bond graph parameters. It is acceptable to fit the kinetic parameters $\alpha_0$ and $\beta_0$ to reduce computation time since the equilibrium points of state transitions are not specified.

\subsection{Channel-specific modelling issues}
\label{sec:channel_specific_issues}
\subsubsection{K\textsuperscript{+} regulation of K\textsuperscript{+} currents}
For the K and K1 channels, Luo and Rudy \cite{luo_dynamic_1994} describe a dependence of the permeability on the square root of extracellular K\textsuperscript{+} concentration. This was incorporated by assigning an additional extracellular K\textsuperscript{+} stoichiometry of 0.5 to both sides of the ion transport reaction.

\subsubsection{Ca\textsuperscript{2+} inactivation of L-type Ca\textsuperscript{2+} current}
Luo and Rudy \cite{luo_dynamic_1994} describe a mechanism whereby the L-type Ca\textsuperscript{2+} channel is inactivated by intracellular Ca\textsuperscript{2+}, using the function:
\begin{align}
f_\mathrm{Ca} = \frac{1}{1 + ([\mathrm{Ca_i^{2+}}]/K_{m,\text{Ca}})^2}
\end{align}
This mechanism was incorporated into the bond graph framework through the reaction:
\begin{align}
A + 2\mathrm{Ca_i} \rightleftharpoons I
\label{eq:Ca_inactivation}
\end{align}
with a dissociation constant equal to $K_{m,\text{Ca}}^2$. It can be shown that at equilibrium:
\begin{align}
\frac{x_A}{x_A + x_I} = \frac{1}{1 + ([\mathrm{Ca_i^{2+}}]/K_{m,\text{Ca}})^2} = f_\mathrm{Ca}
\end{align}
Therefore Ca\textsuperscript{2+} inactivation was incorporated by applying the reaction in Eq. \ref{eq:Ca_inactivation} to each of the states that result from independent $d$ and $f$ gating, using kinetic constants that were sufficiently high to approximate rapid equilibrium.

\subsubsection{f-gate of the L-type Ca\textsuperscript{2+} channel}
\label{sec:f_gate_methods}
Luo and Rudy use the equations from Rasmusson \textit{et al.} \cite{rasmusson_mathematical_1990} for their L-type Ca\textsuperscript{2+} channel $f$-gate, resulting in U-shaped functions for both the steady-state open probability $f_\text{ss}$ and time constant $\tau_f$. Using the exponential dependence in Eq. \ref{eq:gate_trans_rate}, $f_\text{ss}$ must have a monotonic and sigmoidal shape, and $\tau_f$ must either be bell-shaped or monotonic. As neither the $f_\text{ss}$ nor $\tau_f$ could be made U-shaped with the current formulation, we used an alternative mechanism to describe the $f$-gate. We observed that the $f$-gate activated at both negative and positive voltages, and that the minima of $f_\mathrm{ss}$, and $\tau_f$ of the Rasmusson equations appeared to coincide. We modelled the gate using the reaction network $O_1 \xrightleftharpoons[\alpha_1]{\beta_1} C \xrightleftharpoons[\beta_2]{\alpha_2} O_2 \xrightleftharpoons[k_3^+]{k_3^-} O_1$ with the final reaction assumed to be at quasi-equilibrium. The rationale behind using this three-state model was that: (a) there were two open states, one that activated at negative voltages and one that activated at positive voltages, and; (b) the inactivation parameters could be chosen such that the gate inactivated faster than it activated. The initial spike in membrane potential during an action potential implies that the open probability is unable to change, thus we used a reaction in rapid equilibrium to convert between the two open states; without this, the gate would need to pass the closed states to move between the open states.

Similar to the transition parameters in other gates an exponential dependence on voltage was assumed. Since the mechanism involves a biochemical cycle, a detailed balance constraint was used to determine parameters for the third reaction between the two open states:
\begin{align}
\frac{k_3^+(V) } {k_3^- (V)} = \frac{\beta_1 (V) \alpha_2 (V)}{\alpha_1 (V) \beta_2 (V)}
\end{align}

The following information was used to parameterise the $f$-gate:
\begin{enumerate}
	\item The difference between the steady-state open probabilities in the Luo-Rudy model ($f_\text{ss}$) and bond graph model ($f_\text{ss,BG}$) over the range $-90\ \si{mV} \le V \le 50\ \si{mV}$. The open probability of the bond graph formulation was calculated by rapid equilibrium arguments \cite{smith_development_2004}:
	\begin{align}
	f_\text{ss,BG} = \frac{\alpha_1/\beta_1 + \alpha_2/\beta_2}{1 + \alpha_1/\beta_1 + \alpha_2/\beta_2}
	\end{align}
	Differences were taken between the natural logarithms of each of the open probabilities prior to calculating differences to better match lower values.
	\item Simulations of the $f$-gate were run with the voltage held constant. The open probabilities over time were compared to solutions obtained from the Luo-Rudy formulation of the $f$-gate. The conditions for the simulations are summarised in \autoref{tab:f_gate_sim_fit}. For computational efficiency, the third reaction was neglected for the bond graph simulations. All simulations involve either activation/inactiation processes involving one of the open states. It was assumed that very little of the of the other open state would become open.
\end{enumerate} 

\begin{table}[H]
	\caption{\textbf{Summary of conditions used to simulate f-gate for fitting parameters.} $o_1$, $c$ and $o_2$ represent the proportion of the three states representing the inactivation process.}
	\centering
	\begin{tabular}{c c c l}
		\toprule
		\# & Voltage (mV) & Initial conditions & Description \\ \midrule
		1 & $-80$ & $o_1 = 0$, $c=1$, $o_2=0$ & Activation at $-80\ \si{mV}$ \\
		2 & $-40$ & $o_1 = 1$, $c=0$, $o_2=0$ & Inactivation at $-40\ \si{mV}$ \\
		3 & $-40$ & $o_1 = 0$, $c=1$, $o_2=0$  & Activation at $-40\ \si{mV}$\\
		4 & 0 & $o_1 = 1$, $c=0$, $o_2=0$  & Inactivation at $0\ \si{mV}$ from $O_1$\\
		5 &0 & $o_1 = 0$, $c=0$, $o_2=1$  & Inactivation at $0\ \si{mV}$ from $O_2$\\
		6 & 40 & $o_1 = 0$, $c=0$, $o_2=1$ & Inactivation at $40\ \si{mV}$ \\ \bottomrule
	\end{tabular}
	\label{tab:f_gate_sim_fit}
\end{table}

The transition rates for the $f$-gate are
\begin{align}
\alpha_1 (V) &= \alpha_{0,1} \exp \left( \frac{z_{f,1}FV}{RT} \right) \\
\beta_1 (V) &= \beta_{0,1} \exp \left( \frac{z_{r,1}FV}{RT} \right) \\
\alpha_2 (V) &= \alpha_{0,2} \exp \left( \frac{z_{f,2}FV}{RT} \right) \\
\beta_2 (V) &= \beta_{0,2} \exp \left( \frac{z_{r,2}FV}{RT} \right) \\
k_3^+ (V) &= r_3 K_{3,0} \exp \left( \frac{z_{f,3}FV}{RT} \right) \\
k_3^- &=  r_3
\end{align}
with the constants
\begin{align}
&\alpha_{0,1} = 0.8140\ \si{s^{-1}}, \qquad
z_{f,1} = -1.1669\\
&\beta_{0,1} = 36.1898\ \si{s^{-1}} , \qquad
z_{r,1} = 1.6709 \\
&\alpha_{0,2} = 1.6369\ \si{s^{-1}}, \qquad
z_{f,2} = 0.7312 \\
&\beta_{0,2} = 35.5248\ \si{s^{-1}} , \qquad
z_{r,2} = -0.5150 \\
&r_3 = 10000\ \si{s^{-1}} ,\qquad
K_{3,0} = 2.0485 \\
&z_{f,3} = z_{r,1}+z_{f,2}-z_{f,1}-z_{r,2} = 4.0839
\end{align}

\begin{figure}
	\centering
	\begin{tabular}{c c c}
		{\Large \textbf{\textsf{A}}} & 
		{\Large \textbf{\textsf{B}}} & 
		{\Large \textbf{\textsf{C}}} \\
		\includegraphics[width=0.3\linewidth]{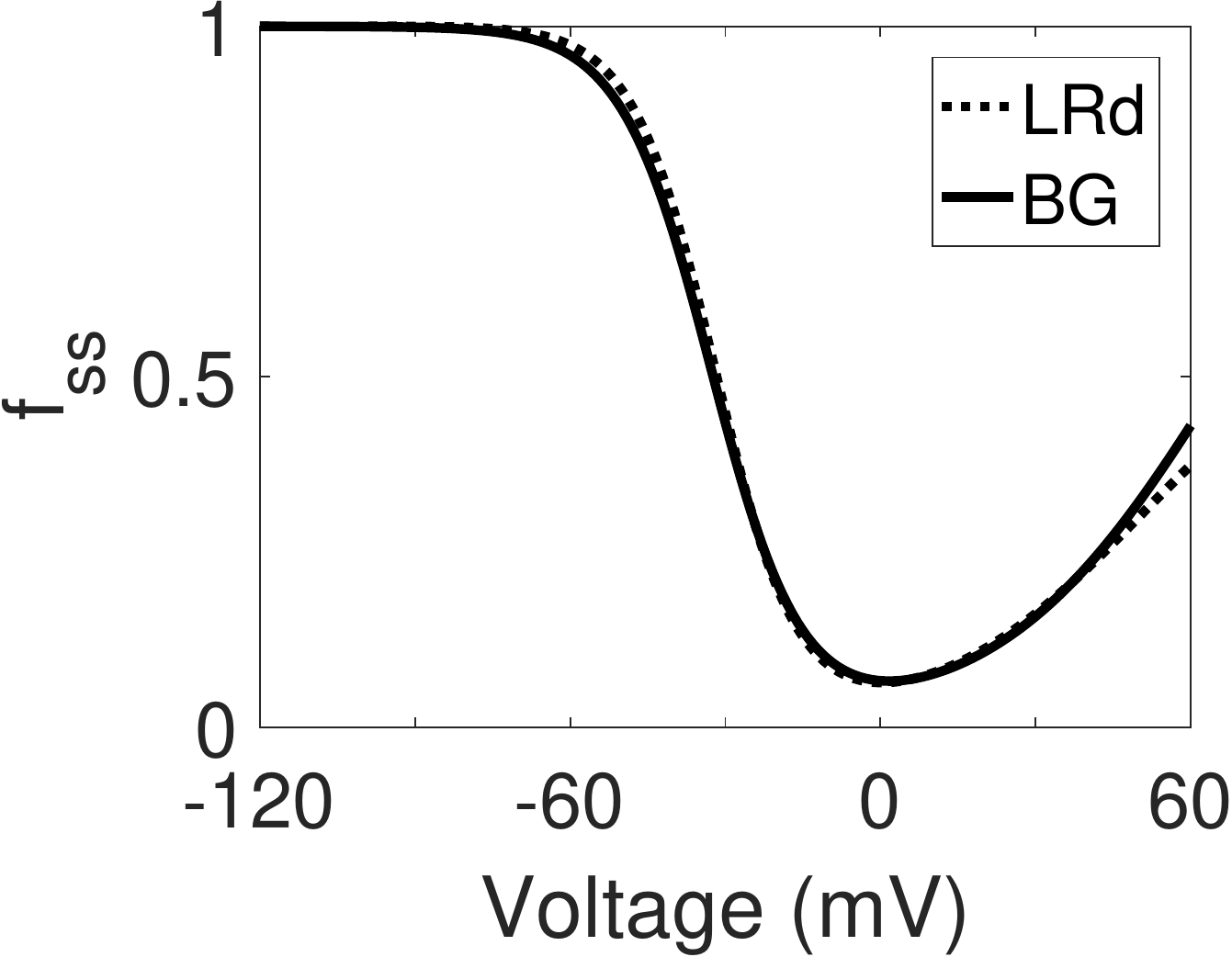} &
		\includegraphics[width=0.3\linewidth]{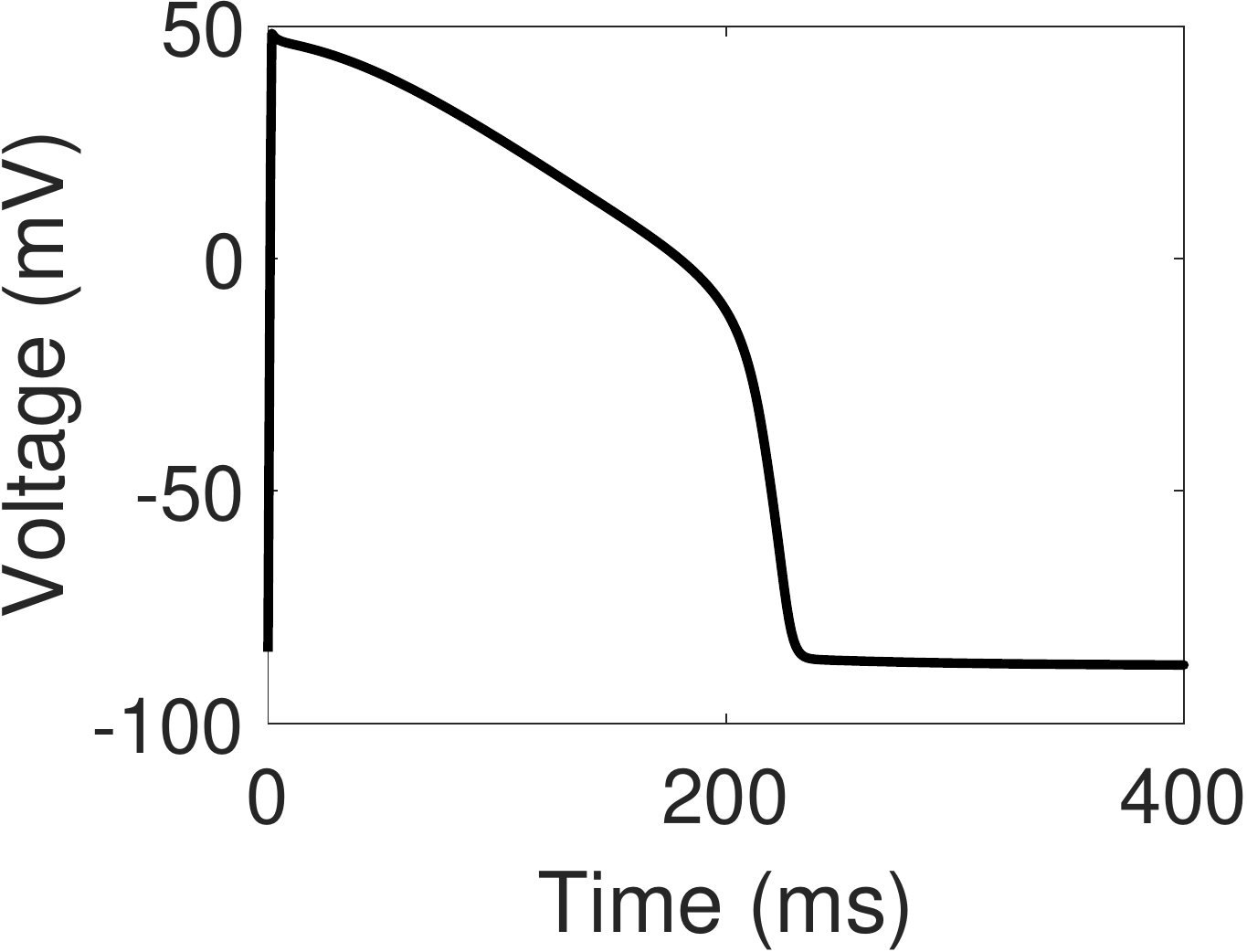} &
		\includegraphics[width=0.3\linewidth]{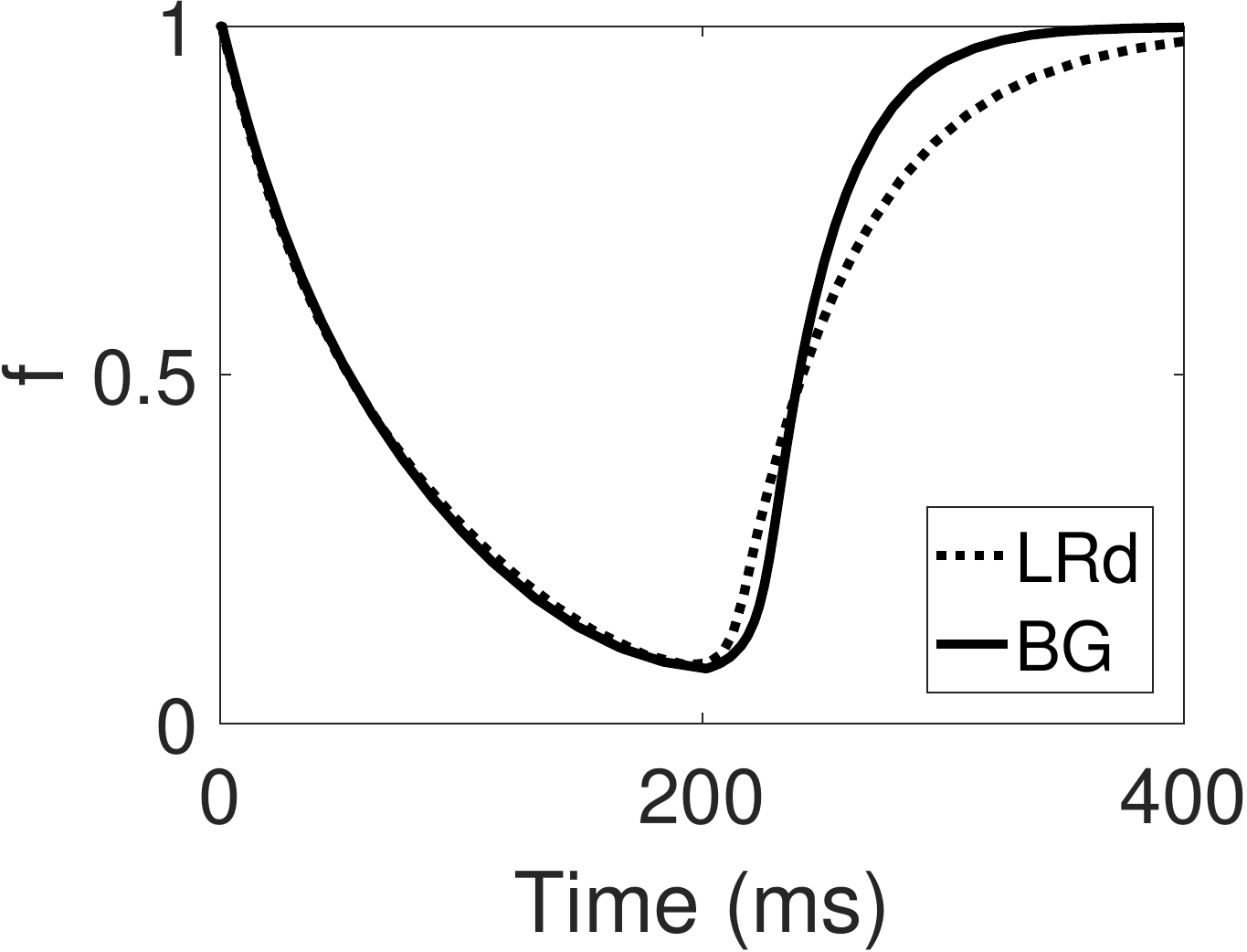}
	\end{tabular}
	\caption{\textbf{Fitting the $f$-gate of the L-type Ca\textsuperscript{2+} channel.} \textbf{(A)} The steady-state open probability of the $f$-gate, calculated by adding the proportion of the two open states. \textbf{(B)} The action potential waveform used to compare the behaviour of the Luo and Rudy (LRd) and bond graph (BG) formulations of the $f$-gate. This was obtained by simulating the Luo-Rudy model with the ion channels used in this study, and holding the ion concentrations constant. \textbf{(C)} The response of the $f$-gates to the voltage signal in B.}
	\label{fig:f_gate}
\end{figure}

The three-state scheme in the bond graph framework produced a similar curve for $f_\mathrm{ss}$ to the $f$-gate of the Luo-Rudy model (\autoref{fig:f_gate}A). Since there is no direct time constant for our three-state model we compared the dynamic behaviour of the $f$-gates by simulating to an action potential-like voltage waveform (\autoref{fig:f_gate}B). During the depolarised phase of the action potential where the $f$-gate steadily inactivates, the bond graph model provides a very good fit to the Luo and Rudy model (\autoref{fig:f_gate}C). In the resting phase the bond graph model reactivates faster, but still provides a reasonable fit.

\section{Fitting ion channel parameters}
\subsection{I-V equations}
\label{sec:IV_fit}
A variety of methods were used to fit permeability constants $P$ for the GHK equations used for the bond graph model. For some channels, $P$ could be determined algebraically (such as the Na\textsuperscript{+} and L-type Ca\textsuperscript{2+} channels). For others, optimisation was required to reduce error between the fitted I-V curve $I_\mathrm{GHK}(V)$ (see Eq. \ref{eq:GHK}) and Luo-Rudy I-V curve $I_\mathrm{LR}(V)$. In these cases, fitting was weighted towards $-90\ \si{mV} \le V \le -30\ \si{mV}$ for $I_\mathrm{K1}$, $-20\ \si{mV} \le V < 30\ \si{mV}$ for $I_\mathrm{K}$, and $0\ \si{mV} \le V \le 60\ \si{mV}$ for $I_\mathrm{Kp}$. These regions were chosen based on when those channels activated. Where applicable, the optimisation problem was carried out by using particle swarm optimisation followed by a local nonlinear optimiser.
The standard concentrations in Luo and Rudy \cite{luo_dynamic_1994} ($\mathrm{[Na_i^+]} = 10\ \si{mM}$, $\mathrm{[Na_e^+]} = 140\ \si{mM}$, $\mathrm{[K_i^+]} = 145\ \si{mM}$, $\mathrm{[K_e^+]} = 5.4\ \si{mM}$, $\mathrm{[Ca_i^+]} = 0.12\ \si{\micro M}$, $\mathrm{[Ca_e^+]} = 1.8\ \si{mM}$) were used to match I-V curves. The permeabilities from fitting I-V curves are summarised in \autoref{tab:GHK_permeabilities}.

\begin{table}[H]
	\caption{\textbf{Permeabilities of the GHK equations used for the bond graph model.}}
	\centering
	\begin{tabular}{c c}
		\toprule
		Permeability &Value (pL/s) \\ \midrule
		$P_\mathrm{Na}$ & 9.0602 \\
		$P_\mathrm{K1}$ & 1.1200 \\
		$P_\mathrm{K}$ & 0.2299 \\
		$P_\mathrm{Kp}$ & 0.0136 \\
		$P_\mathrm{CaL}$ & 28.2471 \\
		$P_\mathrm{KL}$ & 0.0222 \\ \bottomrule
	\end{tabular}
	\label{tab:GHK_permeabilities}
\end{table}

\subsubsection{Sodium current}
The permeability was chosen so match the linear equation at the negative of the Nernst potential \cite{gawthrop_bond_2017-1}:
\begin{align}
P_\mathrm{Na} =  \frac{2 \bar{G}_\mathrm{Na} (1-\exp \left[ FE_\mathrm{Na}/(RT) \right] )}{\mathrm{[Na_i^+]} -\mathrm{[Na_e^+]}\exp \left[ FE_\mathrm{Na}/(RT) \right] } \frac{RT}{F^2}
\end{align}
where
\begin{align}
E_\mathrm{Na} &= \frac{RT}{F} \ln \left(\frac{\mathrm{[Na_e^+]}}{\mathrm{[Na_i^+]}} \right) \\
\bar{G}_\mathrm{Na} &= 2.45\ \si{\micro A / mV}
\end{align}

\subsubsection{Time-independent K\textsuperscript{+} current}
\begin{align}
P_\mathrm{K1} &= \argmin_P \left\{ 
\sum_{V=-90}^{-30}  \left[
I_\mathrm{K1,LR}(V) - I_\mathrm{K1,GHK}(V,P)
\right]^2
\right\} \\
I_\mathrm{K1,LR}(V) &= \bar{G}_\mathrm{K1} (V - E_\mathrm{K}) \\
\bar{G}_\mathrm{K1} &= 1.1505 \times 10^{-4}\ \si{\micro A / mV} \\
E_\mathrm{K} &= \frac{RT}{F} \ln \left( 
\frac{\mathrm{[K_e^+]}}
{\mathrm{[K_i^+]}}
\right) \\
\end{align}

\subsubsection{Time-dependent K\textsuperscript{+} current}
\begin{align}
P_\mathrm{K} &= \argmin_P \left\{ 
\sum_{V=-20}^{29}  \left[
I_\mathrm{K,LR}(V) - I_\mathrm{K,GHK}(V,P)
\right]^2
\right\} \\
I_\mathrm{K,LR}(V) &= \bar{G}_\mathrm{K} (V - E_\mathrm{K,LR}) \\
\bar{G}_\mathrm{K} &= 4.3259 \times 10^{-5}\  \si{\micro A / mV} \\
E_\mathrm{K,LR} &= \frac{RT}{F} \ln \left( 
\frac{\mathrm{[K_e^+]} + P_\mathrm{Na,K} \mathrm{[Na_e^+]}}
{\mathrm{[K_i^+]} + P_\mathrm{Na,K} \mathrm{[Na_i^+]}}
\right) \\
P_\mathrm{Na,K} &= 0.01833
\end{align}

\subsubsection{Plateau K\textsuperscript{+} current}
\begin{align}
P_\mathrm{Kp} &= \argmin_P \left\{ 
\sum_{V=0}^{60}  \left[
I_\mathrm{Kp,LR}(V) - I_\mathrm{Kp,GHK}(V,P)
\right]^2
\right\} \\
I_\mathrm{Kp,LR}(V) &= \bar{G}_\mathrm{Kp} (V - E_\mathrm{K}) \\
\bar{G}_\mathrm{Kp} &= 2.8072 \times 10^{-6}\ \si{\micro A / mV}  \\
\end{align}
$E_\mathrm{K}$ same as for the time-independent K\textsuperscript{+} current.

\subsubsection{L-type Ca\textsuperscript{2+} channel}
For the L-type Ca\textsuperscript{2+} channel, Luo and Rudy \cite{luo_dynamic_1994} use the I-V equation
\begin{align}
I_\mathrm{Ca} = P_\mathrm{Ca} \frac{z^2 F^2 V}{RT} \frac{\gamma_{\text{Cai}}[\mathrm{Ca_i^{2+}}] \exp (zFV/RT) - \gamma_{\text{Cae}}[\mathrm{Ca_e^{2+}}] }{\exp (zFV/RT) - 1}
\end{align}
which resembles the GHK equation, but allows thermodynamic laws to be broken through the use of different partitioning factors $\gamma_{si}$ and $\gamma_{so}$. In the case of the Ca\textsuperscript{2+} component of the current, this was resolved by setting both factors to the value of $\gamma_\mathrm{Cao}$, with little effect on the I-V curve. Thus the permeabilities of the GHK equations are calculated as follows:
\begin{align}
P_\mathrm{CaL} &= P_\mathrm{CaL,LR} \gamma_\mathrm{Cae} \\
P_\mathrm{KL} &= P_\mathrm{KL,LR} \gamma_\mathrm{Ke} = P_\mathrm{KL,LR} \gamma_\mathrm{Ki} 
\end{align}
where
\begin{align}
P_\mathrm{CaL,LR} &=  8.2836 \times 10^{-8}\ \si{cm^3 /s}\\
P_\mathrm{KL,LR}  &= 2.9606 \times 10^{-11}\ \si{cm^3 /s}\\
\gamma_\mathrm{Cae} &= 0.341\\
\gamma_\mathrm{Ke} &= \gamma_\mathrm{Ki} = 0.75
\end{align}

\begin{figure}
	\centering
	\begin{tabular}{c c}
		\textbf{\textsf{(A) Na}} & \textbf{\textsf{(B) K1}} \\
		\includegraphics[width=0.4\linewidth]{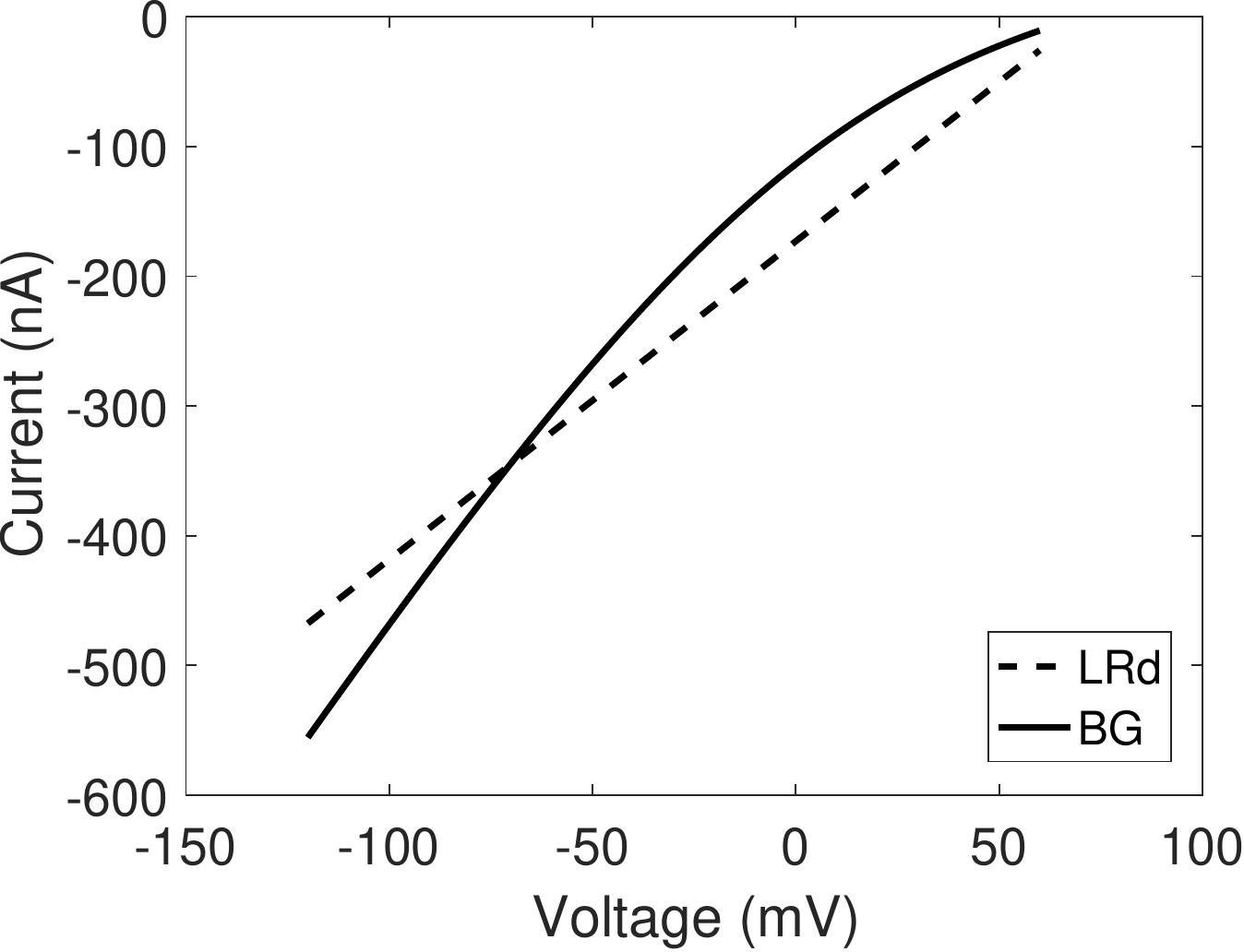} &
		\includegraphics[width=0.4\linewidth]{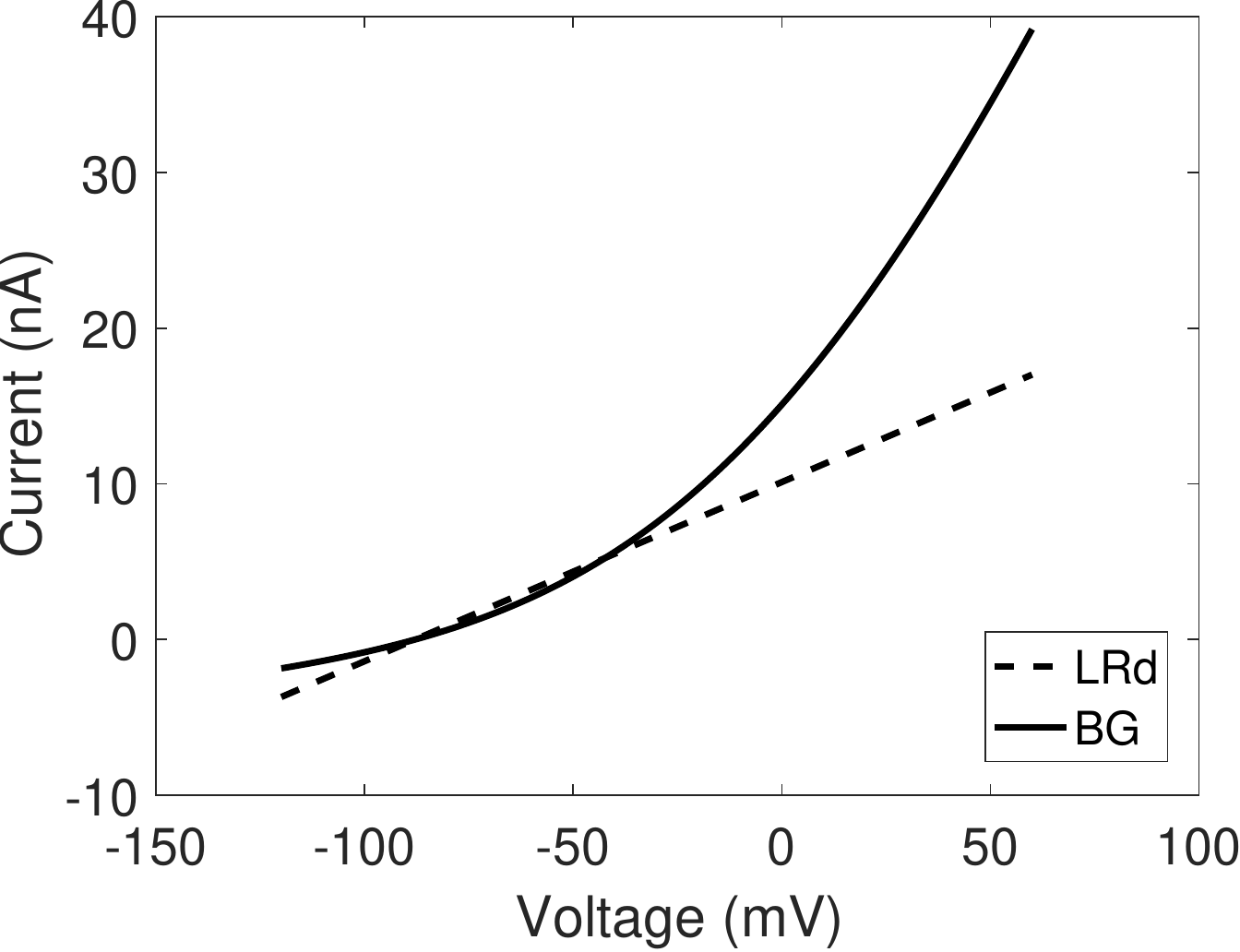} \\
		\textbf{\textsf{(C) K}} & \textbf{\textsf{(D) Kp}} \\
		\includegraphics[width=0.4\linewidth]{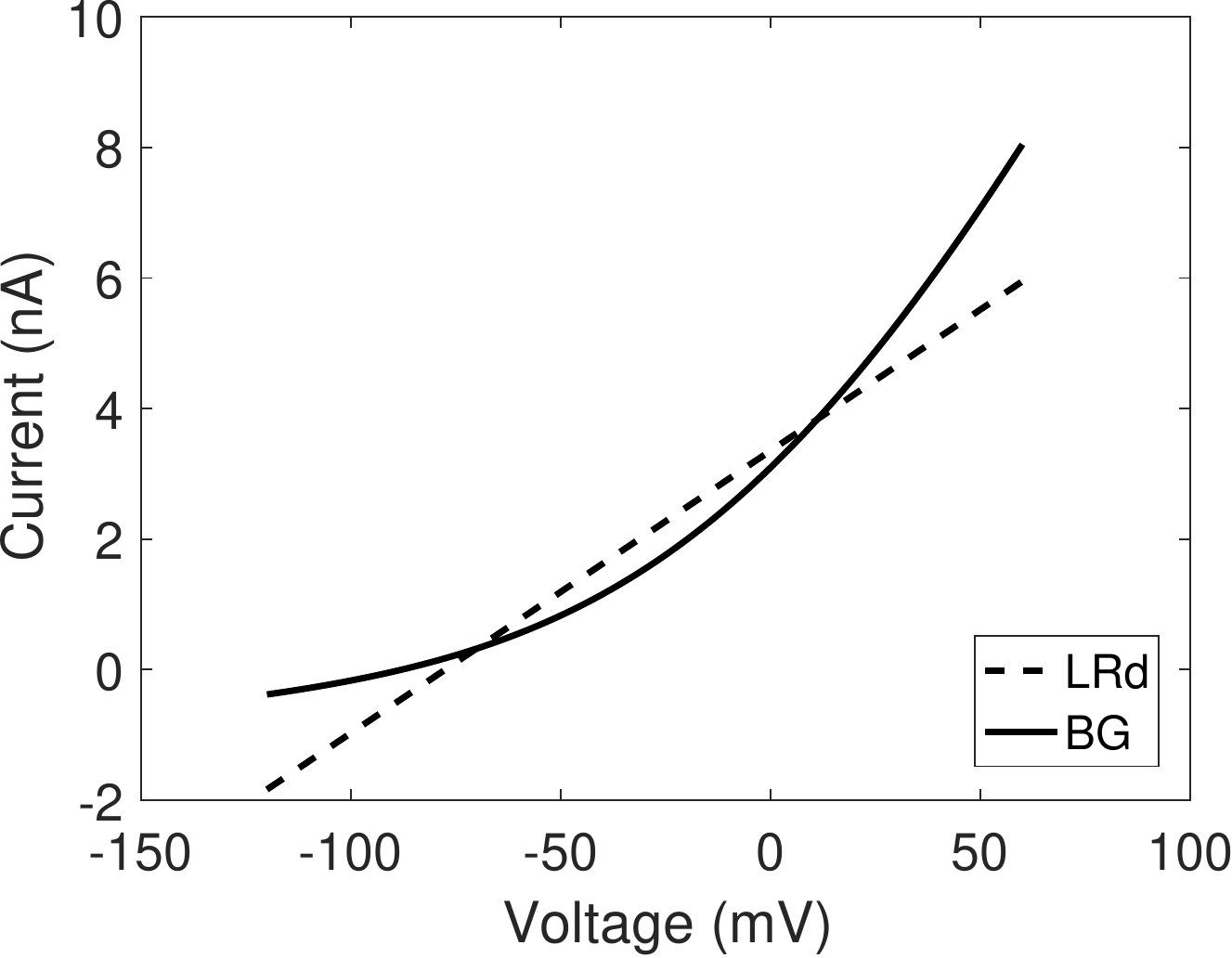} &
		\includegraphics[width=0.4\linewidth]{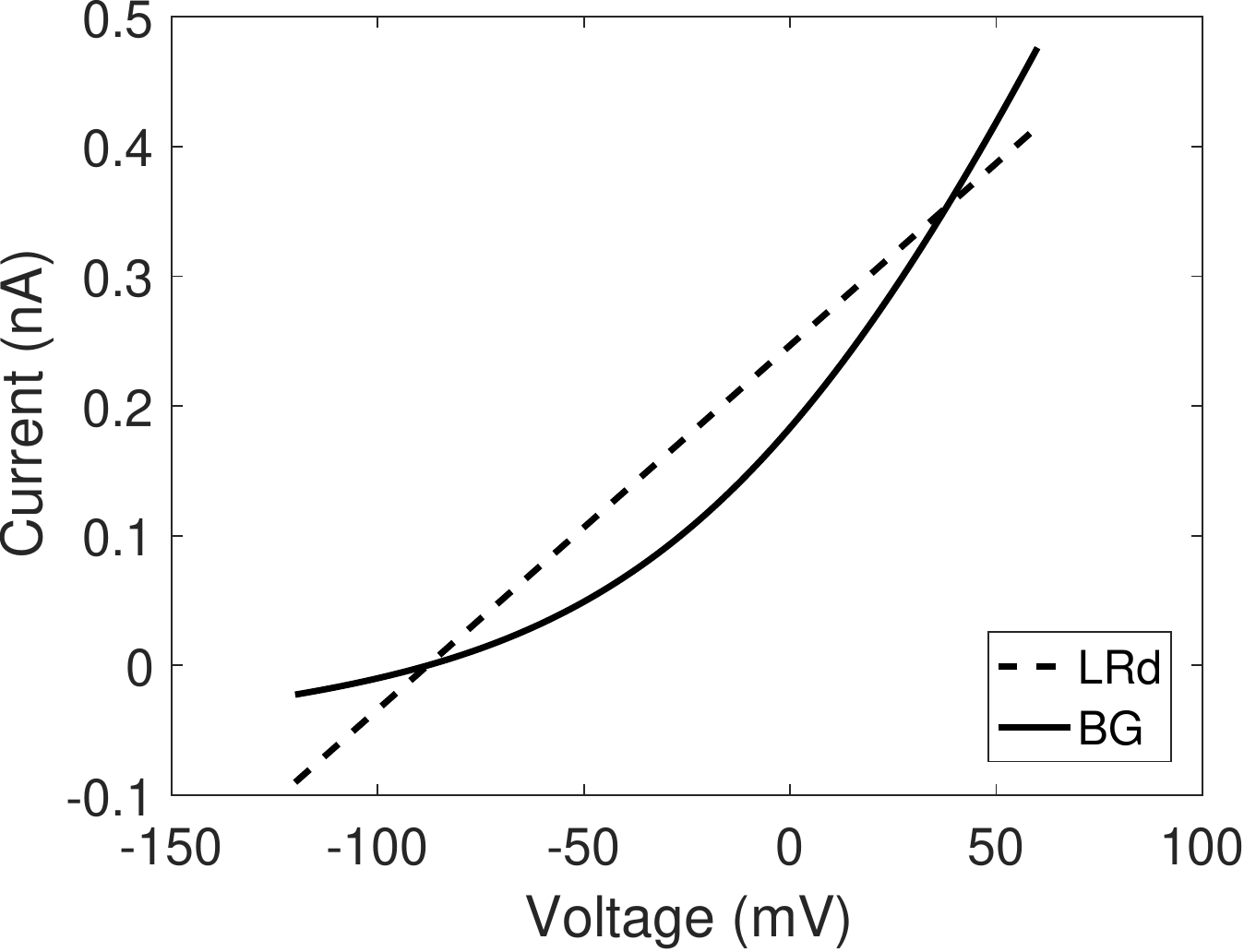} \\
		\textbf{\textsf{(E) LCC, Ca\textsuperscript{2+} current}} & \textbf{\textsf{(F) LCC, K\textsuperscript{+} current}} \\
		\includegraphics[width=0.4\linewidth]{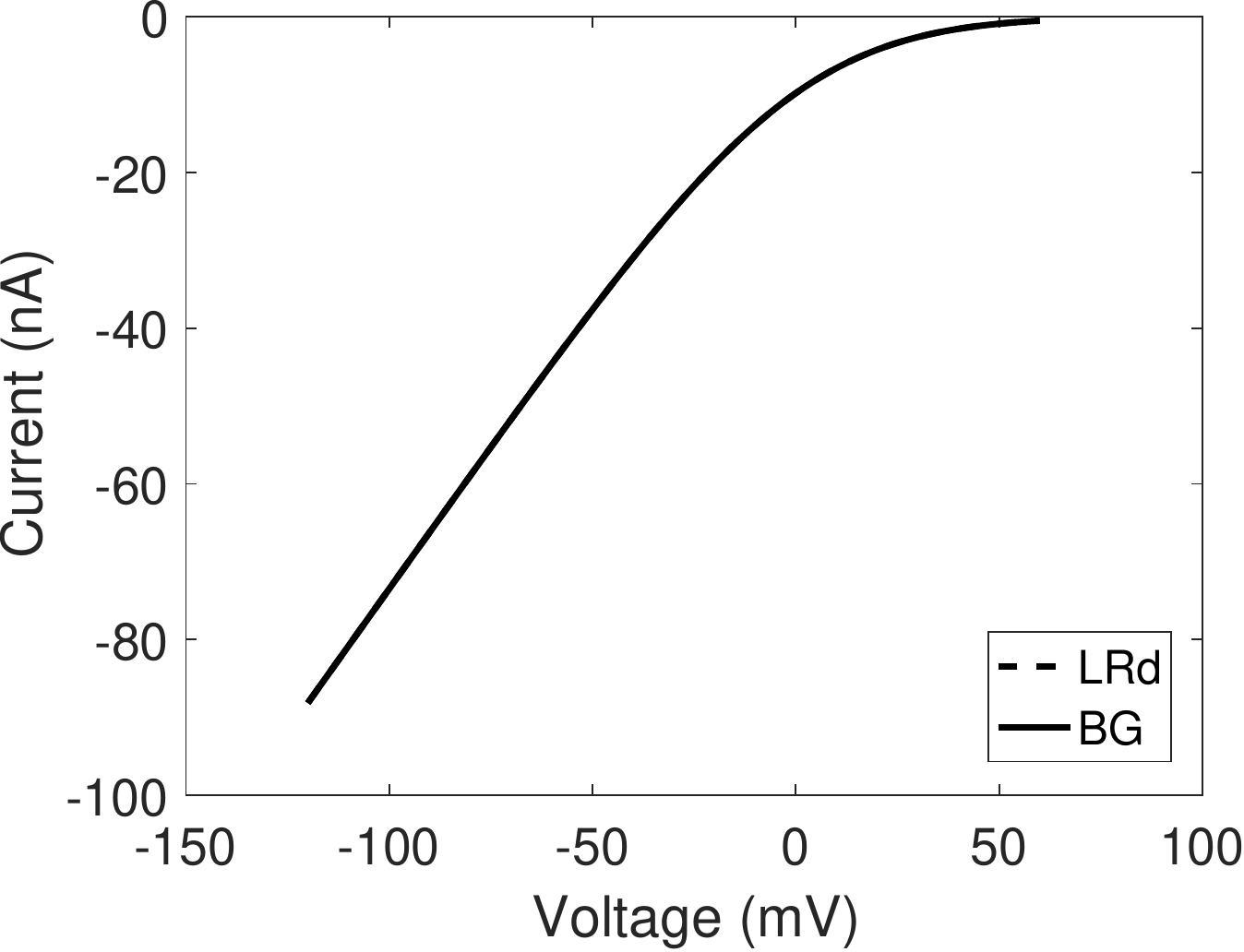} &
		\includegraphics[width=0.4\linewidth]{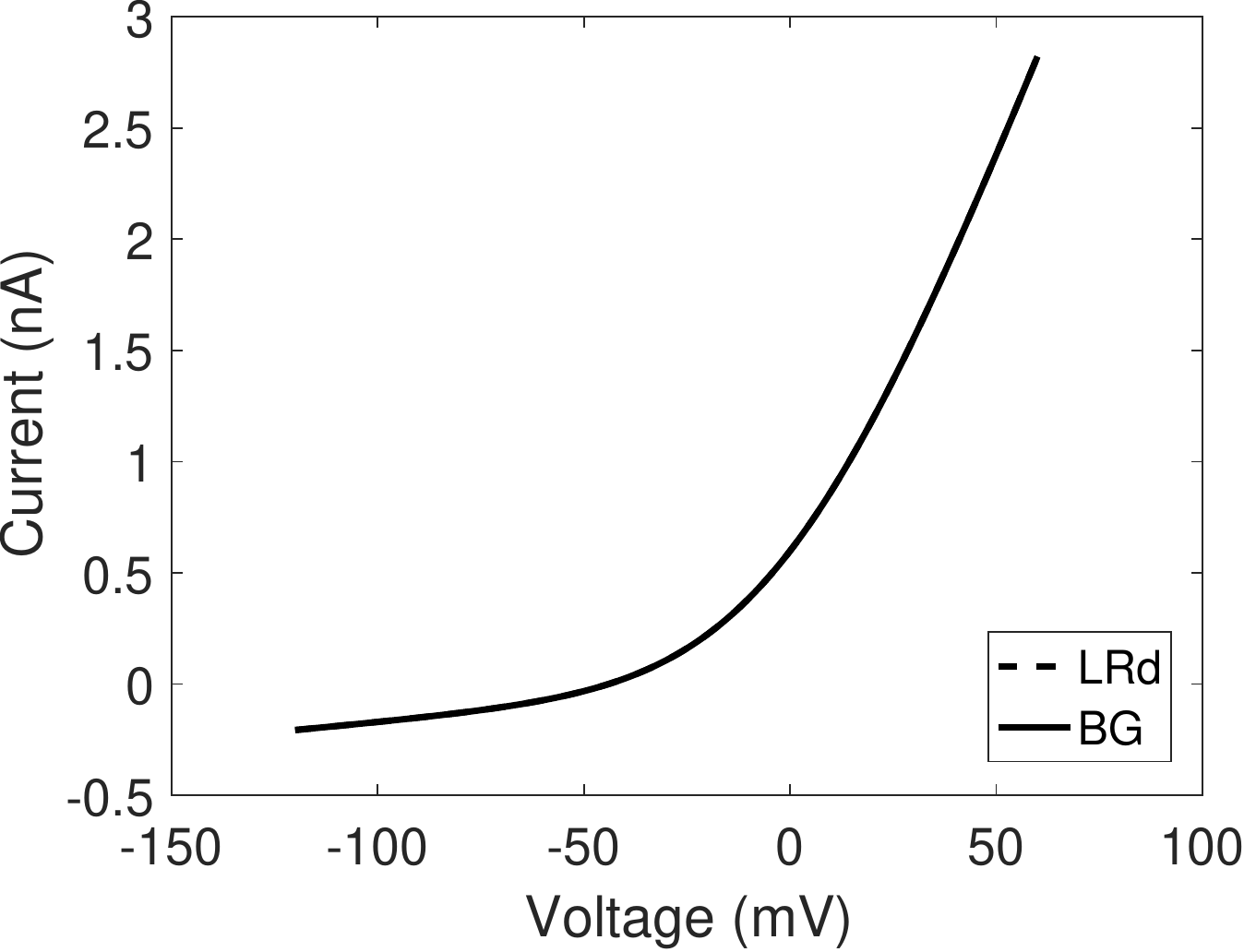}
	\end{tabular}
	\caption{\textbf{Comparison of I-V curves between the Luo-Rudy (LRd) and bond graph (BG) models.} \textbf{(A)} $I_\mathrm{Na}$; \textbf{(B)} $I_\mathrm{K1}$; \textbf{(C)} $I_\mathrm{K}$; \textbf{(D)} $I_\mathrm{Kp}$; \textbf{(E)} $I_\mathrm{Ca,L}$; \textbf{(F)} $I_\mathrm{K,L}$.}
	\label{fig:IV_fitting}
\end{figure}

\subsubsection{Model comparison}
A comparison of the resulting I-V curves is given in \autoref{fig:IV_fitting}. The Na\textsuperscript{+} channel I-V curves appeared to match reasonably well (\autoref{fig:IV_fitting}A), with some discrepancies at positive membrane potentials. For K\textsuperscript{+} channels (\autoref{fig:IV_fitting}B--D), we attempted to optimise the fit across voltages that correspond to their physiological function. Accordingly, for $I_\mathrm{K1}$ ($-90\ \si{mV}\le V < -30\ \si{mV}$), $I_\mathrm{K}$ ($-20\ \si{mV}\le V \le 30\ \si{mV}$) and $I_\mathrm{Kp}$ ($V>0\ \si{mV}$) the I-V curves matched reasonably well in these regions. Discrepancies occurred outside these ranges of voltages, but appeared to only cause minor differences to the currents. In their implementation of $I_\mathrm{K}$, Luo and Rudy \cite{luo_dynamic_1994} use a thermodynamically inconsistent I-V equation where the current is nonzero at the Nernst potential for K\textsuperscript{+}. Despite this, bond graph parameters could still be chosen to give a reasonable fit to this I-V equation (\autoref{fig:IV_fitting}C). Because the Luo-Rudy model based their L-type Ca\textsuperscript{2+} I-V curves on the GHK equation, there was a far closer match between the bond graph and Luo-Rudy models for these currents, (\autoref{fig:IV_fitting}E,F) and the K\textsuperscript{+} curve was matched exactly (\autoref{fig:IV_fitting}F).

\subsection{Gating transition parameters}
\label{sec:gating_fit}
The parameters derived for gate transition are summarised in \autoref{tab:gate_trans_params}, with further detail described below.

\begin{table}[H]
	\caption{\textbf{Gate transition parameters.}}
	\centering
	\begin{tabular}{c c c c c}
		\toprule
		Gate & $\alpha_0$ ($\si{s^{-1}}$) & $z_f$ & $\beta_0$ ($\si{s^{-1}}$) & $z_r$ \\ \midrule
		$m$ & 12516.4361 & 0.4954 & 79.9996 & $-$2.4284 \\
		$h$ & 0.00033539 & $-$4.1892 & 799.9028 & 1.2995 \\
		$j$ & 0.00013079 & $-$4.0381 & 422.7582 & 1.4281 \\
		K1 & 1127.3395 & 0.0336 & 13544806.3586 & 3.1153 \\
		X & 2.2317 & 0.5192 & 0.5750 & $-$0.7317 \\
		Xi & 995.8931 & 0 & 172.6026 & 0.8322 \\
		Kp & 999.8464 & 0 & 3497.4018 & $-$4.4669 \\
		$d$ & 486.7619 & 2.1404 & 98.0239 & $-$2.1404 \\
		$f$ & \multicolumn{4}{c}{See \autoref{sec:channel_specific_issues}}\\ \bottomrule
	\end{tabular}
	\label{tab:gate_trans_params}
\end{table}

\subsubsection{m, h, j, K1 and X-gates}
A vector quantity $\mathbf{p} = (\alpha_0,z_f,\beta_0,z_r) $ was optimised based on the quality of fits to the transition parameters, steady-state open probability and time constant in the range $-120\ \si{mV} \le V \le 60\ \si{mV}$:
\begin{align}
\mathbf{p}_g = \argmin \left\{
\sum_{V=-120}^{60} a(V) \left(
a_\alpha\left[
\alpha_{g,\text{LR}}(V) - \alpha_g (V,\mathbf{p})
\right]^2 + 
a_\beta\left[
\beta_{g,\text{LR}}(V) - \beta_g (V,\mathbf{p})
\right]^2 
\right.
\right. \nonumber \\
\left. \vphantom{\sum_{V=-120}^{60} } \left.
+ a_\text{gss} \left[
g_{ss,\text{LR}}(V) - g_\text{ss} (V,\mathbf{p})
\right]^2 +
a_\tau\left[
\tau_{g,\text{LR}}(V) - \tau_g (V,\mathbf{p})
\right]^2
\right) \right\}
\end{align}
where $g$ is replaced with $m$, $h$, $j$, K1 or $X$ depending on the gate.
$a(V) = 1$ and $a_\alpha=a_\beta=a_\text{gss}=a_\tau = 1$ for the $m$, $h$ and $j$ gates. For the K1 gate, $a(V) = 1$, $a_\alpha=a_\beta=0$, $a_\tau=1$ and $a_\text{gss} = 1000$. For the X gate, 
\begin{align}
a_\text{gss} = 100, \qquad a(V)  = \begin{cases}
1, & V < 0\ \si{mV} \\
25, & V \ge 0\ \si{mV} \\
\end{cases}
\end{align}
The parameters $\alpha_g$ and $\beta_g$ have unit $\si{ms^{-1}}$ and $\tau_g$ has unit ms. Optimisation was carried out using particle swarm optimisation followed by a local optimiser. 

\subsubsection{Xi-gate}
To give a perfect fit for $\mathrm{Xi_{ss}}$, 
\begin{align}
\alpha_0 &= K_\mathrm{Xi} \\
\beta_0 &= K_\mathrm{Xi} e^{56.26/32.1} \\
z_f &= 0 \\
z_r &= \frac{RT}{F} \frac{1000\si{mV/V}}{32.1\si{mV}} = 0.8322
\end{align}
To achieve a time constant of less than 1ms in the range $-120\ \si{mV}\le V \le 60\ \si{mV}$, we chose
\begin{align}
K_\mathrm{Xi} = 0.9959
\end{align}

\subsubsection{Kp-gate}
To give a perfect fit for $\mathrm{Kp_{ss}}$, 
\begin{align}
\alpha_0 &= K_\mathrm{Kp} \\
\beta_0 &= K_\mathrm{Kp}e^{7.488/5.98} \\
z_f &= 0 \\
z_r &= \frac{RT}{F} \frac{1000\si{mV/V}}{5.98\si{mV}} = -4.4669
\end{align}
To achieve a time constant of less than 1ms in the range $-120\ \si{mV}\le V \le 60\ \si{mV}$, we chose
\begin{align}
K_\mathrm{Kp} = 0.9998
\end{align}

\subsubsection{d-gate}
To give a perfect fit for $d_\mathrm{{ss}}$, 
\begin{align}
\alpha_0 &= K_d e^{10/12.48} \\
\beta_0 &= K_d e^{-10/12.48} \\
z_f &= \frac{RT}{F} \frac{1000\si{mV/V}}{12.48\si{mV}} = 2.1404 \\
z_r &= -\frac{RT}{F} \frac{1000\si{mV/V}}{12.48\si{mV}} = -2.1404
\end{align}
$K_d$ was chosen to match the peak time constant because that is where changes would be most likely to make a difference given that the time constant is small:
\begin{align}
K_d = 0.2184
\end{align}

\begin{figure}
	\centering
	\begin{tabular}{c c}
		\imagetop{
			\begin{tabular}{c c}
				\multicolumn{2}{c}{\Large Na current} \\
				\multicolumn{2}{c}{\textbf{(A)} $m$-gate} \\
				\includegraphics[width=0.2\linewidth]{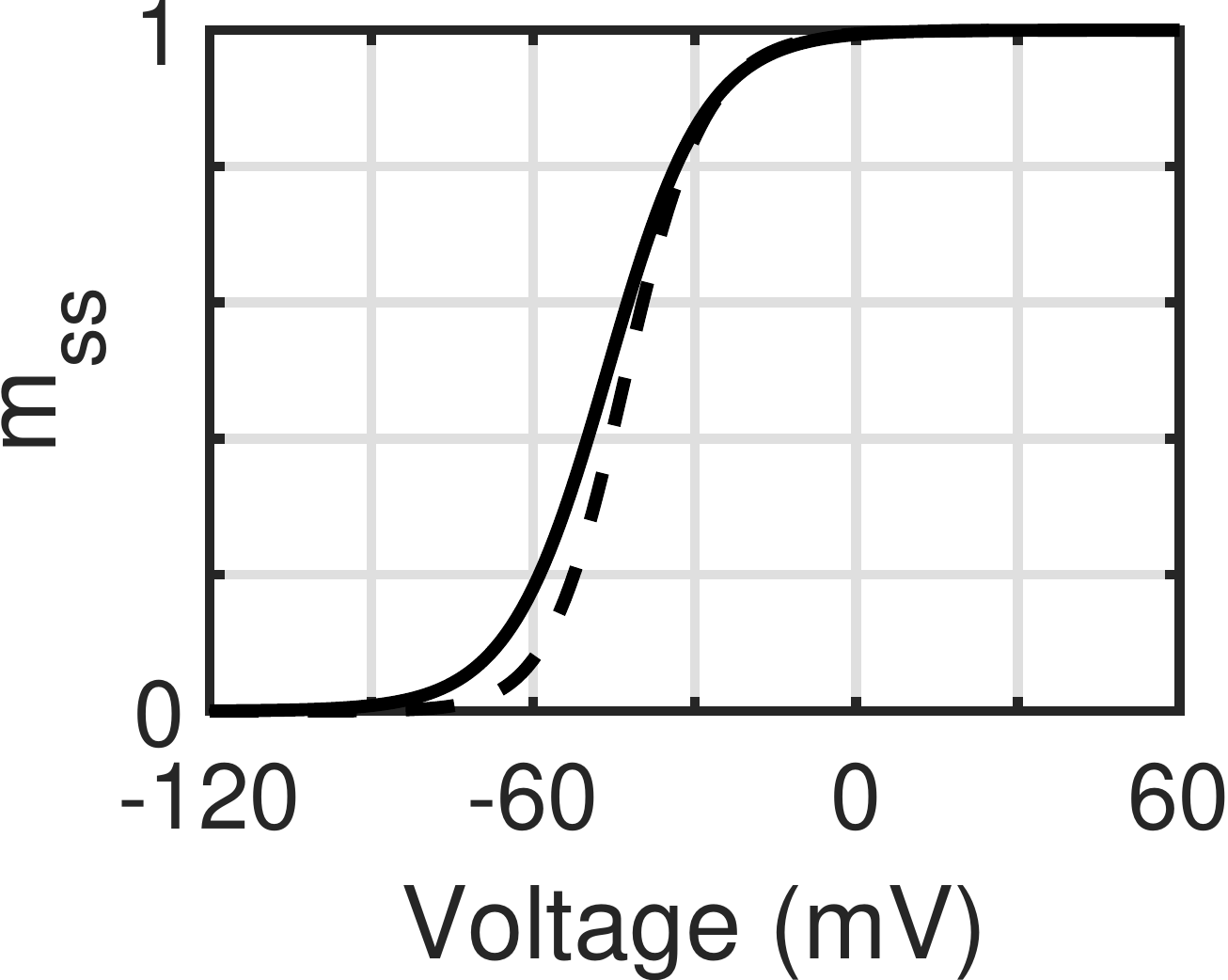} & 
				\includegraphics[width=0.2\linewidth]{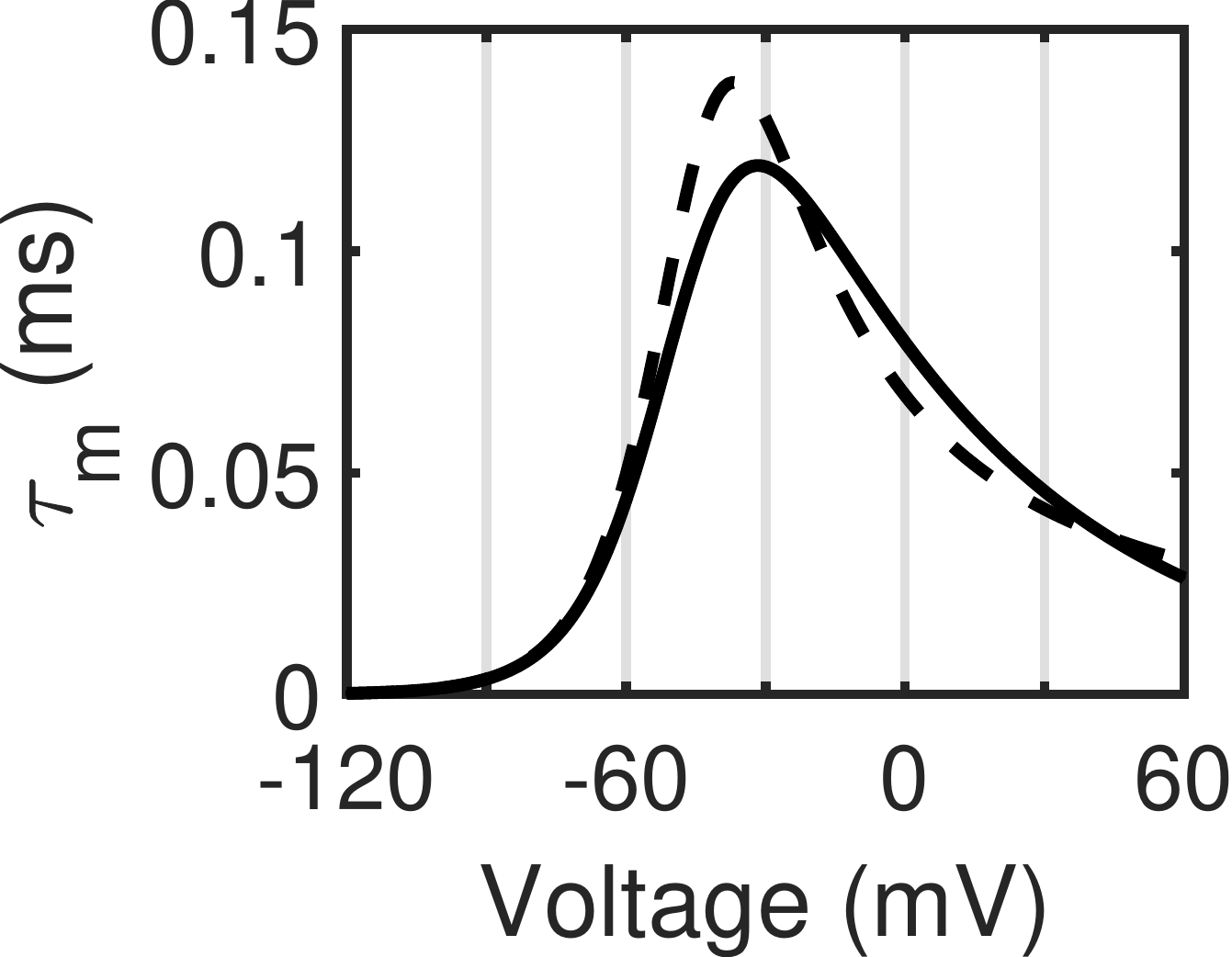} \\
				\multicolumn{2}{c}{\textbf{(B)} $h$-gate} \\
				\includegraphics[width=0.2\linewidth]{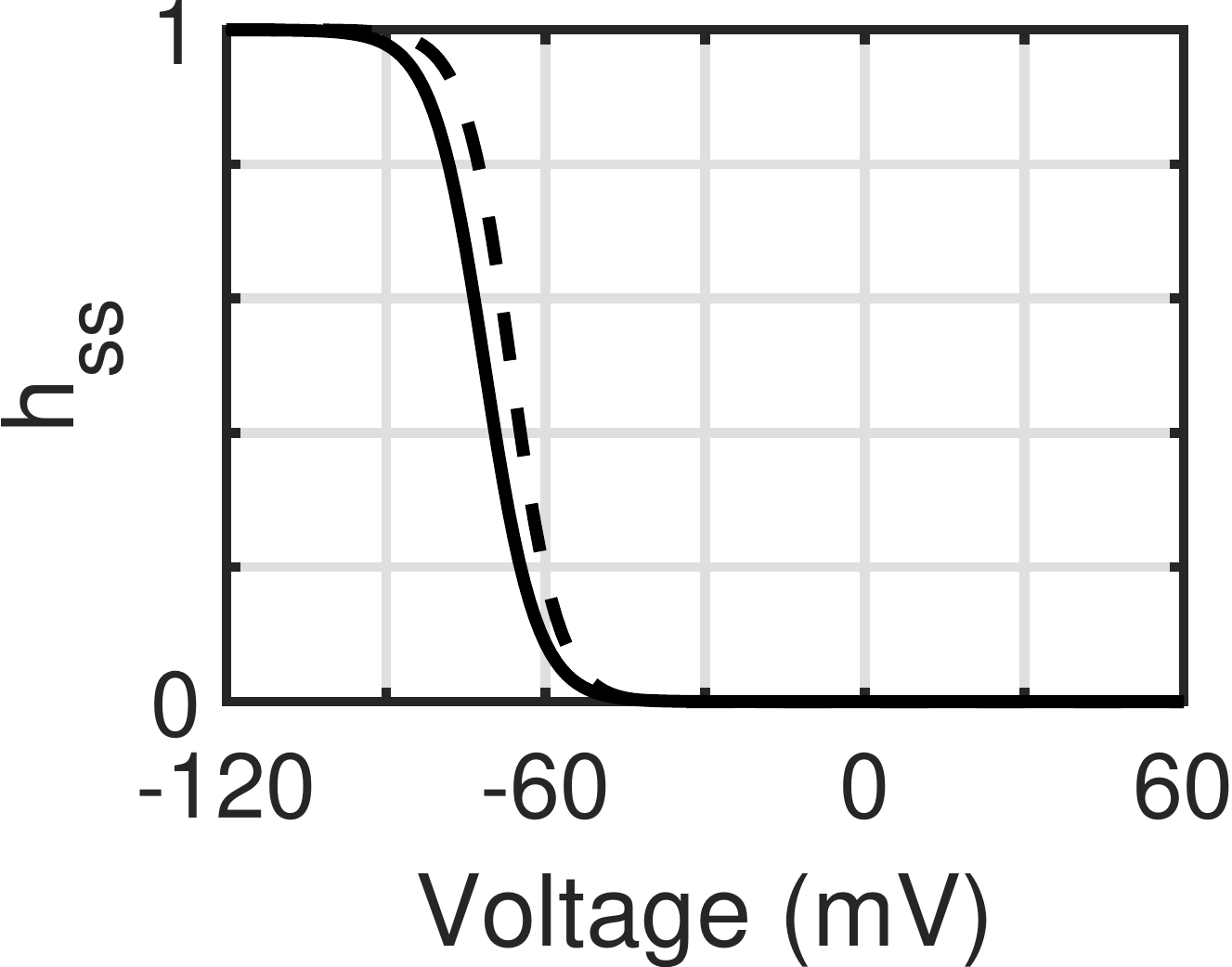} & 
				\includegraphics[width=0.2\linewidth]{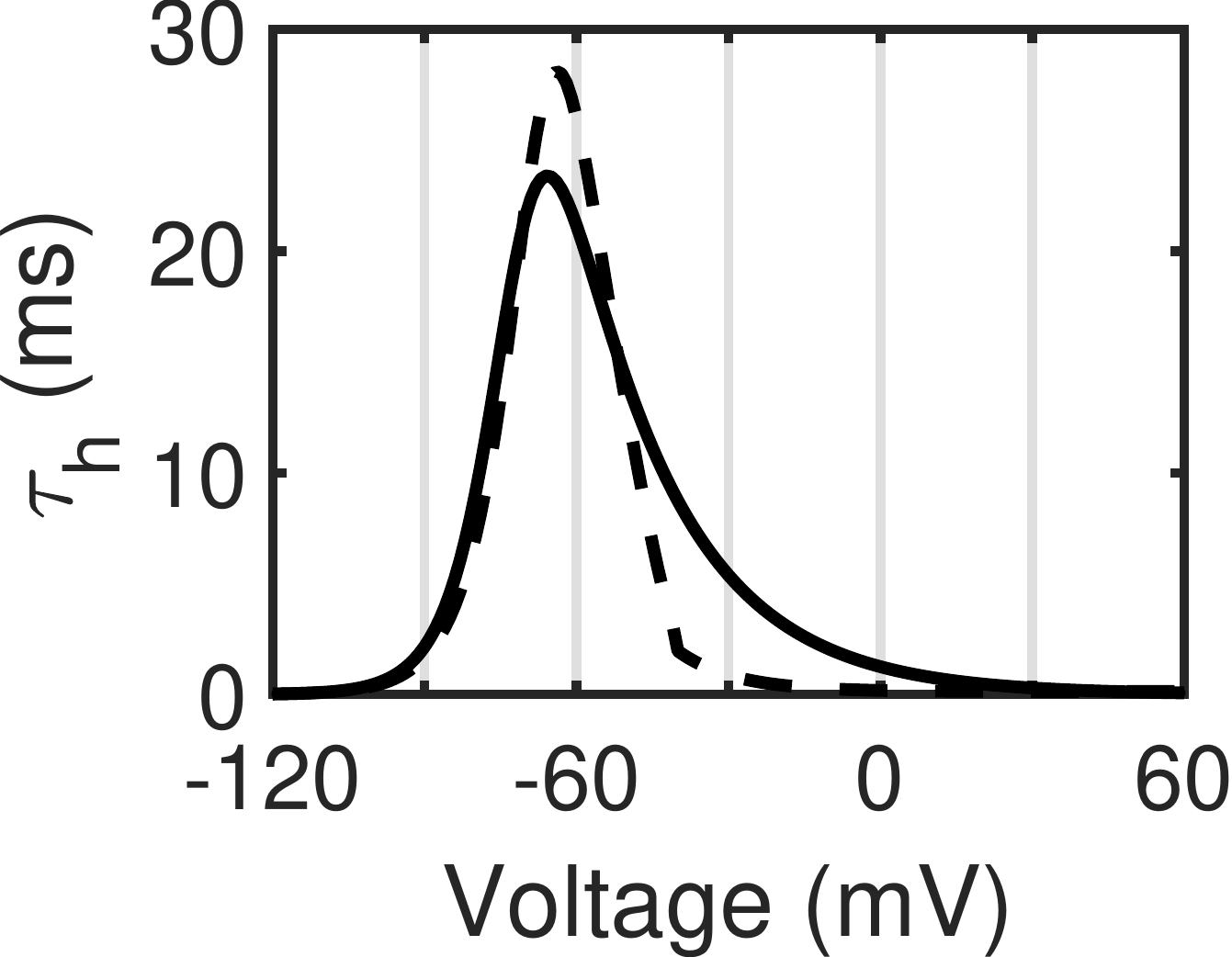} \\
				\multicolumn{2}{c}{\textbf{(C)} $j$-gate} \\
				\includegraphics[width=0.2\linewidth]{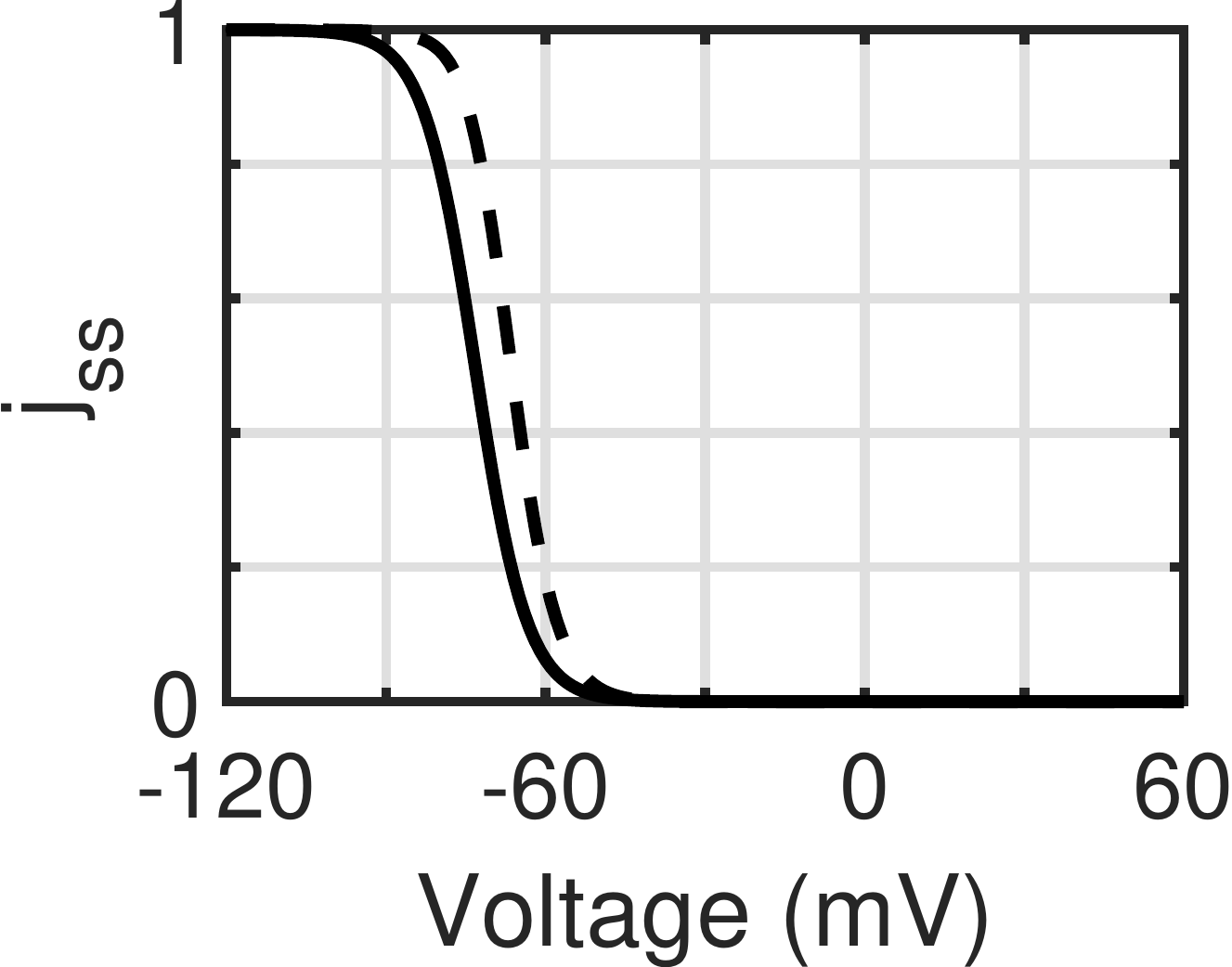} & 
				\includegraphics[width=0.2\linewidth]{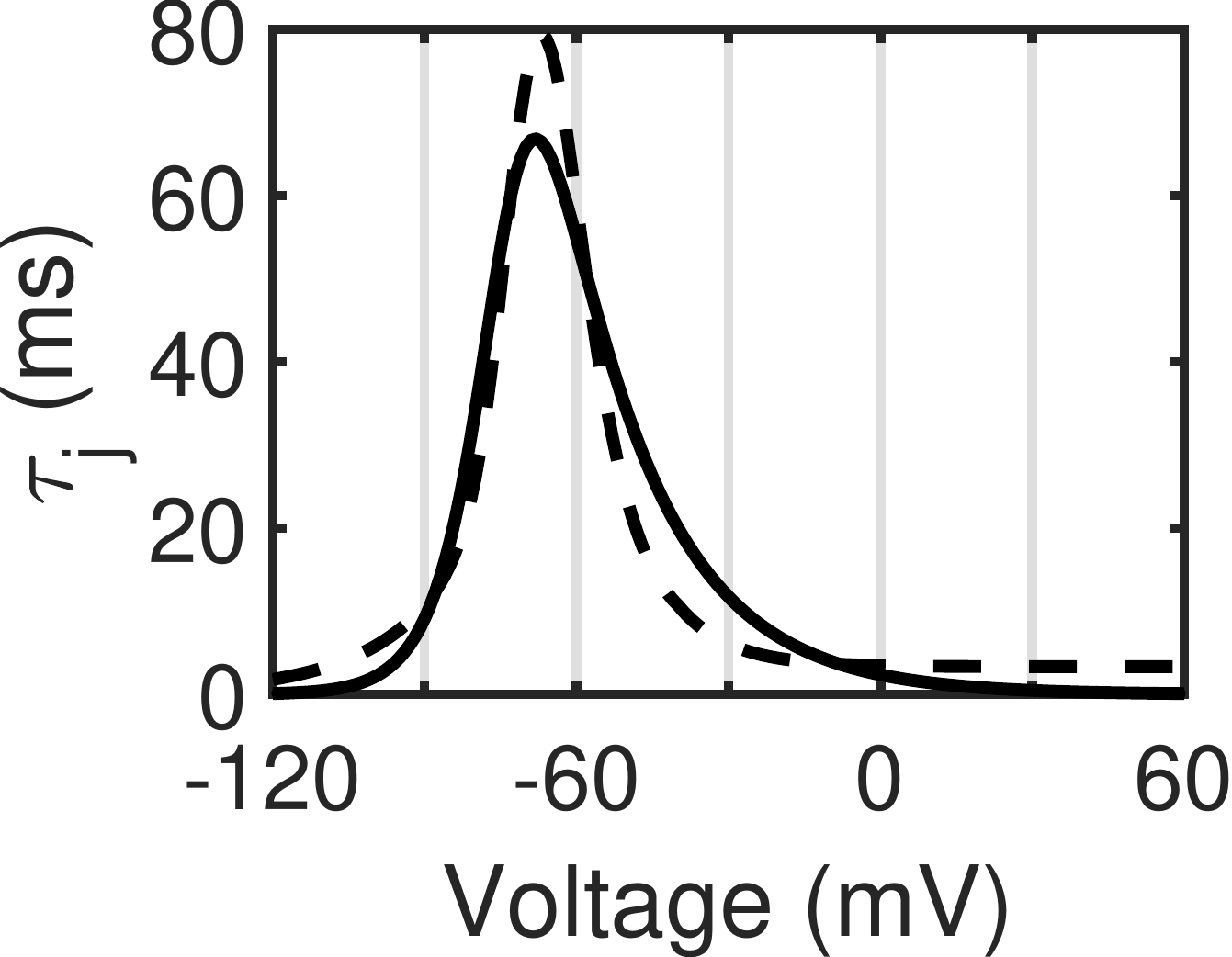} \\[0.5cm]
				\multicolumn{2}{c}{\Large Time-independent $\mathrm{K^+}$ current} \\
				\multicolumn{2}{c}{\textbf{(D)} K1 inactivation gate} \\
				\includegraphics[width=0.2\linewidth]{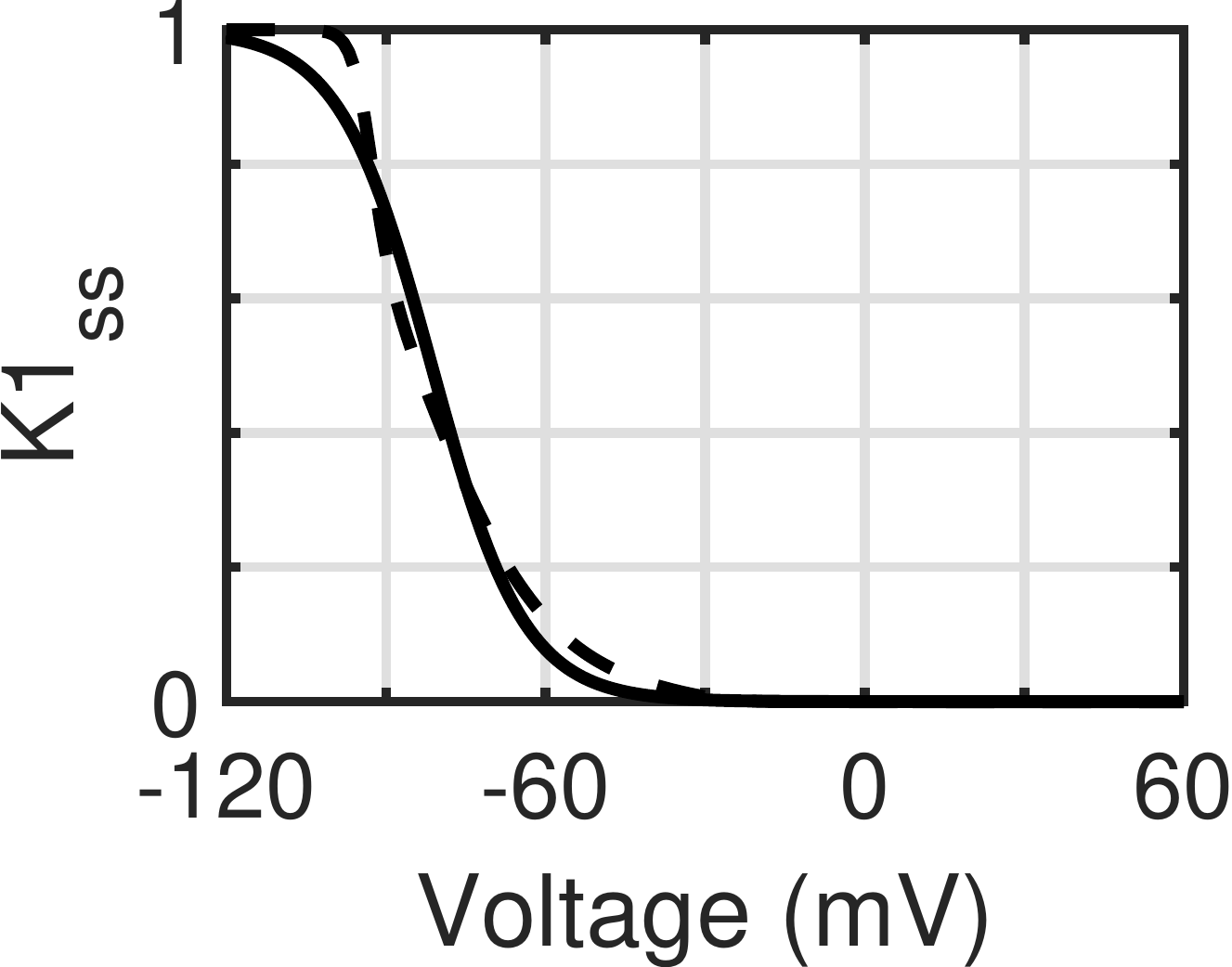} & 
				\includegraphics[width=0.2\linewidth]{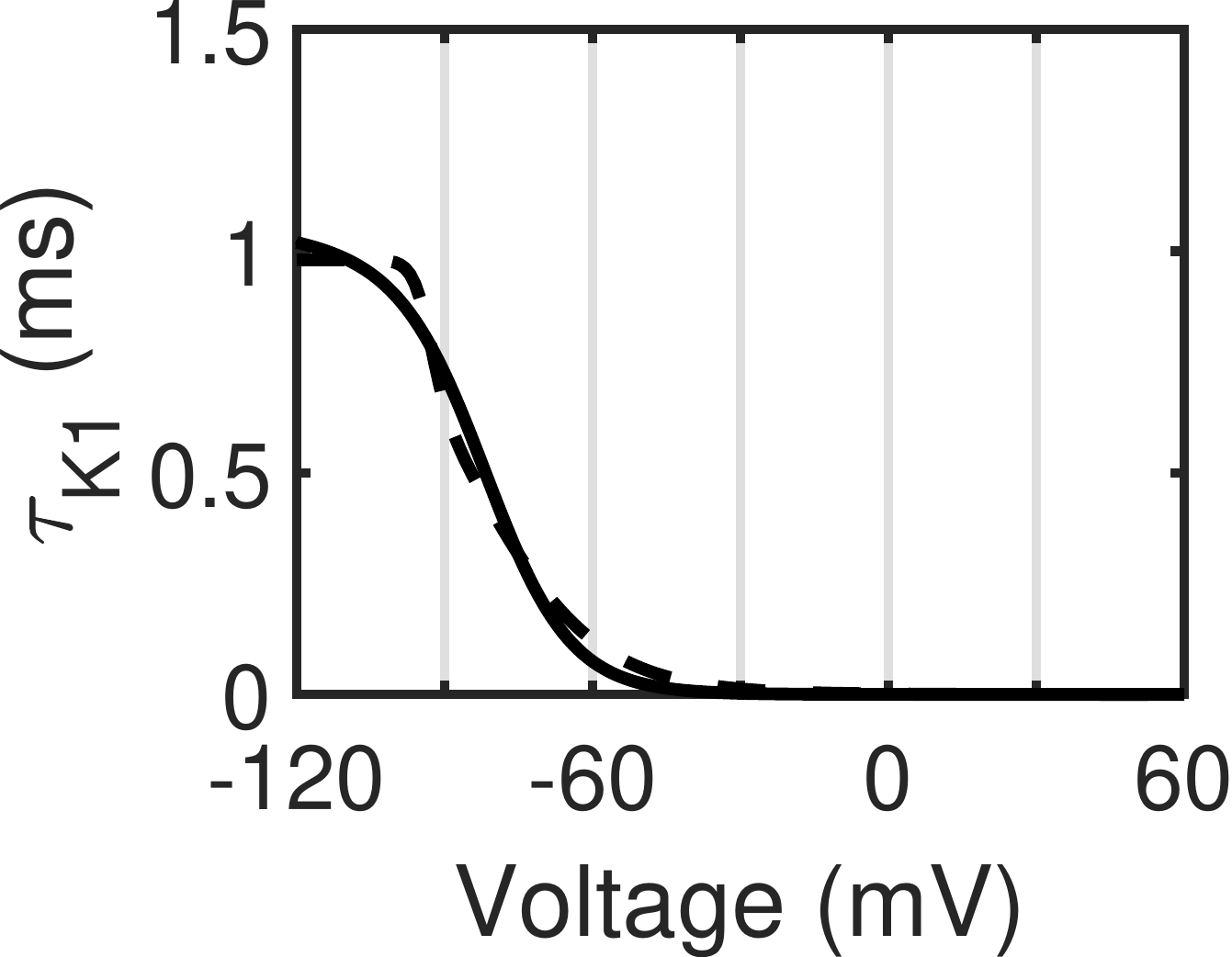} \\
		\end{tabular}}
		&
		\imagetop{
			\begin{tabular}{c c}
				\multicolumn{2}{c}{\Large Time-dependent $\mathrm{K^+}$ current} \\
				\multicolumn{2}{c}{\textbf{(E)} Activation gate (X)} \\
				\includegraphics[width=0.2\linewidth]{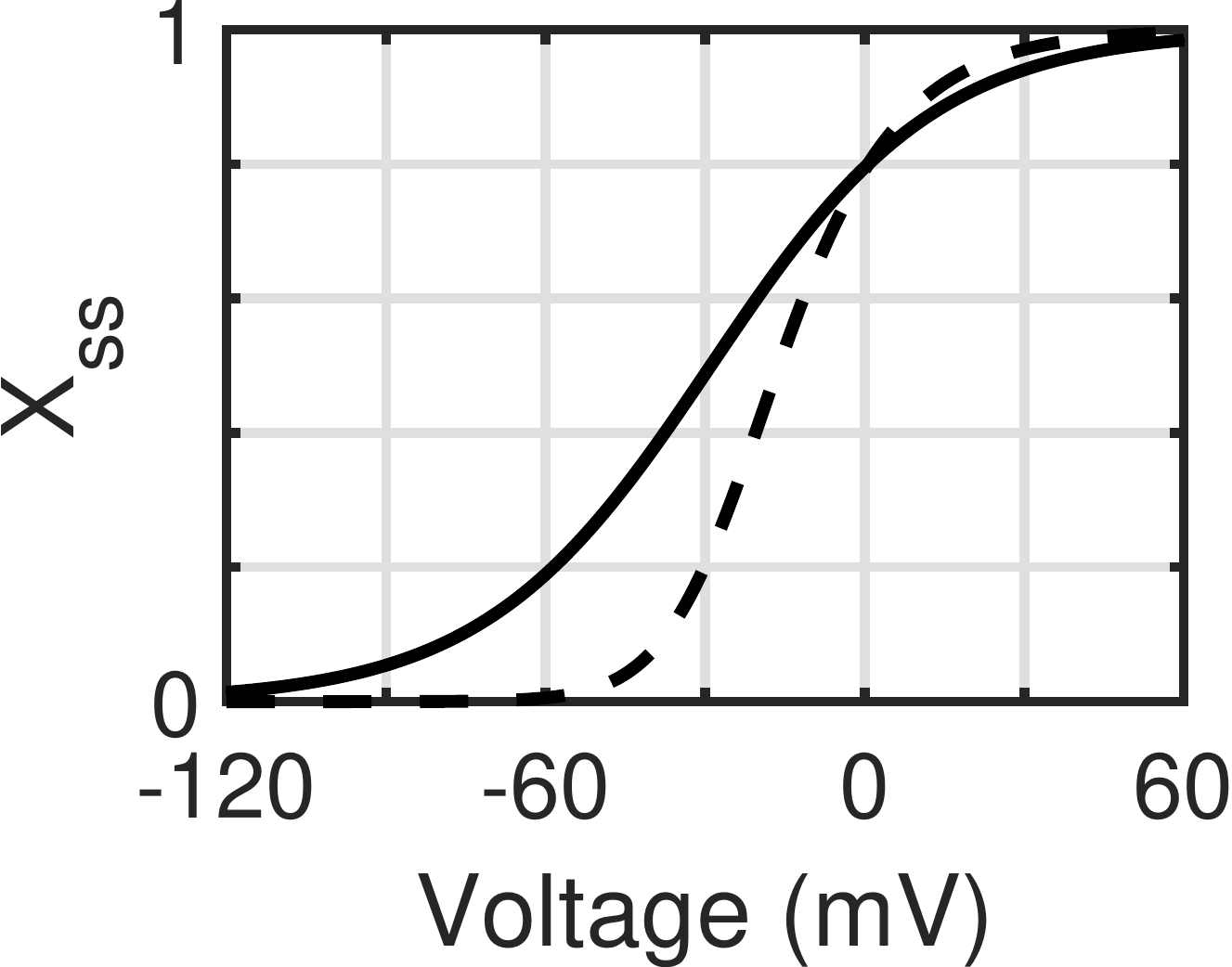} & 
				\includegraphics[width=0.2\linewidth]{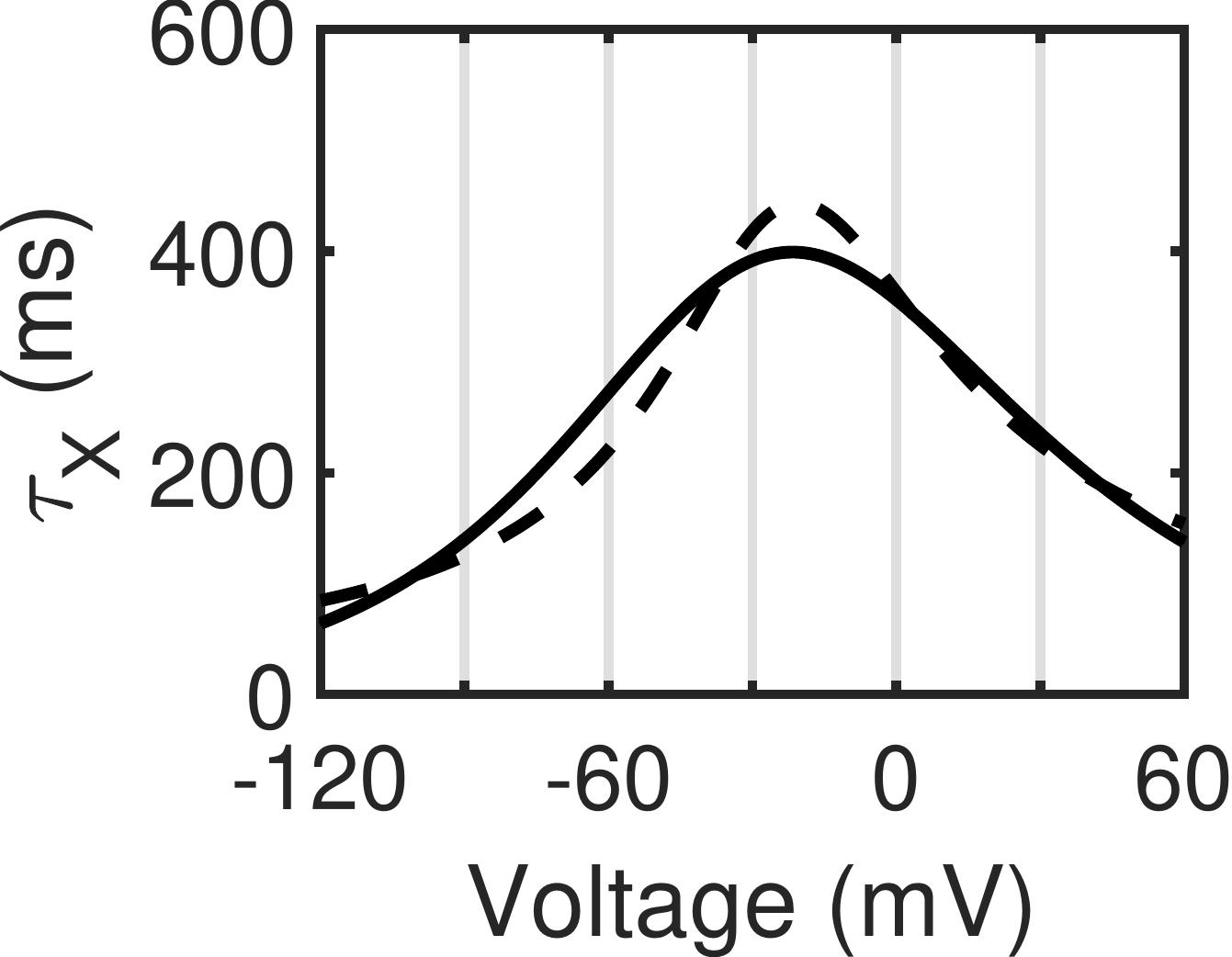} \\
				\multicolumn{2}{c}{\textbf{(F)} Inactivation gate ($\mathrm{X_i}$)} \\
				\includegraphics[width=0.2\linewidth]{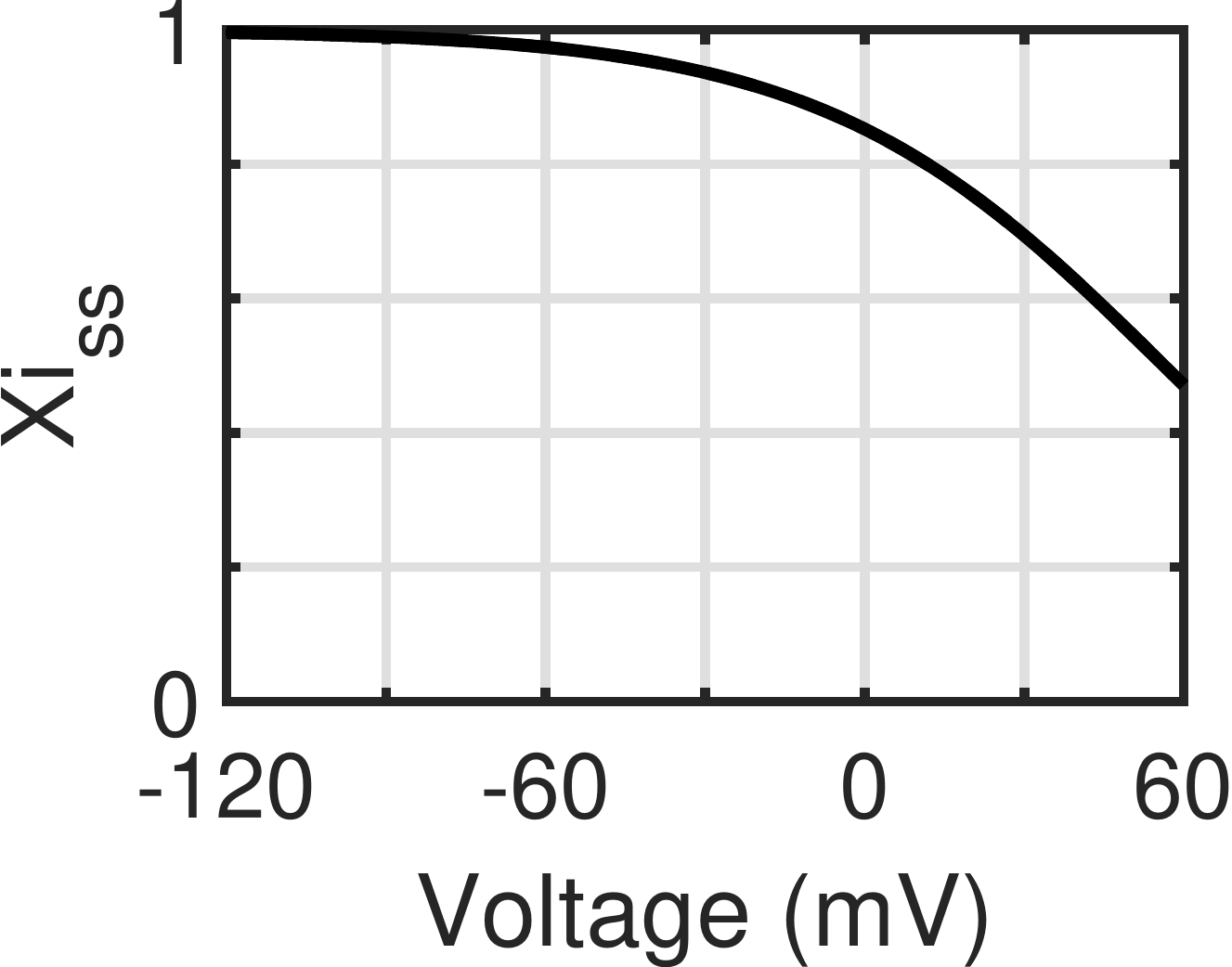} & 
				\includegraphics[width=0.2\linewidth]{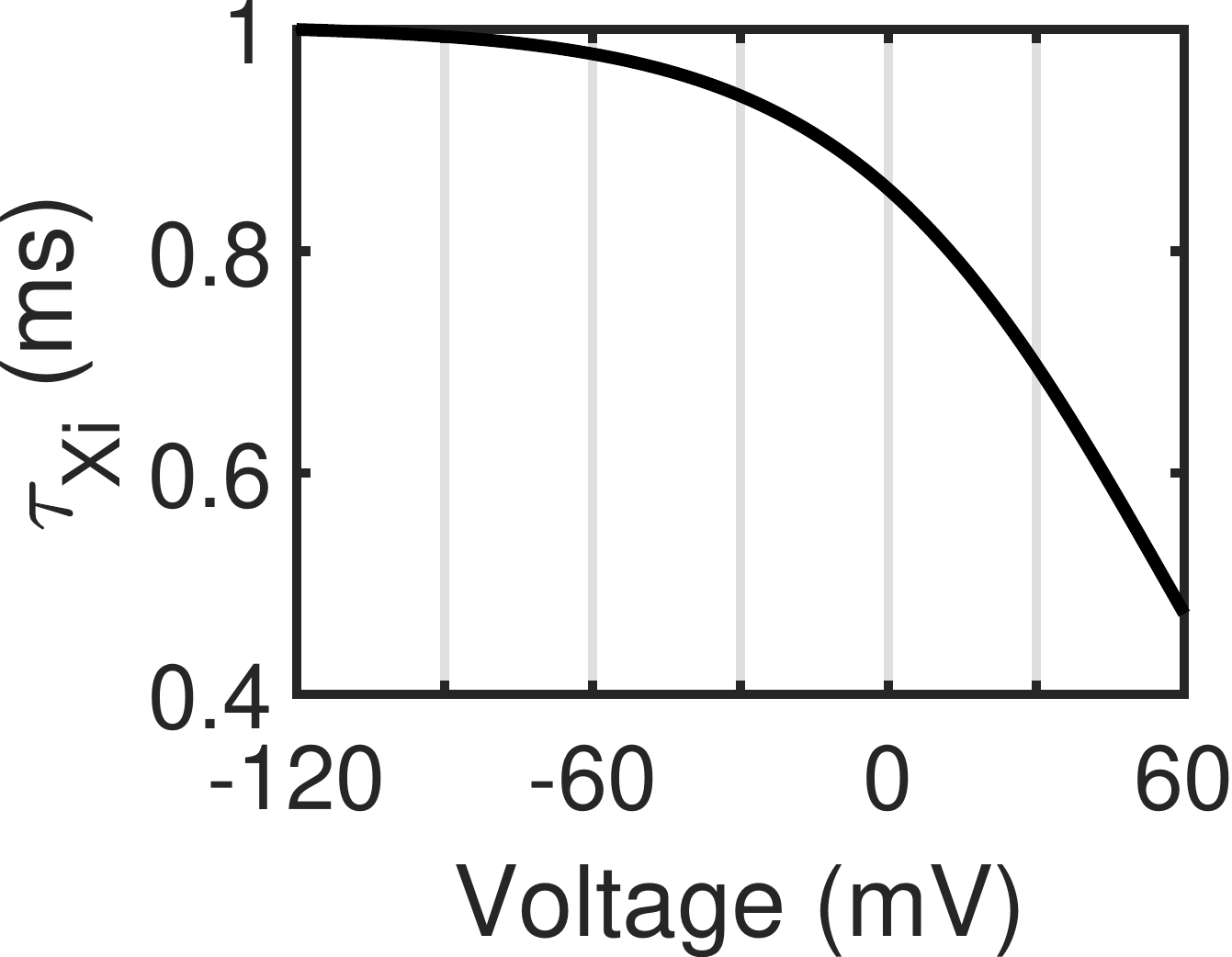} \\[0.5cm]
				\multicolumn{2}{c}{\Large Plateau $\mathrm{K^+}$ current} \\
				\multicolumn{2}{c}{\textbf{(G)} Kp gate} \\
				\includegraphics[width=0.2\linewidth]{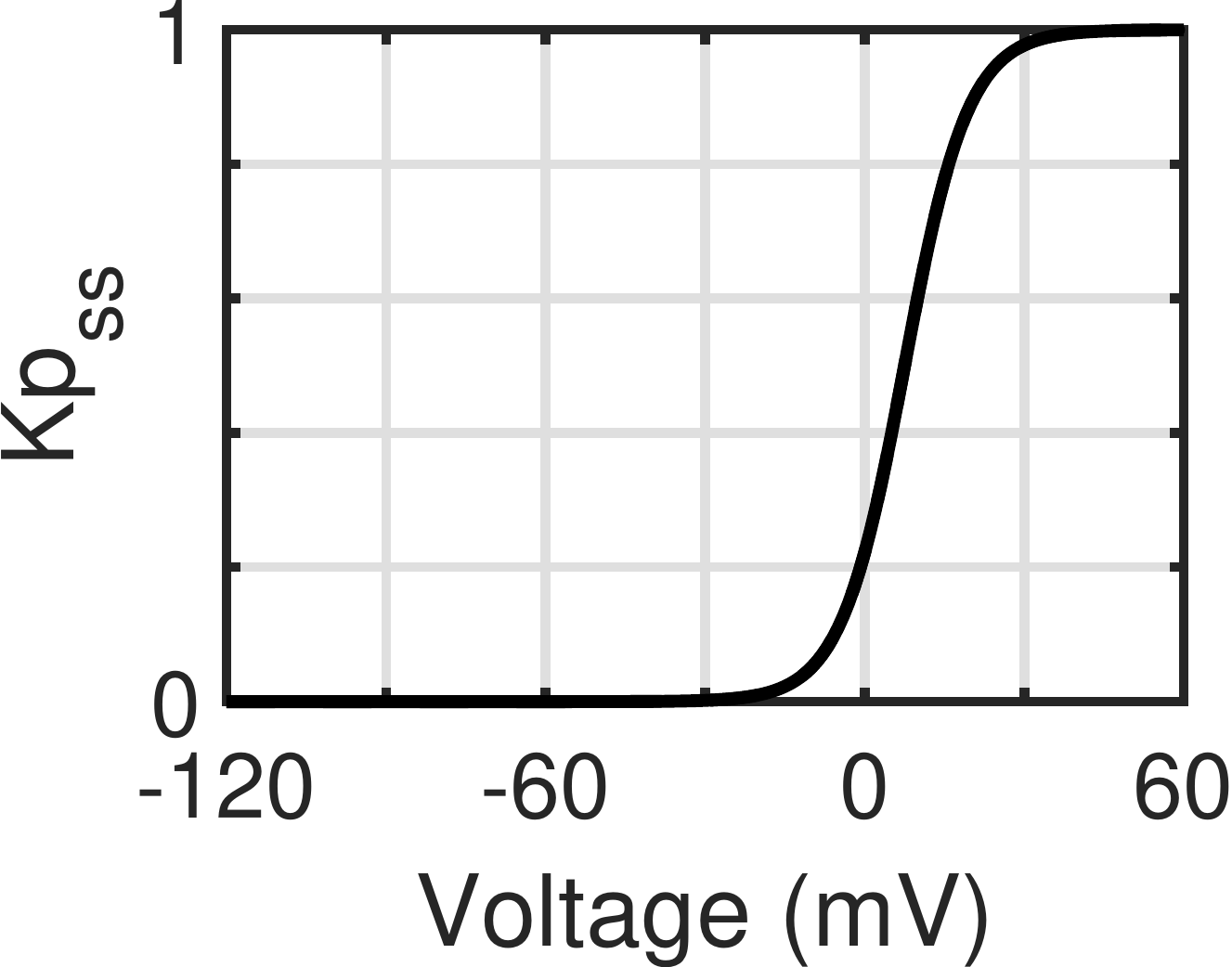} & 
				\includegraphics[width=0.2\linewidth]{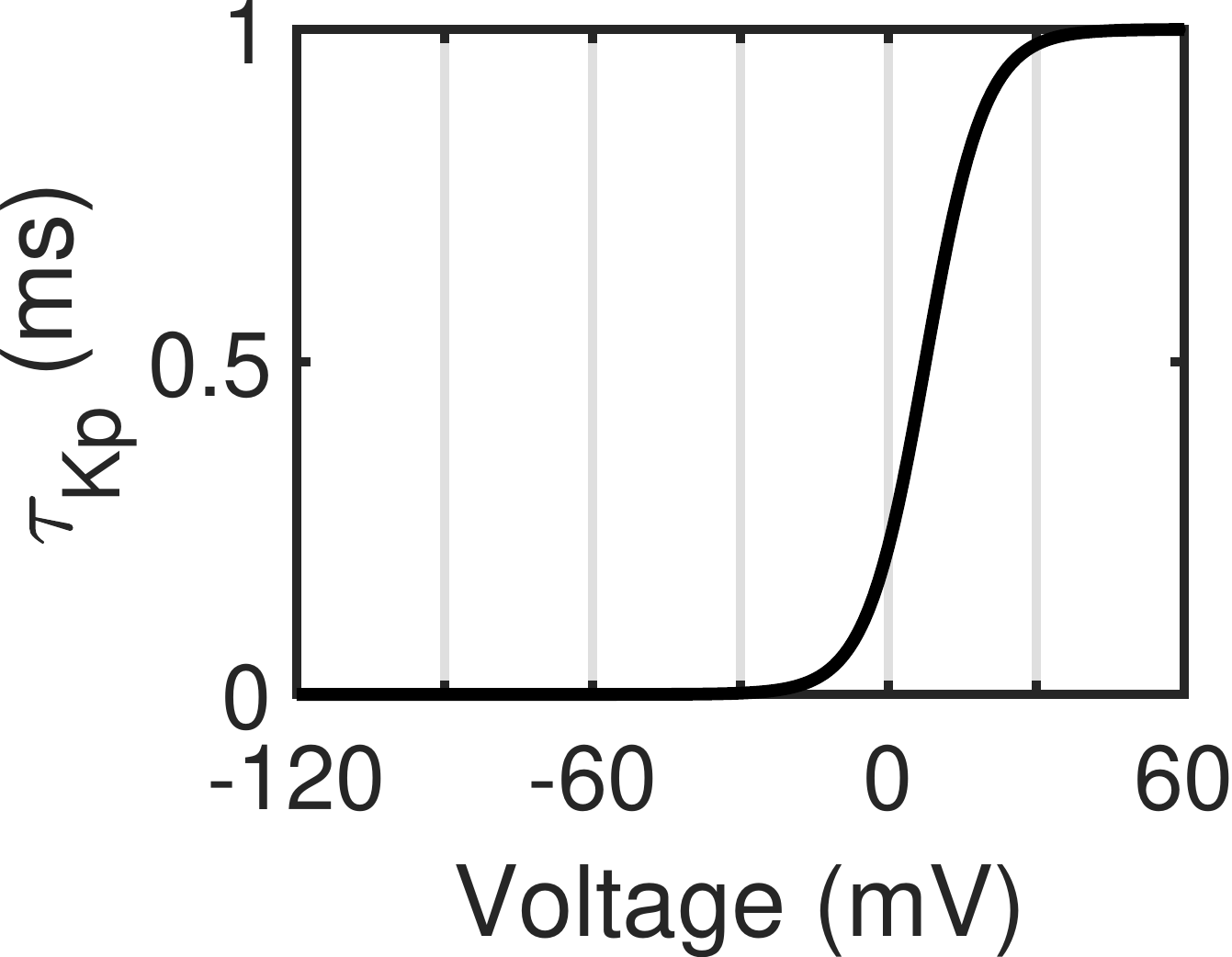} \\[0.5cm]
				\multicolumn{2}{c}{\Large L-type Ca\textsuperscript{2+} channel} \\
				\multicolumn{2}{c}{\textbf{(H)} $d$-gate} \\
				\includegraphics[width=0.2\linewidth]{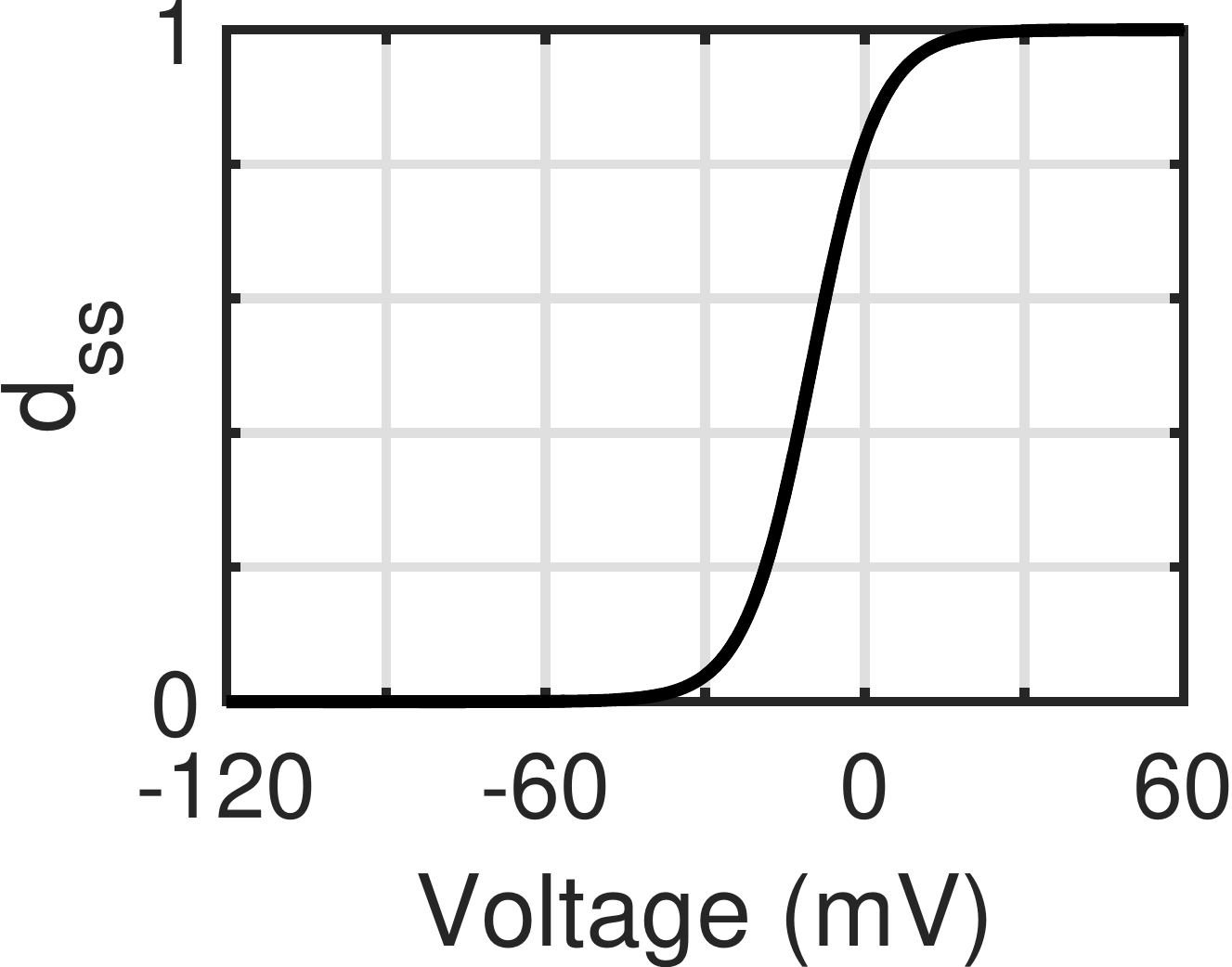} & 
				\includegraphics[width=0.2\linewidth]{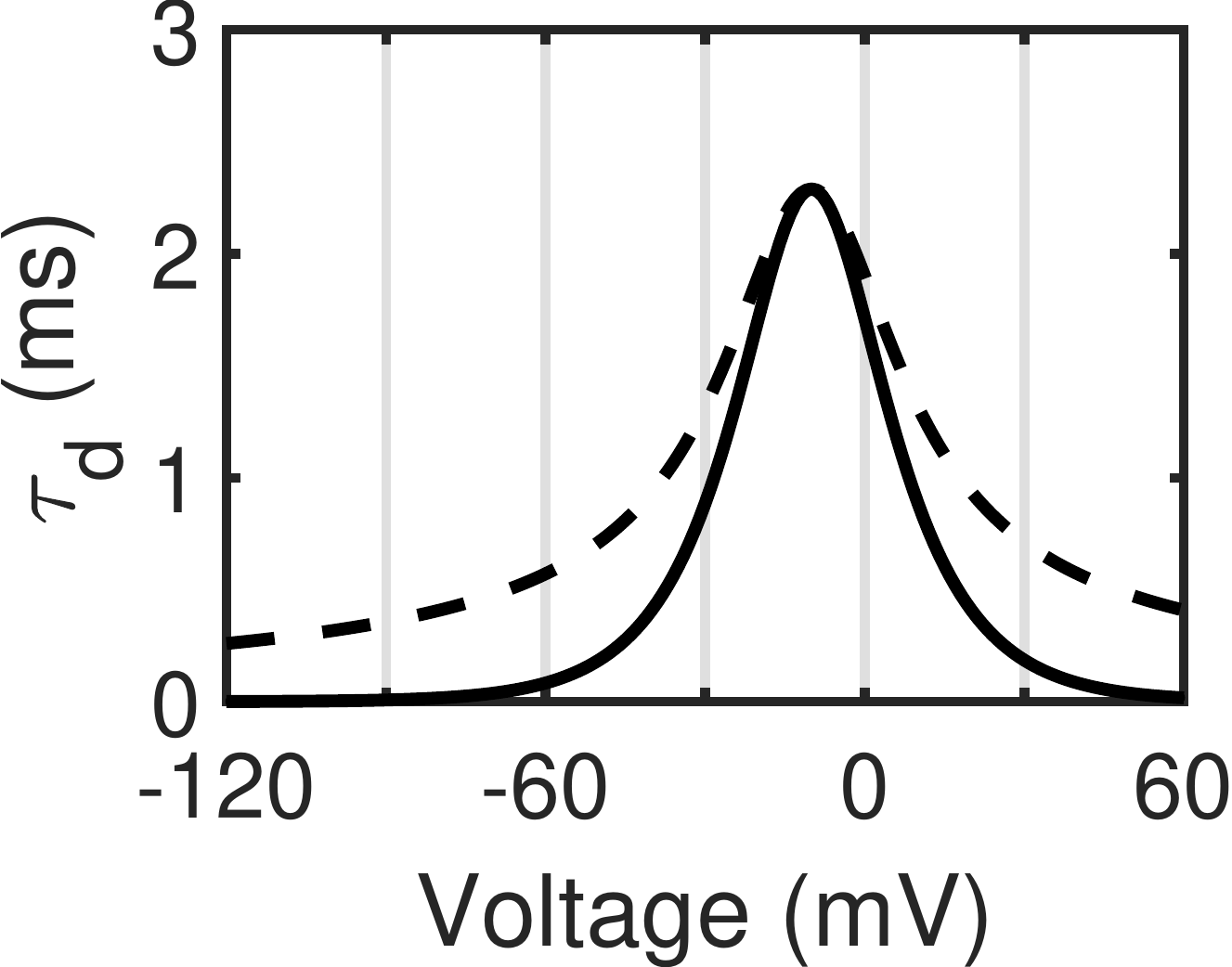} \\
		\end{tabular}}
	\end{tabular}
	\caption{\textbf{Fits for bond graph (BG) parameters against corresponding gating equations from the Luo-Rudy (LRd) model.} Steady-state open probabilities are shown on left panels, and time constants are shown on the right. The bond graph equations are plotted with solid lines, and the Luo and Rudy equations in dashed lines. Gates include \textbf{(A)} $m$, sodium activation; \textbf{(B)} $h$, sodium inactivation; \textbf{(C)} $j$, slow sodium inactivation; \textbf{(D)} K1, time-independent K\textsuperscript{+} activation; \textbf{(E)} X, time-dependent K\textsuperscript{+} activation; \textbf{(F)} $\mathrm{X_i}$, time-dependent K\textsuperscript{+} inactivation; \textbf{(G)} Kp, plateau K\textsuperscript{+} activation; \textbf{(H)} $d$, L-type Ca\textsuperscript{2+} channel activation. Note that the $\mathrm{X_i}$ and Kp gates were originally formulated as steady-state equations, thus time constants are shown only for matched bond graph parameters.}
	\label{fig:gating_fit}
\end{figure}

\subsubsection{Model comparison}
To assess the quality of fit we compare steady-state open probabilities $g_\mathrm{ss} = \alpha(V)/(\alpha(V)+\beta(V))$ and time constants $\tau =1/(\alpha(V)+\beta(V))$ (\autoref{fig:gating_fit}). The curves for $g_{ss}$ and $\tau$ were generally in agreement however there were some exceptions. In particular, time constants for the Na\textsuperscript{+} channel gates have lower peaks in the bond graph model when compared to the Luo-Rudy model (\autoref{fig:gating_fit}A--C), but this did not appear to significantly affect Na\textsuperscript{+} channel function as the peaks were all decreased by a similar proportion, facilitating coordination between opening and closing. Similarly, the time constant $\tau_d$ (\autoref{fig:gating_fit}H) was lower in the bond graph model for some voltages, but given that discrepancies occur at time constants much smaller than the time course of a cardiac action potential we expect that the effects would be negligible. Finally, for the time-dependent K\textsuperscript{+} current $X_\text{ss}$ is substantially higher at negative voltages so that the bond graph model can provide a better match at positive voltages (\autoref{fig:gating_fit}E). The effects of this difference are partially offset by the lower GHK current at negative voltages which are still above the Nernst potential of K\textsuperscript{+} (\autoref{fig:IV_fitting}C).

\section{Ion transporters}
\subsection{Na$^+$/K$^+$ ATPase}
We used the 15-state bond graph model described in Pan \textit{et al.} \cite{pan_cardiac_2017}, with a pump density of 4625 $\si{{\micro}m^{-2}}$ (0.1178 fmol per cell).

\subsection{Na\textsuperscript{+}-Ca\textsuperscript{2+} exchanger}
\begin{figure}
	\centering
	\includegraphics[width=0.8\linewidth]{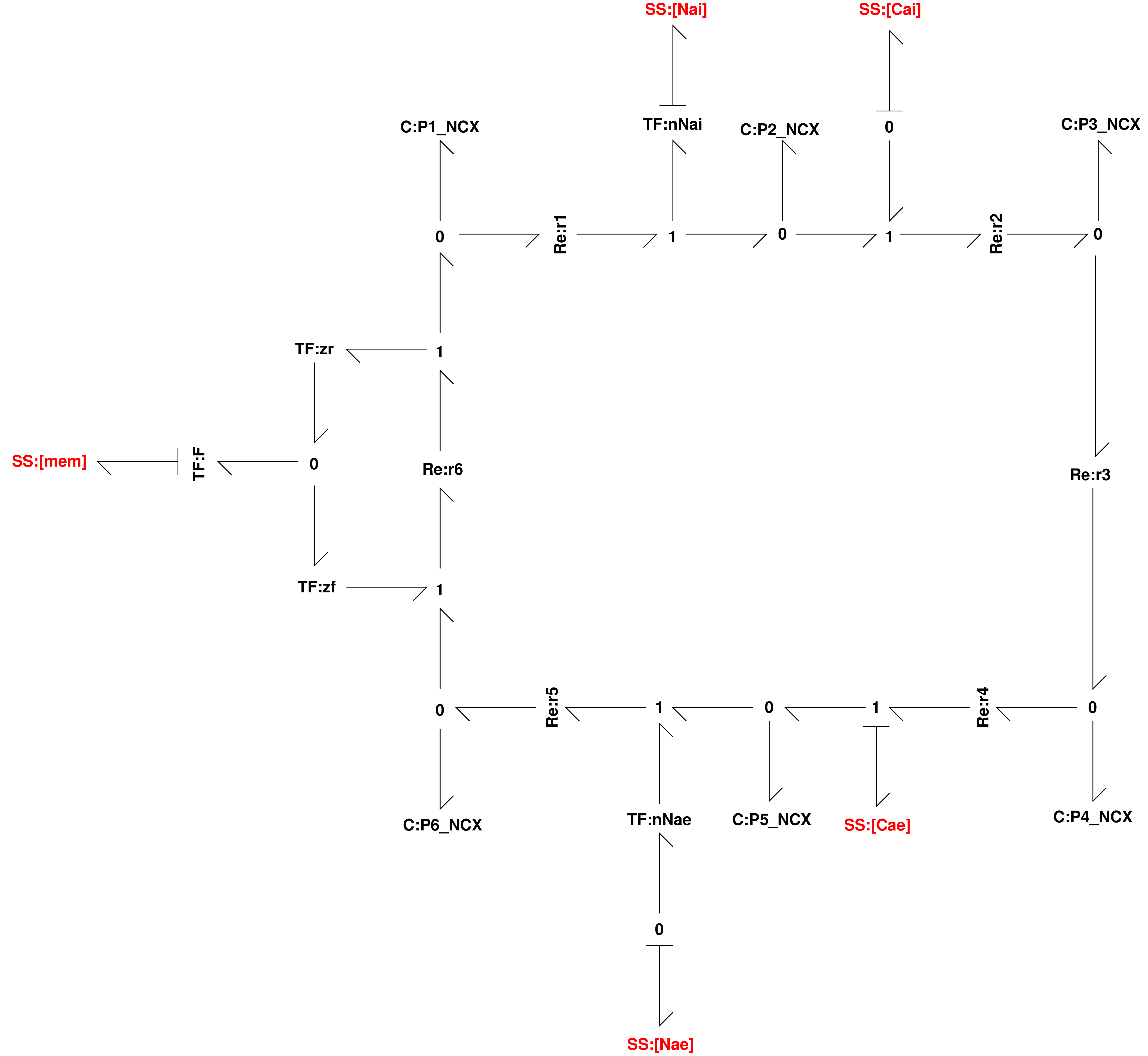}
	\caption{\textbf{The bond graph model of NCX.}}
	\label{fig:NCX_bg}
\end{figure}

The NCX was modelled using the bond graph shown in \autoref{fig:NCX_bg}. The reaction scheme was based on the ping-pong mechanism proposed in Giladi \textit{et al.} \cite{giladi_structure-functional_2016}, with reactions r1, r2, r4 and r5 modelled by fast rate constants to approximate rapid equilibrium. We assigned voltage dependence to translocation of Na\textsuperscript{+}, based on experimental findings from Hilgemann \textit{et al.} \cite{hilgemann_steady-state_1992-1}.

Using similar methods to Luo and Rudy \cite{luo_dynamic_1994}, the NCX model was fitted to the following data, assuming steady-state operation:
\begin{enumerate}
	\item Dependence of cycling rate on extracellular Na\textsuperscript{+} and voltage, from Kimura \textit{et al.} \cite{kimura_identification_1987}.
	\item Dependence of cycling rate on extracellular Ca\textsuperscript{2+}, from Kimura \textit{et al.} \cite{kimura_identification_1987}. Data obtained at $V < -50\ \si{mV}$ and $\mathrm{[Ca^{2+}]_e} = 1\ \si{mM}$ were excluded from the fitting process.
	\item To incorporate behaviour for another intracellular Ca\textsuperscript{2+} concentration, data from Beuckelmann and Wier \cite{beuckelmann_sodium-calcium_1989} were used. Data obtained at $V < -120\ \si{mV}$ were excluded from the fitting process.
\end{enumerate}

\begin{figure}
	\centering
	\begin{tabular}{c c c}
		{\large \textbf{\textsf{A}}} &
		{\large \textbf{\textsf{B}}} &
		{\large \textbf{\textsf{C}}}  \\
		\includegraphics[width=0.3\linewidth]{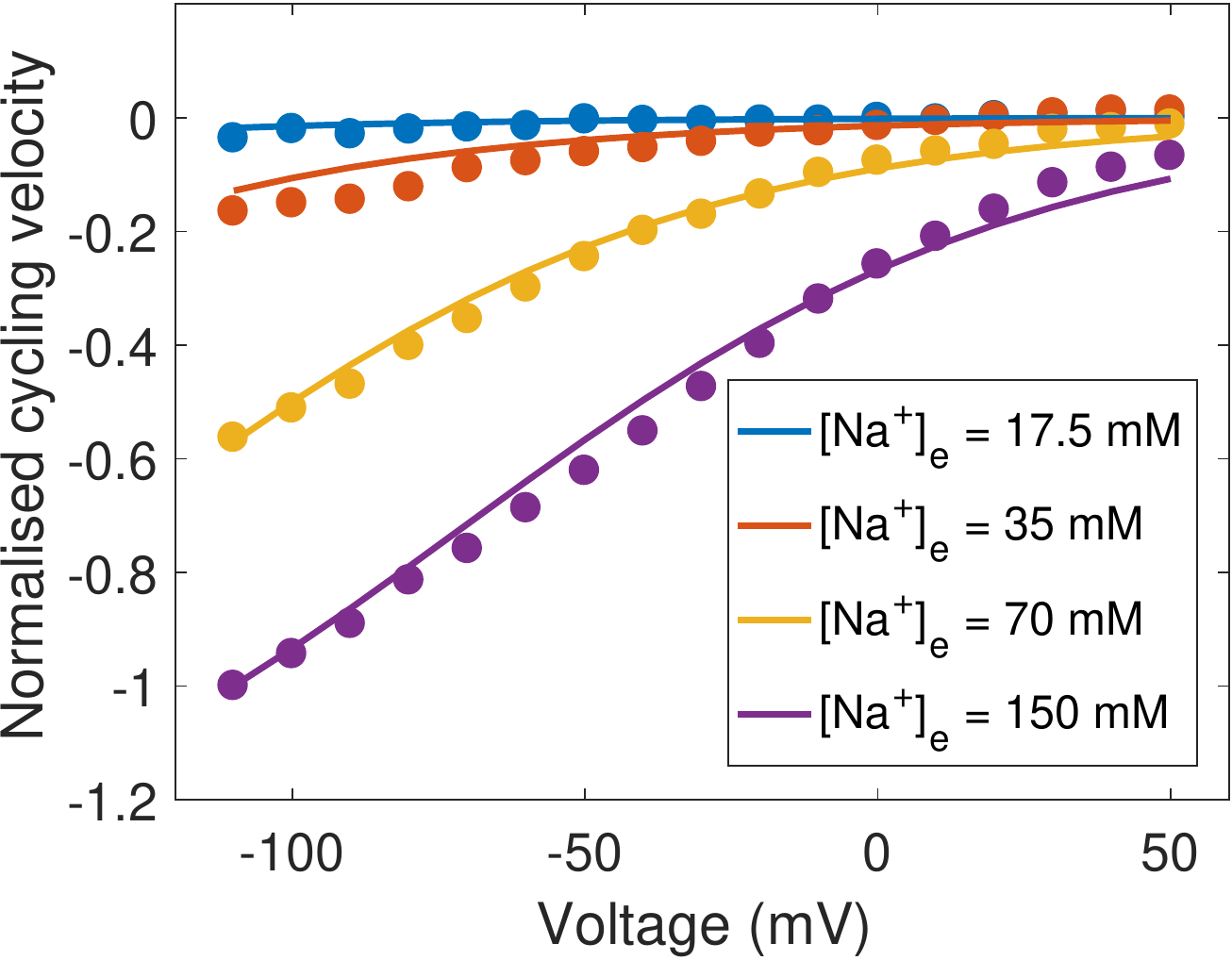} &
		\includegraphics[width=0.3\linewidth]{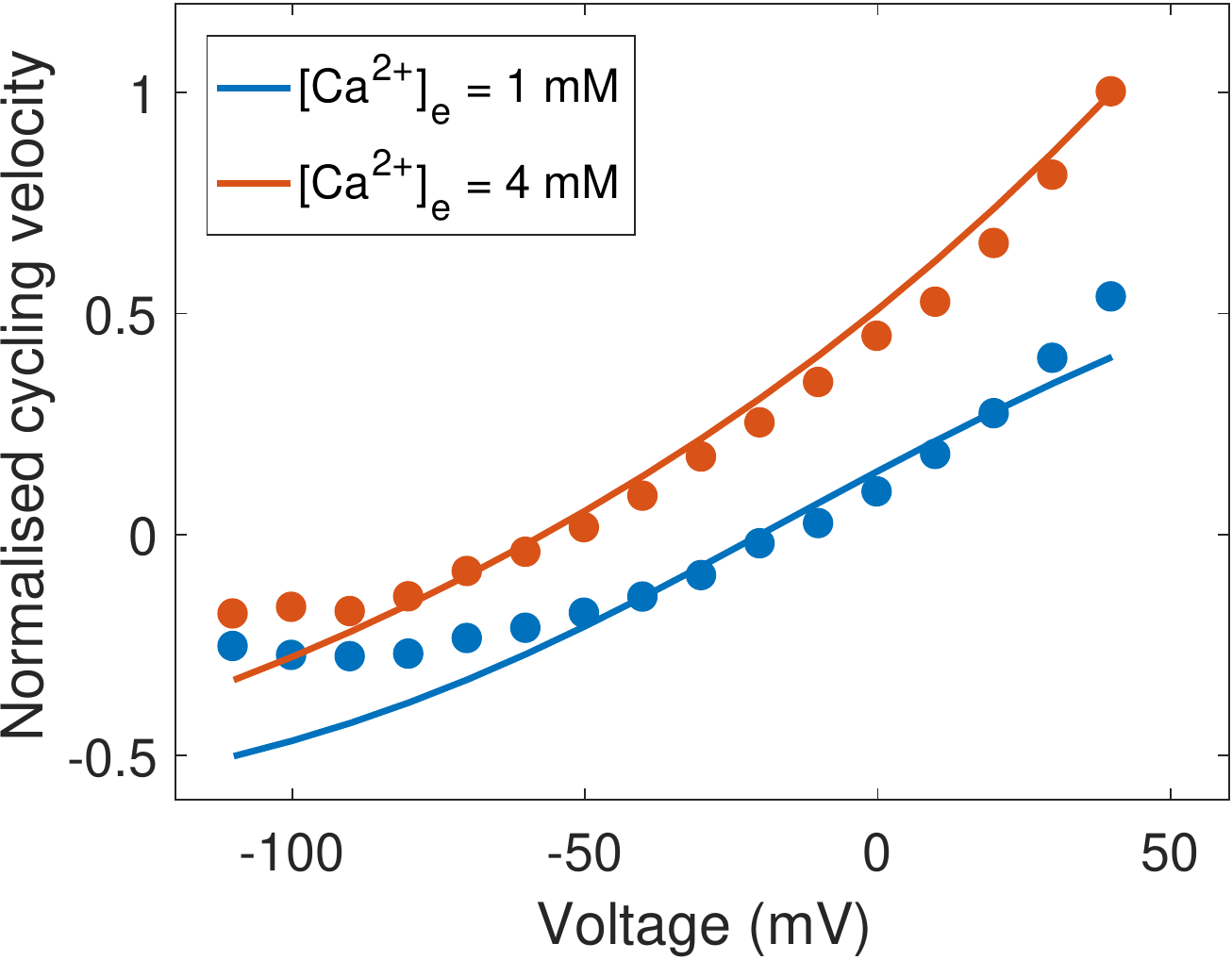} &
		\includegraphics[width=0.3\linewidth]{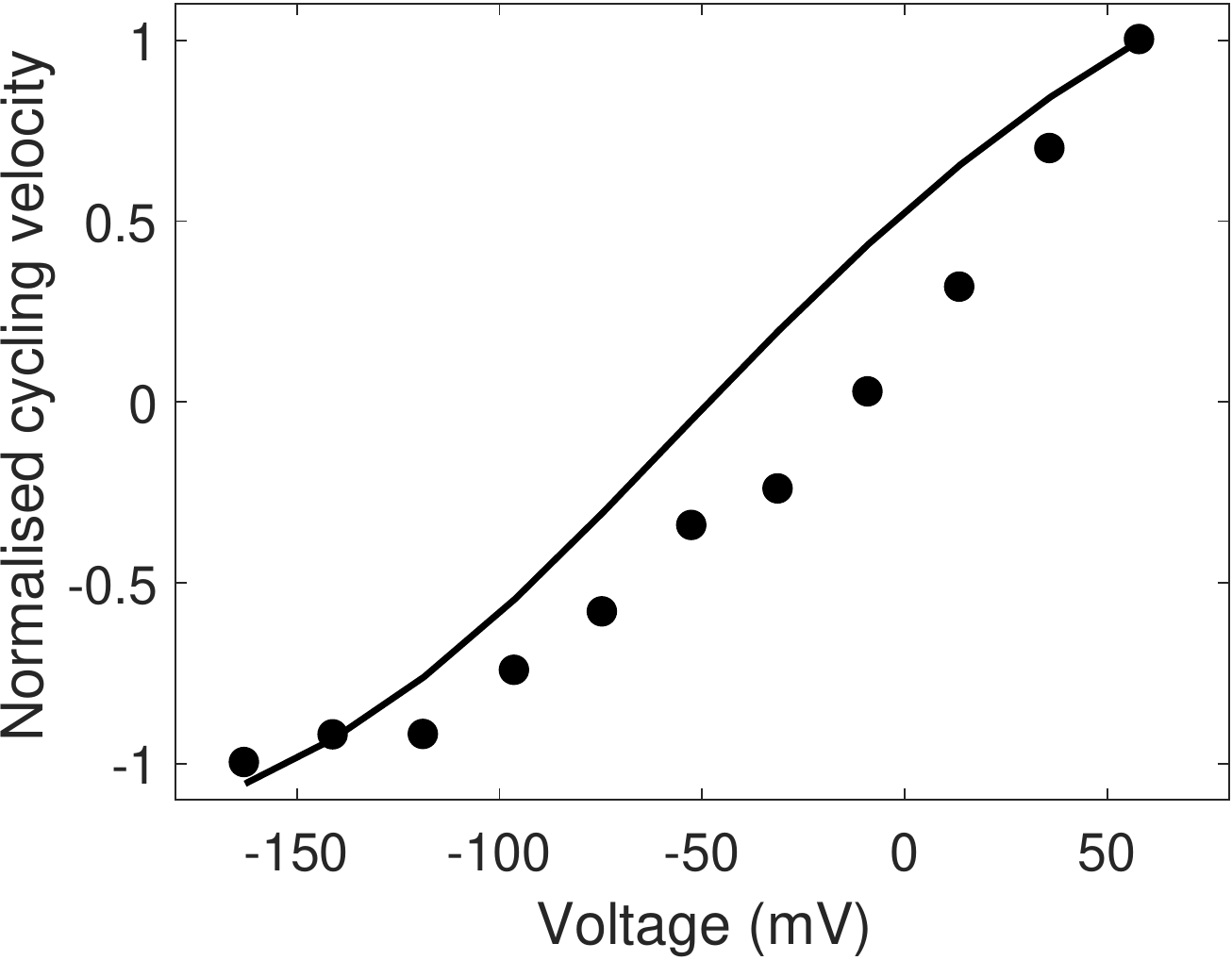} 
	\end{tabular}
	\caption{\textbf{Fit of NCX model to data.} \textbf{(A)} Comparison of model to Fig. 8B of Kimura \textit{et al.} \cite{kimura_identification_1987}. Fluxes were normalised to the value at $\mathrm{[Na^+]_e} = 140\ \si{mM}$ and $V = -110\ \si{mV}$. $\mathrm{[Na^+]_i} = 0\ \si{mM}$, $\mathrm{[Ca^{2+}]_e} = 1\ \si{mM}$, $\mathrm{[Ca^{2+}]_i} = 430\ \si{nM}$. \textbf{(B)} Comparison of model to Fig. 9A of Kimura \textit{et al.} \cite{kimura_identification_1987}. Fluxes were normalised to the value at $\mathrm{[Ca^{2+}]_e} = 4\ \si{mM}$ and $V = 40\ \si{mV}$. $\mathrm{[Na^+]_e} = 140\ \si{mM}$, $\mathrm{[Na^+]_i} = 10\ \si{mM}$, $\mathrm{[Ca^{2+}]_i} = 172\ \si{nM}$. \textbf{(C)} Comparison of model to Fig. 6B of Beuckelmann and Wier \cite{beuckelmann_sodium-calcium_1989}. Fluxes were normalised to the value at $V = 60\ \si{mV}$. $\mathrm{[Na^+]_e} = 135\ \si{mM}$, $\mathrm{[Na^+]_i} = 15\ \si{mM}$, $\mathrm{[Ca^{2+}]_e} = 2\ \si{mM}$, $\mathrm{[Ca^{2+}]_i} = 450\ \si{nM}$.}
	\label{fig:NCX}
\end{figure}

Parameters of the model were identified using particle swarm optimisation followed by a local optimiser, and a comparison between the model and data is shown in \autoref{fig:NCX}. The model closely matched the data describing extracellular Na\textsuperscript{+} dependence (\autoref{fig:NCX}A). Reasonable fits were obtained for the other data, although there was some discrepancy at negative voltages in \autoref{fig:NCX}B. There was some difference between the model and data from Beuckelmann and Wier \cite{beuckelmann_sodium-calcium_1989} (\autoref{fig:NCX}C), although this appears to have resulted from differences in the equilibrium point.

The cycling velocity was normalised to 700 $\si{s^{-1}}$ at the normalisation point of \autoref{fig:NCX}A to approximately match experimental currents at a membrane capacitance of 200 pF and 300 sites per $\si{{\micro}m^{-2}}$. To ensure that the exchanger current had a similar magnitude to that of Luo and Rudy \cite{luo_dynamic_1994}, we used a site density of 170 $\si{{\micro}m^{-2}}$ (0.0043 fmol per cell) in our cardiac action potential model.

\section{Ca\textsuperscript{2+} buffering}
The model of Ca\textsuperscript{2+} buffering was based on the equations described in Luo and Rudy \cite{luo_dynamic_1994}. These equations represent the reactions
\begin{align}
\mathrm{TRPN + Ca_i^{2+} \rlh TRPNCa} \\
\mathrm{CMDN + Ca_i^{2+} \rlh CMDNCa}
\end{align}
with the dissociation constants $K_\text{d,TRPN} = 0.5\ \si{{\micro}M}$ and $K_\text{d,CMDN} = 2.38\ \si{{\micro}M}$. The total concentrations of each buffer were 70 $\si{{\micro}M}$ for troponin and 50 $\si{{\micro}M}$ for calmodulin. The reactions were modelled using sufficiently fast kinetic constants to approximate rapid equilibrium, and the amount of Ca\textsuperscript{2+} bound to each buffer was initialised to the value at equilibrium for the initial intracellular Ca\textsuperscript{2+} concentration of 0.12 $\si{{\micro}M}$.

\section{Bond graph parameters}
\subsection{Calculating bond graph parameters}
\label{sec:bg_params}
Bond graph parameters were found by using an extension of the method presented in Gawthrop \textit{et al.} \cite{gawthrop_hierarchical_2015}. The kinetic parameters and bond graph parameters can be related through the matrix equation
\begin{align}
\textbf{Ln}(\mathbf{k}) = \mathbf{M} \textbf{Ln}(\mathbf{W} \boldsymbol{\lambda}) \label{eq:bg_general}
\end{align}
where
\begin{align}
\textbf{k} = \begin{bmatrix}
k^+ \\ k^-
\end{bmatrix}, \quad
\textbf{M} = \left[ \begin{array}{c | c}
I_{n_r \times n_r} & {N^f}^T \\ \hline
I_{n_r \times n_r} & {N^r}^T
\end{array} \right], \quad
\boldsymbol{\lambda} = \begin{bmatrix}
\kappa \\ K
\end{bmatrix}
\label{eq:standard}
\end{align}
$k^+$ is a column vector consisting of the forward kinetic constants, $k^-$ is a column vector consisting of the reverse kinetic constants, $N^f$ and $N^r$ are the forward and reverse stoichiometric matrices respectively, $\kappa$ is a column vector of bond graph reaction rate constants, and $K$ is column vector of thermodynamic constants. To account for the volumes of each compartment, $\mathbf{W}$ is a diagonal matrix where the $i$-th diagonal element is the volume corresponding to $i$-th bond graph component (either a reaction or species). Depending on compartment, the elements corresponding to each ion were set to either the intracellular volume of  $W_i = 38\ \si{pL}$ or the extracellular volume of $W_e = 5.182\ \si{pL}$. All other diagonal entries were set to 1. Assuming that detailed balance constraints are satisfied, a solution to Eq. \ref{eq:bg_general} is
\begin{align}
\boldsymbol{\lambda_0 } = \mathbf{W}^{-1} \textbf{Exp} (\mathbf{M}^\dagger \textbf{Ln} (\mathbf{k}))
\end{align}
where $\mathbf{M}^\dagger$ is the pseudo-inverse of $\mathbf{M}$. All parameters were identified using $T = 310\ \si{K}$.

For reactions involved in ion transport that use the GHK equation, both the forward and reverse rate constants were set to $P/x_\text{ch}$, where $P$ is the permeability constant found by fitting to Eq. \ref{eq:GHK}, and $x_\text{ch}$ is the total number of channels. The values of $x_\text{ch}$ used for each channel are given in \autoref{tab:x_ch}. Since the bond graph paramters of the NCX model were fitted to kinetic data, the bond graph paramters were converted back to kinetic paramters \cite{gawthrop_hierarchical_2015} to parameterise the action potential model.

\begin{table}[H]
	\caption{\textbf{Amounts of each ion channel per cell.} A geometric area of $0.767 \times 10^{-4}\ \si{cm^2}$ was used to convert between channel density and channels per cell ($x_\text{ch}$). \\
		*Quantity cited from reference.}
	\centering
	\begin{tabular}{l c c c}
		\toprule
		Ion channel & Channel density ($\si{\micro m^{-2}}$) & Channels per cell & Reference \\ \midrule
		Na 	& $16^\text{*}$ 	& 122720 	& \cite{reuter_ion_1984} \\
		K1 	& $1.8^\text{*}$ 	& 4261 		& \cite{sakmann_conductance_1984} \\
		K 	& $0.7^\text{*}$ 	& 5369 		& \cite{shibasaki_conductance_1987} \\
		Kp 	& $0.095$ 	& $725^\text{*}$ 	& \cite{yue_characterization_1996} \\
		LCC & $6.5$ 	& $50000^\text{*}$ & \cite{hinch_simplified_2004} \\ \bottomrule
	\end{tabular} \\[0.1cm]
	
	\label{tab:x_ch}
\end{table}

\section{Charge conserved moiety}
In \autoref{tab:cm} of the main text, $\Sigma$ is defined as
\begin{align}
\Sigma = &+ 3.0818\mathrm{C_{K1}} - 1.6697\mathrm{S_{00,K}} -0.4188\mathrm{S_{10,K}} + 0.8322\mathrm{S_{20,K}} -2.5019\mathrm{S_{01,K}} \notag \\
&-1.2509\mathrm{S_{11,K}} -4.4669\mathrm{C_{Kp}} + 2.1835\mathrm{S_{000,Na}} + 5.1073\mathrm{S_{100,Na}} + 8.0311\mathrm{S_{200,Na}} \notag \\
& + 10.9549\mathrm{S_{300,Na}} -3.3052\mathrm{S_{010,Na}} -0.3814\mathrm{S_{110,Na}} + 2.5424\mathrm{S_{210,Na}} + 5.4662\mathrm{S_{310,Na}} \notag \\ 
&-3.2827\mathrm{S_{001,Na}} -0.3589\mathrm{S_{101,Na}} + 2.5649\mathrm{S_{201,Na}} + 5.4887\mathrm{S_{301,Na}} -8.7714\mathrm{S_{011,Na}} \notag \\ 
& -5.8476\mathrm{S_{111,Na}} -2.9238\mathrm{S_{211,Na}} -1.5253\mathrm{S_{000,LCC}} -4.5742\mathrm{S_{010,LCC}} -0.2808\mathrm{S_{020,LCC}}  \notag \\ 
&+ 2.7555\mathrm{S_{100,LCC}} -0.2933\mathrm{S_{110,LCC}} +4\mathrm{S_{120,LCC}} -5.5253\mathrm{S_{001,LCC}} -8.5742\mathrm{S_{011,LCC}}  \notag \\ 
&-4.2808\mathrm{S_{021,LCC}} -1.2445\mathrm{S_{101,LCC}} -4.2933\mathrm{S_{111,LCC}} + \mathrm{P2_{NaK} + P3_{NaK} + P4_{NaK}} \notag \\ 
& \mathrm{-0.9450P6_{NaK} -0.9450P7_{NaK} -0.9450P8_{NaK} - P1_{NCX}  + 2P2_{NCX}}
\end{align}

\end{document}